\documentclass[letter,preprint,superscriptaddress,nofootinbib,longbibliography]{revtex4-1}
\usepackage[dvipdfmx]{graphicx}
\usepackage{bm}
\usepackage{amssymb}
\usepackage{amsmath}
\allowdisplaybreaks
\usepackage{color}
\usepackage{cancel}
\usepackage{comment}
\usepackage{here}
\usepackage{hyperref}
\usepackage{subfigure}
\usepackage{multirow}
\usepackage[margin=1in, footskip=1.5cm]{geometry}
\usepackage{url}

\begin{document} 

\title{Improved analyses of the electroweak phase transition and its phenomenology in the Georgi--Machacek model}

\author{Cheng-Wei Chiang}
\email{chengwei@phys.ntu.edu.tw}
\affiliation{Department of Physics and Center for Theoretical Physics, National Taiwan University, Taipei, 10617, Taiwan}
\affiliation{Physics Division, National Center for Theoretical Sciences, Taipei, 10617, Taiwan}

\author{Kazuki Enomoto}
\email{k\_enomoto@phys.ntu.edu.tw}
\affiliation{Department of Physics and Center for Theoretical Physics, National Taiwan University, Taipei, 10617, Taiwan}

\begin{abstract}
The Georgi--Machacek model is a promising new physics scenario with isospin-triplet scalar fields, whose effects can qualitatively change the nature of the electroweak phase transition compared with the Standard Model. 
We examine the electroweak phase transition in this model by using the thermally resummed one-loop effective potential.
We also incorporate effects of custodial symmetry breaking due to radiative corrections by evolving the couplings and classical fields according to the renormalization group equations. 
We identify viable parameter regions by Bayesian analysis using theoretical requirements and the latest experimental bounds. 
We then investigate the electroweak phase transition in the resulting parameter regions and discuss its phenomenological implications, such as Higgs boson decays into neutral gauge bosons, di-Higgs production at high-energy colliders, and stochastic gravitational waves associated with a first-order electroweak phase transition.
We find that parameter points for the strong first-order electroweak phase transition can be verified by these observables at next-generation experiments.
\end{abstract}

\maketitle

\section{Introduction}

Although the Standard Model (SM) describes the behavior of elementary particles well, physics beyond the SM (BSM) must exist because of its theoretical incompleteness and the lack of ability to explain some observed phenomena in particle physics and particle cosmology~\cite{PDG2026}. 
Considering BSM, there is plenty of room to extend the Higgs sector because there is no principle to determine it. 
Possibilities of the extended Higgs sectors have been extensively studied so far, for example, additional scalar fields in the isospin singlet, doublet, triplet, and higher representations.

Among additional scalar fields, the isospin triplet with hypercharge $Y=1$ is especially attractive because its vacuum expectation value (VEV) generates neutrino masses via the Type-II seesaw mechanism~\cite{type-II_seesaw}. 
The triplet VEV also induces several other BSM phenomena; however, its size is tightly constrained by electroweak (EW) precision measurements because it breaks the custodial symmetry at tree level. 

The Georgi--Machacek (GM) model~\cite{GM_model} was proposed as a model including the triplets with relatively large VEVs compatible with the results of the EW precision measurements. 
This model includes two triplet fields: $\chi$ with $Y=1$ and $\xi$ with $Y=0$, which form a bi-triplet under the custodial symmetry. 
By assuming the alignment between the triplet VEVs, the Higgs potential respects the custodial symmetry at tree level even after the EW symmetry is spontaneously broken.
Then, the additional scalar bosons are classified into multiplets of the custodial symmetry: one quintet, one triplet, and two singlets, where one of the singlets is the observed 125 GeV Higgs boson. 

Effects of these additional scalar bosons can significantly change the nature of the EW phase transition (EWPT) in the early Universe unless they are too heavy and decouple at the EW scale. 
In Refs.~\cite{Chiang:2014hia, Zhou:2018zli, Bian:2019bsn, Chen:2022zsh}, it has been shown that the EWPT can be strongly first-order, which is phenomenologically favorable, in some parameter regions of the GM model. 
In Refs.~\cite{Garcia-Pepin:2016hvs} and \cite{Lu:2025vif}, supersymmetric and $Z_2$-symmetric extensions of the GM model have been investigated, respectively, and it has been shown that the strong first-order EWPT is also viable in these models. 

However, there exist several points to be improved in the previous analyses of the EWPT in the GM model~\cite{Chiang:2014hia, Zhou:2018zli, Bian:2019bsn, Chen:2022zsh}. 
In some of them~\cite{Zhou:2018zli, Bian:2019bsn, Chen:2022zsh}, only part of the thermal effect is incorporated, and the full one-loop thermal potential is not considered. 
Ref.~\cite{Chiang:2014hia} uses the full one-loop thermal potential; however, thermal corrections are not resummed, although they have a significant effect in the phase transition study~\cite{Dolan:1973qd}. 

In addition, regarding the renormalization of the potential, it is remarkable that custodial-symmetry-breaking counterterms are necessary to absorb all divergences in one-loop corrections in addition to custodial symmetric ones because the symmetry is broken by the gauge and Yukawa interactions~\cite{Chiang:2017vvo, Chiang:2018xpl}. 
Then, even if we assume the symmetric potential at a certain scale, it is no longer valid at other scales. 
This effect should also be included in the EWPT study in the GM model. 

In this paper, we improve the EWPT study in the GM model by addressing the above issues. 
We begin with a general Higgs potential, which includes custodial-symmetry-breaking terms, to construct the effective potential and derive the one-loop potential at zero temperature. 
At the EW scale, we assume some relations among the scalar couplings so that the potential is custodial symmetric. 
Counterterms are determined by requirements that the first and second derivatives of the one-loop effective potential are the same as their tree-level values. 
The scale dependence of the potential symmetry is taken into account by evolving the couplings and the classical fields according to the renormalization group equations derived from the counterterms~\cite{Bando:1992np}. 
Finite temperature effects are also included at the one-loop level. 
Thermal resummation of the bosonic degrees of freedom is performed in the Parwani scheme~\cite{Parwani:1991gq}, where the leading higher-order effect is incorporated as thermal masses. 

The model parameters are restricted by theoretical requirements and experimental bounds. 
We update the Bayesian analysis in Refs.~\cite{Chen:2022zsh, Chiang:2018cgb} by using the latest experimental data through \texttt{HiggsTools}~\cite{Bahl:2022igd}. 
We use the Markov Chain Monte Carlo (MCMC) to find the posterior distributions of the model parameters with the theoretical and experimental constraints and their prior distributions. 

At parameter points generated by the MCMC, we investigate the EWPT by using the aforementioned effective potential. 
We use \texttt{CosmoTransitions}~\cite{Wainwright:2011kj} to determine the critical temperature and the order of the EWPT, to solve the bounce equation, and to find the bounce action and the nucleation temperature. 

We also discuss the phenomenological predictions at the parameter points where the strong first-order EWPT is expected. 
We examine observables closely related to the EWPT: the Higgs boson decays to neutral gauge bosons ($h \to \gamma \gamma$ and $h \to Z \gamma$), the Higgs triple coupling which modifies di-Higgs production at high-energy colliders, and stochastic gravitational waves associated with the first-order EWPT. 
It is shown that the parameter points for the strong first-order EWPT can be verified by these observables at next-generation experiments. 

This paper is organized as follows. 
In Sec.~\ref{sec: model}, the GM model is reviewed. 
In Sec.~\ref{sec: effective_potential}, we explain how to construct the effective potential in detail. 
In Sec.~\ref{sec: global fit}, we perform a global fit by the Bayesian analysis and extract parameter regions preferred by the theoretical and experimental constraints. 
In Sec.~\ref{sec: numerical_studies}, we show results of numerical studies of the EWPT. We also discuss the phenomenological predictions. 
The conclusions are presented in Sec.~\ref{sec: conclusion}. 
Some relevant formulas are displayed in the appendices. 
In Appendix~\ref{app: notation}, the notation of the general Higgs potential in this paper is compared to those in other related works. 
In Appendix~\ref{app: field_dependent_mass}, formulas for field-dependent masses, which are necessary to derive one-loop corrections in the effective potential, are shown. 
In Appendix~\ref{app: beta_function}, beta functions for couplings and anomalous dimensions for scalar fields are derived in the renormalization scheme discussed in Sec.~\ref{sec: effective_potential}. 
Finally, in Appendix~\ref{app: hhh}, we explain how to derive the Higgs triple coupling by using the one-loop effective potential.

\section{The Georgi--Machacek model}
\label{sec: model}

In this section, we review the GM model~\cite{GM_model} and introduce our notation. 
The Higgs sector in this model consists of three isospin multiplets: a doublet $\phi$ with the hypercharge $Y=1/2$, a triplet $\chi$ with $Y=1$, and a triplet $\xi$ with $Y=0$. 
In addition to the electroweak gauge symmetry, we impose a global $SU(2)_R$ symmetry in the Higgs potential. 
The scalar fields are represented as follows in the $SU(2)_L \times SU(2)_R$ covariant form:
\begin{align}
\label{eq: Scalar fields}
\Phi  =
    \begin{pmatrix}
        \left(\phi^0\right)^* & \phi^+\\
        -\phi^- & \phi^0
    \end{pmatrix}, \quad 
\Delta  = 
    \begin{pmatrix}
    \left(\chi^0\right)^* & \xi^+ & \chi^{++}\\
    -\chi^- & \xi^0 & \chi^+\\
    \chi^{--} & -\xi^- & \chi^0\end{pmatrix}. 
\end{align}
The most general Higgs potential symmetric under $SU(2)_L \times SU(2)_R \times U(1)_Y$ is given by
\begin{align}
\label{CS Higgs potential}
V = & \frac{1}{2}m_1^2\mathrm{tr}[\Phi^{\dagger}\Phi] 
    + \frac{1}{2}m_2^2\mathrm{tr}[\Delta^{\dagger}\Delta] 
    + \lambda_1\left(\mathrm{tr}[\Phi^{\dagger}\Phi]\right)^2 
    + \lambda_2\left(\mathrm{tr}[\Delta^{\dagger}\Delta]\right)^2
\nonumber \\[5pt]
    & + \lambda_3\mathrm{tr}\Bigl[\left(\Delta^{\dagger}\Delta\right)^2\Bigr] 
    + \lambda_4\mathrm{tr}\bigl[\Phi^{\dagger}\Phi \bigr]\mathrm{tr}\bigl[\Delta^{\dagger}\Delta \bigr] 
    + \lambda_5 \mathrm{tr} \biggl[\Phi^{\dagger}\frac{\sigma^a}{2}\Phi\frac{\sigma^b}{2} \biggr]
        \mathrm{tr}\Bigl[\Delta^{\dagger}T^a\Delta T^b \Bigr]
\nonumber \\[5pt]
    & + \mu_1 \mathrm{tr}\biggl[\Phi^{\dagger}\frac{\sigma^a}{2}\Phi\frac{\sigma^b}{2} \biggr]
        \left(P^{\dagger}\Delta P\right)_{ab} 
    + \mu_2 \mathrm{tr}\Bigl[\Delta^{\dagger}T^a\Delta T^b\Bigr]
        \left(P^{\dagger}\Delta P\right)_{ab},
\end{align}
where $\sigma^a$ are the Pauli matrices, $T^a$ are defined as 
\begin{align}
    T^1 = \frac{1}{\sqrt{2}}
    \begin{pmatrix}
       0 & 1 & 0\\
       1 & 0 & 1\\
       0 & 1 & 0
    \end{pmatrix},\quad
    T^2 = \frac{1}{\sqrt{2}}
    \begin{pmatrix}
       0 & -i & 0\\
       i & 0 & -i\\
       0 & i & 0
    \end{pmatrix},\quad
    T^3 = 
    \begin{pmatrix}
       1 & 0 & 0\\
       0 & 0 & 0\\
       0 & 0 & -1
    \end{pmatrix}, 
\end{align}
and the unitary matrix $P$: 
\begin{align}
    P = \frac{1}{\sqrt{2}}
    \begin{pmatrix}
        -1 & i & 0\\
        0 & 0 & \sqrt{2}\\
        1 & i & 0
    \end{pmatrix}, 
\end{align}
relates $T^a$ to the adjoint representation of generators of the $SU(2)$ Lie algebra. 

The scalar fields acquire the VEVs as follows:
\begin{align}
\left<\Phi\right>  =\frac{ 1 }{ \sqrt{2} }
    \begin{pmatrix}
        v_\Phi & 0\\
        0 & v_\Phi
    \end{pmatrix}, \quad 
\left<\Delta\right>  = 
    \begin{pmatrix}
   v_\Delta & 0 & 0 \\
    0 & v_\Delta & 0 \\
    0 & 0 & v_\Delta 
    \end{pmatrix}, 
\end{align}
which are constrained by $v_\Phi^2 + 8 v_\Delta^2 = v^2 \simeq (246~\mathrm{GeV})^2$ to satisfy the experimental data. 
For later convenience, we define the angle $\beta$ as $\tan \beta = v_\Phi / (2\sqrt{2} v_\Delta)$. 
These VEVs break the global $SU(2)_L \times SU(2)_R$ symmetry in the Higgs potential, while the custodial symmetry $SU(2)_V$ remains. 
The stationary condition at the vacuum requires 
\begin{align}
\begin{array}{l}
\label{eq: stationary_cond}
\displaystyle{
m_1^2 = -4\lambda_1v_{\Phi}^2 -6\lambda_4v_{\Delta}^2 - 3\lambda_5v_{\Delta}^2 + \frac{3}{4} M_1^2 c_\beta^2
}, \\[5pt]
\displaystyle{
m_2^2 = -12\lambda_2v_{\Delta}^2 - 4\lambda_3v_{\Delta}^2 - 2\lambda_4v_{\Phi}^2 -\lambda_5v_{\Phi}^2 + M_1^2 s_\beta^2 + \frac{ 1 }{ 2 } M_2^2
}, 
\end{array}
\end{align}
where $c_\beta = \cos \beta$ and $s_\beta = \sin \beta$, and 
\begin{align}
\quad M_1^2 = -\frac{v\mu_1}{\sqrt{2}c_\beta},
\quad M_2^2 = -3\sqrt{2}v\mu_2 c_\beta. 
\end{align}

The custodial symmetry in the Higgs potential enables us to classify the physical scalar bosons into the $SU(2)_V$ multiplets: a quintet $H_5$, a triplet $H_3$, and two singlets $H_1$ and $h$.  
By using the original scalar fields in $\Phi$ and $\Delta$, these mass eigenstates are given by
\begin{align}
\begin{split}
& H_5^{\pm\pm}=\chi^{\pm\pm}, 
    \quad H_5^{\pm}=\frac{1}{\sqrt{2}}\left(\chi^{\pm}-\xi^{\pm}\right), 
        \quad H_5^0=\sqrt{\frac{1}{3}} \chi_r-\sqrt{\frac{2}{3}} \xi_r, 
\\[7pt]
& H_3^{\pm}=- c_\beta \phi^{\pm}+ \frac{s_\beta}{\sqrt{2}}\left(\chi^{\pm}+\xi^{\pm}\right), 
    \quad H_3^0=-c_\beta \phi_i+ s_\beta \chi_i,
\\[7pt]
& h= c_\alpha \phi_r-\frac{s_\alpha}{\sqrt{3}}\left(\sqrt{2} \chi_r+\xi_r\right), \quad H_1= s_\alpha \phi_r+\frac{c_\alpha}{\sqrt{3}}\left(\sqrt{2} \chi_r+\xi_r\right),     
\end{split}
\end{align}
where neutral scalars $\phi_r$, $\phi_i$, $\chi_r$, $\chi_i$, and $\xi_r$ are defined as 
\begin{align}
\phi^0 = \frac{ v_\Phi + \phi_r + i \phi_i }{ \sqrt{2} }, \quad 
\chi^0 = v_\Delta + \frac{ \chi_r + i \chi_i }{ \sqrt{2} }, \quad 
\xi^0 = v_\Delta + \xi_r. 
\end{align}
The mixing angle $\alpha$ satisfies $\tan 2\alpha = 2M^2_{12}/(M^2_{22}-M^2_{11})$ with 
\begin{align}
\begin{array}{l}
\displaystyle{
M^2_{11} = 8\lambda_1 v^2 s_\beta^2,
        \quad M^2_{22} = (3\lambda_2 + \lambda_3) v^2 c_\beta^2 + M^2_1s_\beta^2 -\frac{1}{2}M^2_2
}, \\[7pt]
\displaystyle{
M^2_{12} = \sqrt{\frac{3}{2}} \Bigl((2\lambda_4+\lambda_5)v^2-M^2_1 \Bigr)s_\beta c_\beta
}.
\end{array}
\end{align}
Then, the mass formulas of the quintet, the triplet, $H_1$, and $h$ are given by
\begin{align}
\begin{array}{l}
\displaystyle{
m_{H_5}^2 = \biggl(M_1^2-\frac{3}{2} \lambda_5 v^2 \biggr) s_\beta^2 
    +\lambda_3 v^2 c_\beta^2 +M_2^2
}, \\[10pt]
\displaystyle{
m_{H_3}^2 = M_1^2-\frac{1}{2} \lambda_5 v^2
}, \\[10pt]
\displaystyle{
m_{H_1}^2 = M^2_{11} s_\alpha^2 + M^2_{22} c_\alpha^2 + 2M^2_{12} s_\alpha c_\alpha
}, \\[5pt]
\displaystyle{
m_h^2 = M^2_{11} c_\alpha^2 +M^2_{22}s_\alpha^2 -2M^2_{12} s_\alpha c_\alpha
}, 
\end{array}
\end{align}
respectively. In the following, $h$ is identified as the observed 125 GeV Higgs boson. 
The mass degeneracy between the $SU(2)_V$ multiplets is radiatively broken~\cite{Blasi:2017xmc, Keeshan:2018ypw}.

\section{Effective potential in the GM model}
\label{sec: effective_potential}
\setcounter{equation}{0}

In this section, we describe the construction of the one-loop effective potential in the GM model. 
At zero temperature, one-loop corrections are given by the Coleman--Weinberg type potential~\cite{Coleman:1973jx}. 
We impose renormalization conditions on the first and second derivatives of the effective potential. 
To eliminate all the divergences in one-loop corrections, we need to extend the Higgs potential so as to include the custodial-symmetry-breaking terms. 
Coupling relations by which the potential becomes custodial symmetric are imposed at the EW vacuum.  
The finite temperature effect is also included at the one-loop level~\cite{Dolan:1973qd}, and the thermal resummation is performed in the Parwani scheme~\cite{Parwani:1991gq}. 
Finally, to include the effect of the custodial symmetry violation at scales far from the EW vacuum, we perform the renormalization group equation (RGE) improvement of the effective potential~\cite{Bando:1992np}. 

\subsection{General settings and the tree-level effective potential}

Due to the custodial symmetry breaking in the gauge and Yukawa interactions, the Higgs potential in Eq.~(\ref{CS Higgs potential}) is not renormalizable, and we need to consider a general form of the Higgs potential to evaluate the full one-loop effective potential. 
We use the following notation for the general Higgs potential:
\begin{align}
\label{eq: general_Higgs_potential}
V = & \ m_{\phi}^2 |\phi|^2 
	+ m_{\chi}^2 \mathrm{tr}\bigl[\chi^\dagger \chi\bigr]
	+ \frac{ 1 }{ 2 } m_{\xi}^2 \mathrm{tr}\bigl[ \xi^2 \bigr]
\nonumber \\[5pt]
	& - 2 \sqrt{2} \sigma_{1} \phi^\dagger \xi \phi
	+ \sigma_{2} \Bigl( \phi^\dagger \chi \tilde{\phi} + \mathrm{h.c.} \Bigr)
	- \sqrt{2} \sigma_{3} \mathrm{tr}\bigl[ \chi^\dagger \xi \chi \bigr]
\nonumber \\[5pt]
	& + 4 \lambda |\phi|^4 
	+ \rho_{1} \mathrm{tr}\bigl[\chi^\dagger \chi \bigr]^2
	+ \rho_{2} \mathrm{tr} \Bigl[ \bigl( \chi^\dagger \chi \bigr)^2 \Bigr]
	+ 2 \rho_{3} \mathrm{tr}\bigl[ \xi^4 \bigr]
\nonumber \\[5pt]
	& + \rho_{4} \mathrm{tr}\bigl[ \chi^\dagger \chi \bigr] \mathrm{tr} \bigl[ \xi^2 \bigr]
	+ \rho_{5} \Bigl| \mathrm{tr}\bigl[ \chi^\dagger \xi \bigr] \Bigr|^2
	+ 2 \kappa_{1} \mathrm{tr}\bigl[ \chi^\dagger \chi \bigr] |\phi|^2
\nonumber \\[5pt]
	& + 2 \kappa_{2} \phi^\dagger \bigl[ \chi, \chi^\dagger \bigr] \phi
	+ 2 \kappa_{3} \mathrm{tr}\bigl[\xi^2 \bigr] |\phi|^2
	+ \sqrt{2} \kappa_{4} \Bigl( \phi^\dagger \chi  \xi \tilde{\phi} + \mathrm{h.c.} \Bigr), 
\end{align}
where $\phi$ = $(\phi^+, \phi^0)^\mathrm{T}$, $\tilde{\phi} = i \sigma^2 \phi^\ast$, and $\chi$ and $\xi$ are in the bimatrix form:
\begin{align}
\chi = \frac{ 1 }{ \sqrt{2} }
\begin{pmatrix}
\chi^+ & - \sqrt{2} \chi^{++} \\
\sqrt{2} \chi^0 & - \chi^+ \\
\end{pmatrix}, \quad 
\xi = \frac{ 1 }{ \sqrt{2} }
\begin{pmatrix}
\xi^0 & - \sqrt{2} \xi^+ \\
- \sqrt{2} \xi^- & - \xi^0 \\
\end{pmatrix}. 
\end{align}
In Appendix~\ref{app: notation}, this notation is compared with those in other works that study the general Higgs potential in the GM model. 

We neglect the $CP$-violation for the sake of simplicity, and all the scalar couplings are real. 
The relation with the custodial symmetric (CS) couplings is given by
\begin{align}
\label{eq: CS limit}
\begin{array}{l}
\displaystyle{
m_\phi^2 = m_1^2, \quad 
    m_\chi^2 = m_\xi^2 = m_2^2, \quad 
    2 \sigma_1 = \sigma_2 = \frac{ \mu_1 }{ 2 } , \quad 
    \sigma_3 = 6 \mu_2
}, \\[5pt]
\displaystyle{
\lambda = \lambda_1, \quad 
    \rho_1 =  4\lambda_2 + 6 \lambda_3, \quad 
    \rho_2 = - 4 \lambda_3, \quad 
    \rho_3 = \lambda_2 + \lambda_3, \quad 
    \rho_4 = 4 \lambda_2, \quad 
    \rho_5 = 4 \lambda_3
},  \\[5pt]
\displaystyle{
\kappa_1 = 2 \lambda_4, \quad 
    \kappa_2 = \frac{ 1 }{ 2 } \lambda_5, \quad 
    \kappa_3 = \lambda_4, \quad 
    \kappa_4 = \lambda_5
} 
\quad \text{(in the CS limit)}.
\end{array}
\end{align}
We impose these relations at the EW vacuum and use them as the initial condition for the RGE running. 

We define the classical fields $(\varphi_1, \varphi_2, \varphi_3)$ and the dynamical modes around them as follows:
\begin{align}
\label{eq: classical_fields}
& \phi = \frac{1 }{ \sqrt{2} }
\begin{pmatrix}
\sqrt{2} \omega^+_1 \\
\varphi_1 + h_1 + i a_1 \\
\end{pmatrix}, \quad 
\chi = \frac{1 }{ \sqrt{2} }
\begin{pmatrix}
\omega^+_2 & -\sqrt{2}\chi^{++} \\
\sqrt{2}\varphi_2 + h_2 + i a_2 & - \omega_2^+ \\
\end{pmatrix}, \quad 
\nonumber \\[10pt]
& \xi = \frac{ 1 }{ \sqrt{2} }
\begin{pmatrix}
\varphi_3 + h_3 & - \sqrt{2} \omega^+_3 \\
- \sqrt{2} \omega^-_3 & - \varphi_3 - h_3 \\
\end{pmatrix}. 
\end{align}
We define the general vacuum configuration as $(\varphi_1, \varphi_2, \varphi_3) = (v_\phi, v_\chi, v_\xi)$. 
In the CS limit in Eq.~(\ref{eq: CS limit}), these are given by 
\begin{align}
v_\phi = v_\Phi, \quad v_\chi = v_\xi = v_\Delta \quad \text{(in the CS limit)}.
\end{align}
These relations are broken via the RGE running except at the EW vacuum. 

By using the classical fields in Eq.~(\ref{eq: classical_fields}), 
the tree-level effective potential $V_0$ is given by
\begin{align}
\label{eq: V0}
V_0 = & \frac{m_\phi^2}{2} \varphi_1^2 
	+ m_\chi^2  \varphi_2^2 
	+ \frac{ m_\xi^2 }{ 2 } \varphi_3^2 
	+ \sigma_1 \varphi_1^2 \varphi_3 
	+ \sigma_2 \varphi_1^2 \varphi_2
	+ \sigma_3 \varphi_2^2 \varphi_3 
	\nonumber \\[5pt]
	& + \lambda \varphi_1^4
	+ \rho_{12} \varphi_2^4
	+ \rho_3 \varphi_3^4 
	+ \rho_4 \varphi_2^2 \varphi_3^2 
	+ \kappa_{12} \varphi_1^2 \varphi_2^2
	+ \kappa_3 \varphi_1^2 \varphi_3^2
	+ \kappa_4 \varphi_1^2 \varphi_2 \varphi_3, 
\end{align}
where $\kappa_{12} = \kappa_1 + \kappa_2$, and $\rho_{12} = \rho_1 + \rho_2$.
The tree-level potential $V_0$ is symmetric under $\varphi_1 \to - \varphi_1$, while such a reflection symmetry does not exist for $\varphi_2$ and $\varphi_3$. 
This symmetry is broken by one-loop corrections as shown below. 

\subsection{One-loop corrections at zero temperature}

Next, we discuss one-loop corrections given by a sum of the Coleman--Weinberg type potentials~\cite{Coleman:1973jx}:
\begin{align}
V_1 = \sum_i 
\frac{ n_i }{ 64 \pi^2 } \tilde{m}_i^4 
\biggl( \log \frac{ \tilde{m}^2_i }{ \mu^2 } - c_i - D \biggr),  
\end{align}
where $\mu$ is the renormalization scale, $i$ labels particles in loop diagrams, and $n_i$, $\tilde{m}_i$, and $c_i$ are the number of degrees of freedom including the statistical factor, the field-dependent mass, and a coefficient for the particle $i$. 
For the gauge boson, $c_i = 5/6$, while $c_i = 3/2$ for other particles. 
In this formula, we keep the divergent part:
\begin{align}
D = \frac{ 2 }{ 4 - d } + \gamma_E + \log 4 \pi, 
\end{align}
where $d$ is the spacetime dimension and $\gamma_E$ is the Euler--Mascheroni constant, to discuss the cancellation of the divergence with counterterms and to derive the beta function.  

In computing the loop diagrams, we employ the Landau gauge, where the Faddeev--Popov ghosts do not contribute at the one-loop level.\footnote{Although the effective potential is gauge dependent, an effect of changing gauge fixing conditions can be canceled by shifting classical fields according to the Nielsen--Fukuda--Kugo identity~\cite{Nielsen:1975fs, Fukuda:1975di} at a fixed loop order. 
In this paper, we use only the Landau gauge and do not discuss the gauge dependence of the potential further.} 
Then, the SM fermions, the EW gauge bosons, and the scalar bosons play a part in $V_1$. 
The field-dependent masses for each particle are presented in Appendix~\ref{app: field_dependent_mass}. 

\subsection{Renormalization and counterterms}
\label{sec: renormalization}

Here, we discuss the counterterm of the effective potential $\delta V$ and the renormalization procedure to fix the parameters in the Higgs potential.  
In addition to the counterterms for the scalar couplings in Eq.~(\ref{eq: general_Higgs_potential}), we prepare the VEV counterterms $\delta v_\phi$, $\delta v_\chi$, and $\delta v_\xi$ for later convenience, following the method introduced in Ref.~\cite{ref:KE_preparation}.
Then, the bare classical fields $\varphi_{iB}$ ($i=1$, 2, 3) are represented by 
\begin{align}
\varphi_{1B} = Z_{\phi}^{1/2} (\varphi_1 - v_\phi) + v_\phi + \delta v_\phi, \\[10pt] 
\varphi_{2B} = Z_{\chi}^{1/2} (\varphi_2 - v_\chi) + v_\chi + \delta v_\chi, \\[10pt]
\varphi_{3B} = Z_\xi^{1/2} (\varphi_3 - v_\xi) + v_\xi + \delta v_\xi, 
\end{align}
where $Z_\phi$, $Z_\chi$, and $Z_\xi$ are the wave function renormalizations. 
The wave function renormalizations are determined by the $\overline{\mathrm{MS}}$ scheme. 
We assume that $\delta v_\phi$ is also given by the $\overline{\mathrm{MS}}$ scheme because its finite part is not necessary in analyses below. 

We note that the wave function renormalizations are optional in the renormalization of the effective potential, {\it i.e.}, correlation functions at zero external momenta. 
These counterterms can be effectively absorbed into definitions of the coupling counterterms, which do not change the form of the one-loop effective potential. 
However, when we discuss the RGE improvement of the effective potential, we need to consider the wave function renormalizations because they induce the running of the classical fields~\cite{Bando:1992np}. 

The counterterm for the coupling $\lambda$ is defined as $\lambda_B = \lambda + \delta \lambda$, and the counterterms for the other scalar couplings are defined analogously. 
By using these counterterms and the wave function renormalizations, we can obtain the counterterm of the effective potential $\delta V$. 

The divergent parts of the counterterms are found such that the terms proportional to $D$ in $V_1$ are canceled.\footnote{In this way, divergent parts of $\delta \rho_1$, $\delta \rho_2$, $\delta \kappa_1$, $\delta \kappa_2$ are not uniquely determined, nor $\delta \rho_5$ at all. These are fixed by vertex renormalization at the one-loop level.}
They are used to derive the beta function of the couplings below. 
It is worth emphasizing that the custodial symmetric counterterms are not sufficient to eliminate all the divergences in $V_1$. 
The custodial symmetry is broken except at the initial scale due to the $U(1)_Y$ gauge interaction at the one-loop level. 
There is no symmetry-breaking contribution from the Yukawa interactions at the one-loop level. 

The finite part of $\delta V$ is given by
\begin{align}
\mathrm{Fin}\bigl[ \delta V \bigr] 
    = & \ 2 m_\chi^2 \delta v_\chi \varphi_2 
	+ m_\xi^2 \delta v_\xi \varphi_3 
	+ \frac{ 1 }{ 2 } \varphi_1^2 \bigl( \delta m_\phi^2 
		+ 2 \sigma_1 \delta v_\xi 
		+ 2 \sigma_2 \delta v_\chi \bigr)
	+ \varphi_2^2 \bigl( \delta m_\chi^2 
		+ \sigma_3 \delta v_\xi \bigr)
\nonumber \\[5pt]
	& + \frac{ \delta m_\xi^2 }{ 2 } \varphi_3^2 
	+ 2 \sigma_3 \delta v_\chi \varphi_2 \varphi_3
    +  \varphi_1^2 \varphi_2 \bigl( 
		\delta \sigma_2 
		+ 2 \kappa_{12} \delta v_\chi
		+ \kappa_4 \delta v_\xi \bigr)
\nonumber \\[5pt]
	& + \varphi_1^2 \varphi_3 \bigl(
		\delta \sigma_1 
		+ 2 \kappa_3 \delta v_\xi 
		+ \kappa_4 \delta v_\chi \bigr)
    + 4 \rho_{12} \delta v_\chi \varphi_2^3
    +  \varphi_2^2 \varphi_3 \bigl(
		 \delta \sigma_3 
		 + 2 \rho_4 \delta v_\xi \bigr)
\nonumber \\[5pt]
	& + 2 \rho_4 \delta v_\chi \varphi_2 \varphi_3^2
	+ 4 \rho_3 \delta v_\xi \varphi_3^3
    + \delta \lambda \varphi_1^4
	+ \delta \rho_{12} \varphi_2^4 
	+ \delta \rho_{3} \varphi_3^4 
	+ \delta \rho_{4} \varphi_2^2 \varphi_3^2
\nonumber \\[5pt]
	& + \delta \kappa_{12} \varphi_1^2 \varphi_2^2 
	+ \delta \kappa_{3} \varphi_1^2 \varphi_3^2 
	+ \delta \kappa_{4} \varphi_1^2 \varphi_2 \varphi_3, 
\end{align}
where $\delta \kappa_{12} = \delta \kappa_1 + \delta \kappa_2$, $\delta \rho_{12} = \delta \rho_1 + \delta \rho_2$, and $\mathrm{Fin}[\cdots]$ represents the finite part. 
The wave function renormalizations and $\delta v_\phi$ are absent because they consist of only the divergent terms.

The finite part is determined by imposing renormalization conditions on first and second derivatives of the effective potential: 
\begin{align}
\label{eq: renormalization_cond}
\left. \frac{ \partial V_\mathrm{T=0} }{ \partial \varphi_i } \right|_0 
    = \left. \frac{ \partial V_0 }{ \partial \varphi_i } \right|_0, \quad 
\left. \frac{ \partial^2 V_\mathrm{T=0} }{ \partial \varphi_i \partial \varphi_j } \right|_0 
    = \left. \frac{ \partial^2 V_0 }{ \partial \varphi_i \partial \varphi_j } \right|_0, 
\quad (i,j=1,2,3), 
\end{align}
where $V_\mathrm{T=0} = V_0 + V_1 + \delta V$, and $X|_0$ means that $X$ is evaluated at the EW vacuum. 
By these conditions, nine counterterms are fixed, while $\delta V$ has fifteen counterterms. 
Thus, the finite part of six of them cannot be determined by these conditions, and we assume that they consist of only the divergent part as in the $\overline{\mathrm{MS}}$ scheme.

We decide which counterterms are determined by these conditions as follows. 
Since small triplet VEVs $v_\chi, v_\xi \ll v$ are included in phenomenologically favorable parameter space, the counterterms are preferred to be analytic in the limit $v_\chi, v_\xi \to 0$. 
Thus, we first consider the coupled equations Eq.~(\ref{eq: renormalization_cond}) with vanishing triplet VEVs. 
Then, we find that the nine counterterms $\delta m_\phi^2$, $\delta m_\chi^2$, $\delta m_\xi^2$, $\delta v_\xi$, $\delta v_\chi$, $\delta \lambda$, $\delta \sigma_1$, $\delta \sigma_2$, $\delta \kappa_4$ can be chosen to possess a nonzero finite part so that these simplified coupled equations can be solved.\footnote{This choice of these couplings is not unique. We can use $\delta \kappa_{12}$ and $\delta \kappa_3$ instead of $\delta m_\chi^2$ and $\delta m_\xi^2$, respectively.} 
Therefore, we solve the full renormalization conditions for this minimum set of counterterms to avoid non-analytic terms in the limit of $v_\chi$, $v_\xi \to 0$, for example, $v_\phi/v_\chi$. 

Then, the renormalization conditions are reduced to the following equations. 
\begin{align}
& \delta v_\chi \bigl( 2\sigma_2 v_\phi 
		+ 4 \kappa_{12} v_\phi v_\chi 
		+ 2 \kappa_4 v_\phi v_\xi \bigr)
	+ \delta v_\xi \bigl( 2 \sigma_1 v_\phi 
		+ 2 \kappa_4 v_\phi v_\chi 
		+ 4 \kappa_3 v_\phi v_\xi \bigr)
	\nonumber \\[5pt]
	& \hspace{55pt} + v_\phi \delta m_\phi^2
	+ 2 v_\phi v_\xi \delta \sigma_1 
	+ 2 v_\phi v_\chi \delta \sigma_2
	+ 4 v_\phi^3 \delta \lambda
	+ 2 v_\phi v_\chi v_\xi \delta \kappa_4 = - V_1^\prime,  \\[10pt]
& \delta v_\chi \bigl( 2 m_\chi^2 
		+ 2 \sigma_3 v_\xi 
		+ 2 \kappa_{12} v_\phi^2
		+ 12 \rho_{12} v_\chi^2 
		+ 2 \rho_4 v_\xi^2 \bigr)
	\nonumber \\[5pt]
	& \hspace{25pt} + \delta v_\xi \bigl( 
		2 \sigma_3 v_\chi 
		+ \kappa_4 v_\phi^2
		+ 4 \rho_4 v_\chi v_\xi \bigr)
	+ 2 v_\chi \delta m_\chi^2 
	+ v_\phi^2 \delta \sigma_2
	+ v_\phi^2 v_\xi \delta \kappa_4 = - V_2^\prime,  \\[10pt]
& \delta v_\chi \bigl( 
		2 \sigma_3 v_\chi 
		+ \kappa_4 v_\phi^2
		+ 4 \rho_4 v_\chi v_\xi  \bigr)
	+ \delta v_\xi \bigl( 
		m_\xi^2 + 2 \kappa_3 v_\phi^2 
		+ 2 \rho_4 v_\chi^2 
		+ 12 \rho_3 v_\xi^2 \bigr)
	\nonumber \\[5pt]
	& \hspace{190pt} 
	+ v_\xi \delta m_\xi^2 
	+ v_\phi^2 \delta \sigma_1 
	+ v_\phi^2 v_\chi \delta \kappa_4 = - V_3^\prime,  \\[10pt]
& \delta v_\chi \bigl( 
		2 \sigma_2 + 4 \kappa_{12} v_\chi 
		+ 2 \kappa_4 v_\xi \bigr)
	+ \delta v_\xi \bigl( 
		2 \sigma_1 + 2 \kappa_4 v_\chi 
		+ 4 \kappa_3 v_\xi \bigr)
	\nonumber \\[5pt]
	& \hspace{90pt} + \delta m_\phi^2 
	+ 2 v_\xi \delta \sigma_1 
	+ 2 v_\chi \delta \sigma_2 
	+ 12 v_\phi^2 \delta \lambda
	+ 2 v_\chi v_\xi \delta \kappa_4 = - V_{11}^{\prime \prime}, \\[10pt]
& 4 \kappa_{12} v_\phi \delta v_\chi 
	+ 2 \kappa_4 v_\phi \delta v_\xi 
	+ 2 v_\phi \delta \sigma_2 
	+ 2 v_\phi v_\xi \delta \kappa_4 = - V_{12}^{\prime \prime}, \\[10pt]
& 2 \kappa_4 v_\phi \delta v_\chi
	+ 4 \kappa_{3} v_\phi \delta v_\xi 
	+ 2 v_\phi \delta \sigma_1 
	+ 2 v_\phi v_\chi \delta \kappa_4 = - V_{13}^{\prime \prime}, \\[10pt]
& 24 \rho_{12} v_\chi \delta v_\chi 
	+ \delta v_\xi \bigl( 
		2 \sigma_3 + 4 \rho_4 v_\xi \bigr)
	+ 2 \delta m_\chi^2
	 = - V_{22}^{\prime \prime}, \\[10pt]
& \delta v_\chi \bigl( 
		2 \sigma_3 + 4 \rho_4 v_\xi \bigr)
	+ 4 \rho_4 v_\chi \delta v_\xi 
	+ v_\phi^2 \delta \kappa_4 = - V_{23}^{\prime \prime},  \\[10pt]
& 4  \rho_4 v_\chi \delta v_\chi 
	+ 24 \rho_3 v_\xi \delta v_\xi
	+ \delta m_\xi^2 
	= - V_{33}^{\prime \prime},
\end{align}
where all the counterterms represent only their finite part, and $V_i^\prime$ and $V_{ij}^{\prime \prime}$ ($i$, $j = 1$, 2, 3) are defined by
\begin{align}
V_{i}^\prime = \mathrm{Fin} \Biggl[ \left. \frac{ \partial V_1 }{ \partial \varphi_i } \right|_0 \Biggr], \quad 
V_{ij}^{\prime\prime} = \mathrm{Fin} \Biggl[ \left. \frac{ \partial^2 V_1 }{ \partial \varphi_i \partial \varphi_j } \right|_0  \Biggr]. 
\end{align}
The solution of the coupled equations is given by
\begin{align}
\label{eq: CT1}
& \delta v_\chi = \frac{ 1 }{ 2 (m_\chi^2 + \sigma_3 v_\xi - 6\rho_{12} v_\chi^2 + \rho_4 v_\xi^2) }
\biggl( - V_2^\prime + \frac{ v_\phi }{ 2 } V_{12}^{\prime \prime} + v_\chi V_{22}^{\prime \prime} \biggr), \\[10pt]
& \delta v_\xi = \frac{ 1 }{ m_\xi^2 - 12 \rho_3 v_\xi^2 + 2 \rho_4 v_\chi^2 } \biggl( - 2 \sigma_3 v_\chi \delta v_\chi 
	- V_3^\prime + \frac{ v_\phi }{ 2 } V_{13}^{\prime \prime} 
	+ v_\xi V_{33}^{\prime \prime} \biggr), \\[10pt]
& \delta m_\phi^2 = - 2 \sigma_2 \delta v_\chi 
	- 2 \sigma_1 \delta v_\xi
	+ 2 v_\chi v_\xi \delta \kappa_4 
	- \frac{ 3 }{ 2 v_\phi } V_1^\prime 
	+ \frac{ 1 }{ 2 } V_{11}^{\prime \prime}
	+ \frac{ v_\xi }{ v_\phi } V_{13}^{\prime \prime}
	+ \frac{ v_\chi }{ v_\phi } V_{12}^{\prime \prime}, \\[10pt]
& \delta m_\chi^2 = - \frac{ V_{22}^{\prime \prime} }{ 2 } 
	- 12 \rho_{12} v_\chi \delta v_\chi 
	- (\sigma_3 + 2 \rho_4 v_\xi) \delta v_\xi, \\[10pt]
& \delta m_\xi^2 = - V_{33}^{\prime \prime} 
	- 4 \rho_4 v_\chi \delta v_\chi 
	-24 \rho_3 v_\xi \delta v_\xi, \\[10pt]
& \delta \sigma_1 = -\frac{ V_{13}^{\prime \prime} }{ 2 v_\phi } 
	- \kappa_4 \delta v_\chi 
	-2 \kappa_3\delta v_\xi 
	- v_\chi \delta\kappa_4, \\[10pt]
& \delta \sigma_2 = - \frac{ V_{12}^{\prime\prime} }{ 2 v_\phi }	
	- 2 \kappa_{12} \delta v_\chi 
	- \kappa_4 \delta v_\xi 
	- v_\xi \delta \kappa_4, \\[10pt]
& \delta \lambda = \frac{ 1 }{ 8 v_\phi^3 } \Bigl( V_1^\prime - v_\phi V_{11}^{\prime \prime} \Bigr), \\[10pt]
\label{eq: CT9}
& \delta \kappa_4 = - \frac{ 1 }{ v_\phi^2 } 
	\biggl( V_{23}^{\prime \prime} 
		+ 2 (\sigma_3 + 2\rho_4 v_\xi) \delta v_\chi 
		+ 4 \rho_4 v_\chi \delta v_\xi
		\biggr). 
\end{align}
We note that all the counterterms are well defined in the limit of vanishing triplet VEVs. 
If $\delta v_\chi$ and $\delta v_\xi$ do not exist, the counterterms given by the renormalization conditions must contain non-analytic terms of the triplet VEVs, which leads to a huge radiative correction for small $v_\chi$ and $v_\xi$. 
To avoid this issue, we have introduced the triplet-VEV counterterms. 

\subsection{The finite-temperature effect}

Next, we discuss thermal corrections at the one-loop level. 
In quantum field theory at finite temperature, we need to resum a part of higher-order effects to improve the perturbative expansion and avoid the infrared (IR) divergence due to the bosonic zero mode of the Matsubara frequency~\cite{Dolan:1973qd}. 
We adopt the Parwani scheme~\cite{Parwani:1991gq}, where the resummation results in the inclusion of thermal corrections to the field-dependent masses in the entire effective potential formula:
\begin{align}
\tilde{m}^2_i \to \hat{m}^2_i = \tilde{m}_i^2 + \Delta m^2_i, 
\end{align}
where $\Delta m^2_i$ represents terms proportional to $T^2$. 
Formulas for the thermal masses of each particle are shown in Appendix~\ref{app: field_dependent_mass}. 

By using the thermal masses, thermal one-loop corrections are given by~\cite{Dolan:1973qd}
\begin{align}
V_{1T}  = 
	\sum_{i \in \mathrm{bosons}} n_i J_B\biggl(\frac{ \hat{m}_i^2 }{ T^2 } \biggr) 
	+ \sum_{ i \in \mathrm{fermions}} n_i J_F \biggl(\frac{\hat{m}_i^2}{T^2} \biggr), 
\end{align}
where $J_B$ and $J_F$ are given by the following integrals: 
\begin{align}
& J_B (x) = \frac{ T^4 }{ 2\pi^2 } \int_0^\infty \mathrm{d}z\, 
	z^2 \log \Bigl( 1 - e^{-\sqrt{z^2 + x }} \Bigr), \\[10pt]
& J_F (x) = \frac{ T^4 }{ 2\pi^2 } \int_0^\infty \mathrm{d}z\, 
	z^2 \log \Bigl( 1 + e^{-\sqrt{z^2 + x }}\Bigr).  
\end{align}
In addition, we need to replace $\tilde{m}^2_i$ in $V_1$ with the thermal mass $\hat{m}^2_i$ to evaluate thermal corrections in the effective potential consistently. 

By summarizing all the above results, the full one-loop effective potential at finite temperature is given by
\begin{align}
V_\mathrm{tot}(\varphi_1,\varphi_2,\varphi_3; T) = 
    V_0 + V_1 + \delta V + V_{1T}. 
\end{align}

\subsection{Improvement by the RGE running}
\label{sec: RGE}

As discussed in Sec.~\ref{sec: renormalization}, the custodial symmetry in the potential is broken at loop levels.  
The effect of the custodial-symmetry violation propagates to the tree-level couplings via their scale dependence, {\it i.e.}, the renormalization group flow.
Here, we evaluate the RGE running of the scalar couplings and the classical fields to include this effect in the effective potential. 

The RGE running of a coupling $g$ is described by the beta function $\beta_g$, 
\begin{align}
\label{eq: coupling_RGE}
\mu \frac{ \mathrm{d} }{ \mathrm{d} \mu } g = \beta_g, 
\end{align}
where $\mu$ is the renormalization scale. 
From the fact that the bare coupling is independent of $\mu$, the beta function $\beta_g$ is evaluated by 
\begin{align}
\label{eq: beta_function}
\beta_g (\mu)
	 = - \rho_g \delta_\mathrm{res} g
	+ \sum_{g^\prime} \rho_{g^\prime} g^\prime \frac{ \partial }{ \partial g^\prime } \delta_\mathrm{res} g
	- \mu \frac{ \partial }{ \partial \mu } \delta g, 
\end{align}
where $\delta_\mathrm{res} g$ is the residue of $\delta g$ in the $d \to 4$ limit, and $\rho_g$ is defined such that the coupling in $d$ dimensions is given by $g \mu^{\rho_g (4-d)}$.  

The first and second terms in Eq.~(\ref{eq: beta_function}) are given by the divergent part of couplings.
Thus, this part is universal for any renormalization scheme as long as counterterms are prepared in the same way.
This part is denoted by $\bar{\beta}_g$ for each coupling in the following.

The remaining part of the beta function originates from the explicit $\mu$ dependence of the counterterms. 
This part depends on the choice of the renormalization scheme and is denoted by $\delta \beta_g$ in the following. 
At the one-loop level, the divergent part does not depend on $\mu$ explicitly, and only the finite part of $\delta g$ contributes to $\delta \beta_g$. 
Therefore, the contributions from the divergent part and the finite part can be completely separated. 
For example, in the $\overline{\mathrm{MS}}$ scheme, the counterterms consist of only the divergent part, and the beta function is given by the universal part $\bar{\beta}_g$.
When other renormalization conditions are imposed as in the current analysis, a counterterm may have a finite part, and the additional contribution $\delta \beta_g$ exists in the RGE running. 

Explicit formulas for the beta functions are shown in Appendix~\ref{app: beta_function}. 
By using them, we can solve the RGEs as a system of differential equations to obtain their running values depending on $\mu$.
The initial condition for the differential equations is the coupling relations in Eq.~(\ref{eq: CS limit}) and input values for the CS parameters. 

The running of the shifted classical fields $h_i = \varphi_i - v_i$ is determined by the anomalous dimensions:
\begin{align}
\label{eq: field_RGE}
\mu \frac{ \mathrm{d} h_i }{ \mathrm{d} \mu }  = - \gamma_i h_i, 
\end{align}
where $i = \phi$, $\chi$, or $\xi$, which corresponds to $i=1$, 2, or 3, respectively, for $\varphi_i$ and $h_i$. 
Thus, the scale-dependent classical fields are given by~\cite{ref:KE_preparation}
\begin{align}
\varphi_i(\mu) = C_i(\mu) h_i(\mu_0) + v_i(\mu), \quad 
C_i = \mathrm{exp} \biggl( - \int_{\mu_0}^{\mu} \mathrm{d}\mu^\prime\, \gamma_i(\mu^\prime) \biggr), 
\end{align}
where $\mu_0$ is the initial scale. 

The one-loop-level RGE improvement of the effective potential can be performed by replacing the couplings and the classical fields in the tree-level potential $V_0$ with the running ones truncated at the one-loop level: 
\begin{align}
\begin{array}{l}
\displaystyle{
    g (\mu) \simeq g + \beta_g \log \frac{ \mu }{ \mu_0 }
}, \\[7pt]
\displaystyle{
    \varphi_i(\mu) \simeq \biggl( 1 - \gamma_i \log \frac{\mu}{\mu_0} \biggr) h_i
    + v_i + \beta_{v_i} \log \frac{ \mu }{ \mu_0 } 
}. 
\end{array}
\end{align}
Let $V_0^\mathrm{RGE}$ be a potential given by this procedure. 
Then, the $\mu$-dependence in $V_0^\mathrm{RGE}$ via the running is completely canceled by that in the loop corrections $V_1 + \delta V$.\footnote{The running of the vacuum energy $\Omega$ must also be included for the cancellation of the $\mu$-dependence~\cite{Bando:1992np}. The beta function is given by the field-independent divergence in $V_1$ and shown in Appendix~\ref{app: beta_function}. }
We have numerically checked the $\mu$-dependence cancellation by using the beta functions in Appendix~\ref{app: beta_function}. 

In the practical computation, we solve the RGEs~(\ref{eq: coupling_RGE}) and (\ref{eq: field_RGE}) as differential equations with the one-loop beta functions and anomalous dimensions and use them to evaluate the entire potential, not only $V_0$. 
This leads to the $\mu$-dependence in the total effective potential at orders higher than or equal to the two-loop level. 
It should be canceled by contribution from higher-order loop diagrams.  

In addition to the scalar couplings, we include the RGE running of the SM Yukawa couplings and the gauge couplings.
Their beta functions are also presented in Appendix~\ref{app: beta_function}. 
These are at least two-loop effects in the effective potential because these couplings appear only in loop corrections. 
Thus, they do not play a role in the $\mu$-dependence cancellation at the one-loop level. 

As the scale $\mu$ where the values of the tree-level couplings are evaluated, we employ the simple choice:
\begin{align}
\label{eq: dynamical_scale}
\mu = \sqrt{ m_Z^2 + T^2 }, 
\end{align}
so that the renormalization scale at zero temperature is the $Z$ boson mass. 
At finite temperature, the RGE flow leads to the custodial symmetry breaking in the effective potential.
From the point of view of the resummation of the leading-log terms, it would be more natural to use the scale $\mu$ depending on the classical fields~\cite{Bando:1992np}. 
Despite this fact, we employ Eq.~(\ref{eq: dynamical_scale}) for the following three reasons.  

First, the $\varphi_i$-dependent choice of $\mu$ generally breaks Eq.~(\ref{eq: renormalization_cond}), which is imposed as the renormalization condition at the vacuum, because of its nonzero first and second derivatives with respect to $\varphi_i$. 
Although it may be possible to choose $\mu$ such that it equals $m_Z$ and whose first and second derivatives are both zero at the zero-temperature vacuum, such a $\mu$ is a complicated function of $\varphi_i$ and would make the phase transition analysis complicated. 
The current $\mu$ in Eq.~(\ref{eq: dynamical_scale}) avoids this concern while keeping the preferred features above. 

Second, the phase transition dynamics is determined by the potential in the region where both $|\varphi_i|$ and $T$ are less than or close to the EW scale, and is insensitive to that at large scales as long as the potential is bounded from below, and no unwanted vacuum exists. 
Therefore, we expect that Eq.~(\ref{eq: dynamical_scale}) is a good approximation of the typical scale at the EWPT. 

Finally, the GM model has multiple mass scales in the Higgs potential, and it is impossible to set all the log terms to zero by choosing a single scale $\mu$. 
In such a case, we need to consider the decoupling of heavier modes and matching to the effective field theory so that a single log factor dominates the result at each scale~\cite{Bando:1992wy}. 
However, this is impractically complicated in the current model and also has nontrivial aspects in the case of multiple classical fields. 
Such a comprehensive analysis is beyond the scope of this work, which is the first attempt to include the effect of the custodial symmetry breaking via the RGE running in the effective potential of the GM model.

\section{Global fit by experimental and theoretical constraints}
\label{sec: global fit}
\setcounter{equation}{0}

In this section, we discuss the experimental and theoretical constraints on the GM model. 
We perform the global fit for the model parameters by using the posterior distribution given by the Bayesian analysis. 

\subsection{Theoretical constraints}
\label{sec: theoretical_constraint}

As theoretical constraints on the Higgs potential of the GM model, we consider the perturbative unitarity, the vacuum stability, and the perturbativity bound. 

The tree-level perturbative unitarity demands that an eigenvalue $a_0$ of tree-level $s$-wave amplitudes for two-to-two scalar boson scatterings satisfies
\begin{align}
\bigl| a_0 \bigr| < 16 \pi c, 
\end{align}
where $c$ is normally set to $1$~\cite{Lee:1977yc} or $1/2$~\cite{Luscher:1988gc}. 
In this paper, we employ the latter criterion $c=1/2$. 

In the GM model, there are nineteen eigenvalues of the $s$-wave amplitude: $x_a$, $x_b^\pm$, $x_{12}^i$, where $a=1$--6, $b=7$--11, and $i=1$--3.
The eigenvalues are given by~\cite{Chen:2023ins}
\begin{align}
& x_1 = 2 (\rho_1+ \rho_2), \\[5pt]
& x_2 = 2 \rho_1 - \rho_2, \\[5pt]
& x_3 = 2 \rho_4 + \rho_5, \\[5pt]
& x_4 = 2 (\rho_4 + 2 \rho_5), \\[5pt]
& x_5 = 2 (\kappa_1 + \kappa_2), \\[5pt]
& x_6 = 2 (\kappa_1 - 2 \kappa_2), \\[5pt]
& x_7^\pm = \rho_1 + 4 \rho_3 \pm \sqrt{(\rho_1 - 4 \rho_3)^2 + 2 \rho_5^2}, \\[5pt]
& x_8^\pm = 4 \lambda + \rho_1 + 2 \rho_2 \pm \sqrt{ (4 \lambda - \rho_1 - 2 \rho_2)^2 + 16 \kappa_2^2}, \\[5pt]
& x_9^\pm = \kappa_1 - \kappa_2 + 2 \kappa_3 \pm \frac{ 1 }{ 2 } \sqrt{ (2 \kappa_1 - 2 \kappa_2 - 4 \kappa_3)^2 + 8 \kappa_4^2 }, \\[5pt]
& x_{10}^\pm = 4 \lambda + \rho_4 - \frac{ \rho_5 }{ 2 } \pm \frac{ 1 }{ 2 } \sqrt{ (8 \lambda - 2 \rho_4 + \rho_5)^2 + 16 \kappa_4^2 }, \\[5pt]
& x_{11}^\pm = \kappa_1 +2 \kappa_2 + 2 \kappa_3 \pm \frac{ 1 }{ 4 } \sqrt{ (4 \kappa_1 + 8 \kappa_2 - 8 \kappa_3)^2 + 128 \kappa_4^2 }, 
\end{align}
and $x_{12}^i$ ($i=1$, 2, 3) are the eigenvalues of the matrix: 
\begin{align}
X_{12} = 
\begin{pmatrix}
20 \rho_3 & 4 \sqrt{3} \kappa_3 & \sqrt{2} (3 \rho_4 + \rho_5) \\
4 \sqrt{3} \kappa_3 & 24 \lambda & 2 \sqrt{6} \kappa_1 \\
\sqrt{2}(3 \rho_4 + \rho_5) & 2\sqrt{6} \kappa_1 & 8 \rho_1 + 6 \rho_2 \\
\end{pmatrix}. 
\end{align}
These eigenvalues coincide with those given in Ref.~\cite{Aoki:2007ah} in the CS limit in Eq.~(\ref{eq: CS limit}).

For the vacuum stability condition, we consider two conditions on the potential. 
First, we require that the potential is bounded from below in any direction of scalar field space, which is often called the BFB condition. 
This imposes the following inequalities for the general Higgs potential~\cite{Chen:2023ins}\footnote{In Ref.~\cite{Chen:2023ins}, the sixth inequality is given in a different form as a solution of that given in Eq.~(\ref{eq: BFB_cond}). The formula in Ref.~\cite{Chen:2023ins} is derived by assuming that the quadratic function of $t^2$ on the left-hand side has the minimum in $t^2 \in [0,1]$, which is not always valid.}:
\begin{align}
\label{eq: BFB_cond}
& \lambda > 0, \quad \rho_3 > 0, \quad
\rho_1 + \mathrm{min}\Bigl( \frac{ \rho_2 }{ 2 } , \rho_2 \Bigr) > 0, \quad 
4\lambda \rho_3 - \kappa_3^2 > 0, 
\nonumber \\[5pt]
& 4\lambda (\rho_1 + \zeta \rho_2) > 
    \Bigl\{ \kappa_1 
        - |\kappa_2| \sqrt{2\zeta-1} \Bigr\}^2, 
\nonumber \\[5pt]
& (\rho_1 + \zeta\rho_2 + \rho_3 - \rho_4 - \eta \rho_5 ) t^4 
    + (\eta \rho_5 - 2 \rho_3 + \rho_4) t^2 + \rho_3 > 0, 
 \\[5pt]
& 4 \lambda \Bigl\{ (\rho_1 + \rho_2 \zeta + \rho_3 - \rho_4 - \eta \rho_5 ) t^4
    - (2\rho_3 - \rho_4 - \eta \rho_5) t^2 + \rho_3 \Bigr\}
    \nonumber \\
    & \hspace{50pt} > \Bigl\{ \Bigl( \kappa_1 - |\kappa_2| \sqrt{2\zeta-1} - \kappa_3 \Bigr) t^2 - |\kappa_4| \sqrt{1-\eta} t \sqrt{1-t^2} + \kappa_3 \Bigr\}^2, \nonumber 
\end{align}
where the last three inequalities should be satisfied for any values of $t$, $\zeta$, and $\eta$ in the domains:
\begin{align}
t \in [0, 1], \quad \zeta \in \Bigl[\frac{1}{2}, 1\Bigr], \quad \eta \in [0,1]. 
\end{align}
We note that these conditions are sufficient conditions and are not identical to the BFB condition for the CS potential given in Ref.~\cite{Hartling:2014zca} even in the CS limit~(\ref{eq: CS limit}). 
We numerically confirmed that Eq.~(\ref{eq: BFB_cond}) is stronger than that in Ref.~\cite{Hartling:2014zca} in the CS limit. 

As the second condition for the vacuum stability, we require that the electroweak vacuum is the global minimum of the Higgs potential. 
We impose this condition by numerically finding the minimum configurations of the tree-level potential in Eq.~(\ref{eq: V0}). 
The one-loop corrections $V_1$, $\delta V$, and the RGE running are not included to reduce the computational cost. 
We expect that these neglected contributions do not create a new minimum unless the tree-level potential has a mostly flat direction, which would not be the case in almost all the parameter regions. 

Finally, we impose the perturbativity bound:
\begin{align}
|\lambda_i| < 4 \pi, 
\end{align}
for all the quartic scalar couplings $\lambda_i$. 

In the global fit, all the above constraints are imposed on the input values of the couplings at $\mu = m_Z$, and we do not consider the running. 
On the other hand, in the phase transition analysis in the next section, we examine the theoretical constraints using the RGE to select preferred parameter points. 
In this case, we can find the cutoff scale $\Lambda$ such that the theoretical constraints are no longer satisfied at scales larger than that.

\subsection{Experimental constraint}
\label{subsec: experimental_constraint}

Due to the mixing between the $CP$-even scalar bosons, the couplings of the 125 GeV Higgs boson $h$ can deviate from the SM predictions even at the tree level. 
Such deviations are constrained by the current LHC data. 
We implement this constraint by using \texttt{HiggsSignals} in \texttt{HiggsTools}~\cite{Bahl:2022igd}. 
We define effective coupling factors $\kappa$ for each coupling, which represent deviations from the SM values. 
Then, \texttt{HiggsTools} automatically evaluates the production cross section and the decay rates via the gauge and Yukawa couplings. 
Since the effective couplings between $h$ and other scalars are not prepared in \texttt{HiggsTools}, we evaluate the decay rates for $h \to \gamma \gamma$ and $h \to Z \gamma$, where charged-scalar-loop diagrams play a significant role, by using the analytical formulas in Ref.~\cite{Degrande:2017naf}\footnote{There is a typo in Eq.~(35) of Ref.~\cite{Degrande:2017naf}, which shows the $W$ loop contribution in the scalar boson decay into $Z \gamma$. We alternatively use a formula in Eq.~(2.54) of Ref.~\cite{Djouadi:2005gi}.} and pass their values to \texttt{HiggsTools}. 

In addition to the Higgs couplings, we consider the constraint from direct searches for the additional scalar bosons at the LHC, which is implemented by \texttt{HiggsBounds} in \texttt{HiggsTools}. 
For the neutral scalar bosons $H_1$, $H_3^0$, and $H_5^0$, we use the effective coupling factors $\kappa$ to define their interactions with the SM particles, like those for $h$. 
Then, the main production cross section and the decay rates to the SM particles are automatically evaluated. 
The other decay channels listed in Table~\ref{table: decay_rates} are individually evaluated by using the analytical formulas, and their values are passed to \texttt{HiggsTools}. 
The cross sections for the additional production channels such as $p p \to Z^\ast \to h H_3^0/H_1H_3^0$ are implemented by interpolating reference values calculated by using \texttt{MadGraph5}~\cite{Alwall:2014hca}. 

For the charged scalars, we use \texttt{EffectiveCouplingCxns} for the $tb$ associated production for the singly charged scalars $p p \to H_3^\pm tb$, where the effective vertex is defined as in Ref.~\cite{Bechtle:2020pkv}. 
The cross sections for other production channels shown in Table~\ref{table: prod_xsec} are implemented by interpolating values computed by using \texttt{MadGraph5}. 

All the decay rates for the charged scalar bosons are evaluated by using the analytical formulas.\footnote{For decays via the Yukawa interaction between $\chi$ and lepton doublets, we estimate the Yukawa couplings by using the neutrino oscillation data for the normal mass hierarchy~\cite{PDG2026} with the assumptions of the massless lightest neutrino and $CP$-conserving Yukawa couplings.} 
We consider only the tree-level decays and neglect loop-induced ones. 
The implemented decay channels are summarized in Table~\ref{table: decay_rates}. 

\begin{table}[H]
\begin{center}
\caption{The decay channels implemented by using the analytical formulas. $\ell$ and $\ell^\prime$ represent the charged leptons: $e$, $\mu$, or $\tau$.}
\label{table: decay_rates}
\begin{tabular}{|c|l|} \hline
$h$ & $h \to \gamma \gamma$, \quad $h \to Z \gamma$ \\[5pt]
$H_1$ & $H_1 \to Z H_3^0$, \quad $H_1 \to W^\pm H_3^\mp$, \quad $H_1 \to hh$, \quad $H_1 \to \nu\nu$ \\[5pt]
$H_3^0$ & $H_3^0 \to Zh$, \quad $H_3^0 \to Z H_1$, \quad $H_3^0 \to ZH_5^0$, \quad $H_3^0 \to W^\pm H_5^\mp$, \quad $H_3^0 \to \nu\nu$ \\[5pt]
$H_3^\pm$ & $H_3^\pm \to ud$, \quad $H_3^\pm \to cs$, \quad $H_3^\pm \to tb$, \quad $H_3^\pm \to \ell^\pm \nu$ \\[5pt]
$H_5^0$ & $H_5^0 \to Z H_3^0$, \quad $H_5^0 \to W^\pm H_3^\mp$, \quad $H_5^0 \to \nu\nu$ \\[5pt]
$H_5^\pm$ & $H_5^\pm \to W^\pm H_3^0$, \quad $H_5^\pm \to Z H_3^\pm$, \quad $H_5^\pm \to W^\pm Z$, \quad $H_5^\pm \to \ell^\pm \nu$ \\[5pt]
$H_5^{\pm\pm}$ & $H_5^{\pm\pm} \to W^\pm H_3^\pm$, \quad $H_5^{\pm\pm} \to W^\pm W^\pm$, \quad $H_5^{\pm\pm} \to \ell^\pm \ell^{\prime \pm}$ \\ \hline
\end{tabular}
\end{center}
\end{table}

\begin{table}[H]
\begin{center}
\caption{The production cross sections whose values are implemented by using \texttt{MadGraph5}. The channel in the parentheses means that it is already listed in the row above. The cross sections are evaluated for $\sqrt{s} = 8$ TeV and $13$ TeV.}
\label{table: prod_xsec}
\begin{tabular}{|l|l|} \hline
$H_1$ & $p p \to H_1 H_3^0$ \\[5pt]
$H_3$ & $p p \to h H_3^0$, \quad ($p p \to H_1 H_3^0$), \quad $p p \to H_3^0 H_5^0$ \\[5pt]
$H_5$ & $p p \to H_5^\pm j j$ via VBF, \quad $p p \to H_5^{++} H_5^-$, \quad $p p \to H_5^{++} H_5^{--}$ \\ \hline
\end{tabular}
\end{center}
\end{table}

\subsection{The Bayesian analysis}
\label{sec: bayesian_analysis}

Here, we perform the global fit for the CS scalar couplings by using the Bayesian analysis to examine the preferred parameter regions by the theoretical and experimental constraints discussed above. 
From the Bayes theorem, the posterior distribution for the model parameters $\vec{p}$ under the observed data $\vec{d}$ and the prior knowledge $m$ is given by
\begin{align}
\label{eq: Bayes_theorem}
p \bigl(\vec{p} \bigm|\vec{d}, m \bigr) = 
\frac{ p\bigl(\vec{d}\bigm|\vec{p},m\bigr) \times p\bigl(\vec{p} \bigm|m\bigr) }{ p\bigl(\vec{d}\bigm| m\bigr) },
\end{align}
where $p\bigl(\vec{p}\bigm|m\bigr)$ is the prior distribution, and $p\bigl(\vec{d}\, \bigm|\vec{p},m\bigr)$ is the likelihood. 
The prior knowledge $m$ provides the information on the prior distribution of the input parameters like the mean values and variances, which is summarized in Table~\ref{table: prior_knowledge}. 
We also impose auxiliary one-sided
Gaussian constraints on new particle masses in Table~\ref{table: prior_knowledge} to focus on the mass ranges accessible at next-generation high-energy colliders. 
The prior and auxiliary distributions are based on those in Ref.~\cite{Chen:2022zsh}.
\begin{table}[b]
\begin{center}
\caption{The information on the prior distribution of the input parameters and auxiliary distribution for the new scalar boson masses.}
\label{table: prior_knowledge}
\begin{tabular}{|ccccc|} \hline 
Parameters & Feature & Shape & Mean & Error/range \\ \hline
Input      &         &       & Priors & \\ \hline
$m_2^2/\mathrm{GeV^2}$ & Log & Gaussian & $10^2$ & $(10^{-4}, 10^8)$ \\ 
$\lambda_2$ & Linear & Uniform & $\cdots$ & $(-\pi, \pi)$ \\ 
$\lambda_3$ & Linear & Uniform & $\cdots$ & $(-\pi, \pi)$ \\ 
$\lambda_4$ & Linear & Uniform & $\cdots$ & $(-\pi, \pi)$ \\ 
$\lambda_5$ & Linear & Uniform & $\cdots$ & $(-\pi, \pi)$ \\ 
$\mu_1/\mathrm{GeV}$ & Linear & Gaussian &  $0$ & $(-5\times 10^3, 5\times 10^3)$ \\ 
$\mu_2/\mathrm{GeV}$ & Linear & Gaussian &  $0$ & $(-5\times 10^3, 5\times 10^3)$ \\ \hline
Auxiliary & & & Priors & \\ 
$m_{H_1,H_3,H_5}/\mathrm{GeV}$ & $\mathbb{R}_+$ & Asymmetric Gaussian & $10^{2}$ & $(50,10^3)$ \\ \hline 
\end{tabular}
\end{center}
\end{table}

The input parameters for the fitting are $m_2^2$, $\lambda_2$, $\lambda_3$, $\lambda_4$, $\lambda_5$, $\mu_1$, and $\mu_2$ in the CS Higgs potential in Eq.~(\ref{CS Higgs potential}). 
The other scalar couplings are determined as follows. 
First, we solve the second stationary condition in Eq.~(\ref{eq: stationary_cond}) for $v_\Delta$ with $v_\Phi^2 = v^2 - 8 v_\Delta^2$.\footnote{The equation is cubic in $v_\Delta$ and has one or three real solutions. We note that the latter case does not always mean that the potential has another local minimum because we determine $\lambda_1$ for a given value of $v_\Delta$. Different values of $v_\Delta$ generally lead to different $\lambda_1$, {\it i.e.}, different Higgs potentials. We do not consider the possibility that $v_\Delta$ is complex and spontaneously breaks $CP$.}
By using the obtained $v_\Delta$, we solve the first stationary condition to find $m_1^2$. 
Finally, $\lambda_1$ is determined such that the mass of $h$ is given by the observed value $\simeq 125.1$ GeV~\cite{PDG2026}. 

The likelihood function is given by the theoretical and experimental constraints on a given parameter point $\vec{p}$. 
We use $\mathcal{L}$ to represent the logarithm of the likelihood in the following:
\begin{align}
\mathcal{L} = \log p\bigl(\vec{d}\, \bigm|\vec{p},m\bigr). 
\end{align}
When the likelihood is given by a product of each probability, $\mathcal{L}$ is the sum of their logarithms.

For the theoretical constraints, we set the likelihood as the step function:
\begin{align}
\mathcal{L}_\mathrm{th} = \left\{
\begin{array}{ll}
0 & \text{Theoretical constraints are satisfied}, \\[5pt]
-\infty \quad & \text{Theoretical constraints are not satisfied}. \\
\end{array}
\right.
\end{align}
In the global fit, we evaluate the theoretical constraint in the CS limit and do not include the RGE running of the couplings because the constraint based on the running parameters is sensitive to the choice of the lower bound of the cutoff scale $\Lambda$. 

The direct search constraint for the new scalar bosons is evaluated by using \texttt{HiggsBounds}, and the likelihood $\mathcal{L}_\mathrm{HB}$ is defined in the same way as the theoretical constraint. 
On the other hand, by using \texttt{HiggsSignals}, we can obtain the $\chi^2$ for the observations of the 125 GeV Higgs couplings. 
We assume the Gaussian distribution and determine the likelihood $\mathcal{L}_\mathrm{HS}$ by 
\begin{align}
\mathcal{L}_\mathrm{HS} = - \frac{ \chi^2 }{ 2 }. 
\end{align}

In addition to the above theoretical and experimental constraints, we include the auxiliary distributions for the
new particle masses shown in Table~\ref{table: prior_knowledge} in the likelihood, which is denoted by $\mathcal{L}_\mathrm{aux}$ in the following. 

By using the likelihood and the prior distribution, we can evaluate the numerator of Eq.~(\ref{eq: Bayes_theorem}). 
Then, a chain of parameter points following the posterior distribution is given by the Markov Chain Monte Carlo (MCMC), where the normalization factor in Eq.~(\ref{eq: Bayes_theorem}) need not be evaluated. 
We implement the MCMC by the Metropolis--Hastings algorithm~\cite{Numerical_Recipes}, where the initial point is randomly selected by using the prior distribution in Table~\ref{table: prior_knowledge}. 

\begin{figure}[t]
\begin{center}
\subfigure[\ The prior distribution]{
\includegraphics[width=0.48\textwidth]{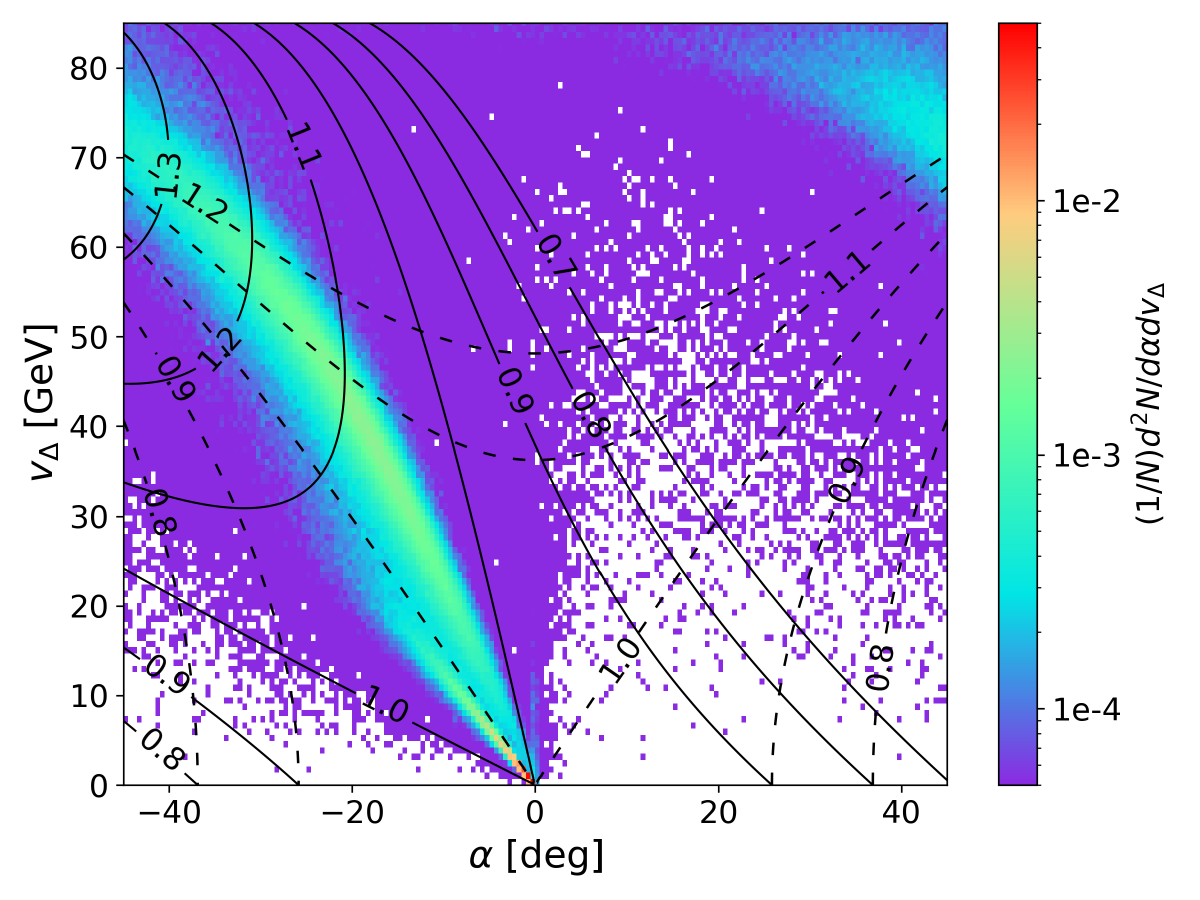}
\label{fig: prior}
}
\subfigure[\ $\mathcal{L} = \mathcal{L}_\mathrm{aux} + \mathcal{L}_\mathrm{th}$]{
\includegraphics[width=0.48\textwidth]{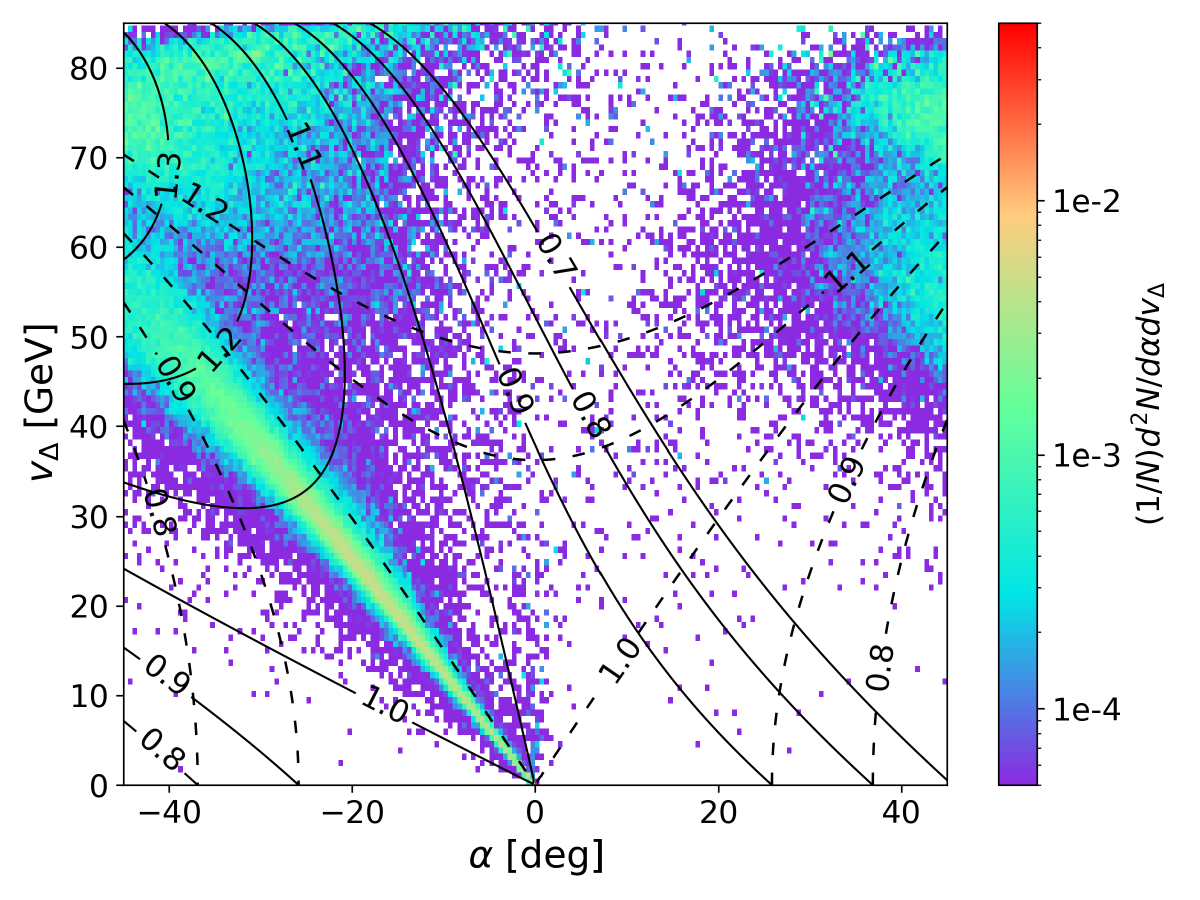}
\label{fig: ThC}
}
\subfigure[\ $\mathcal{L} = \mathcal{L}_\mathrm{aux} + \mathcal{L}_\mathrm{th} + \mathcal{L}_\mathrm{HS}$]{
\includegraphics[width=0.48\textwidth]{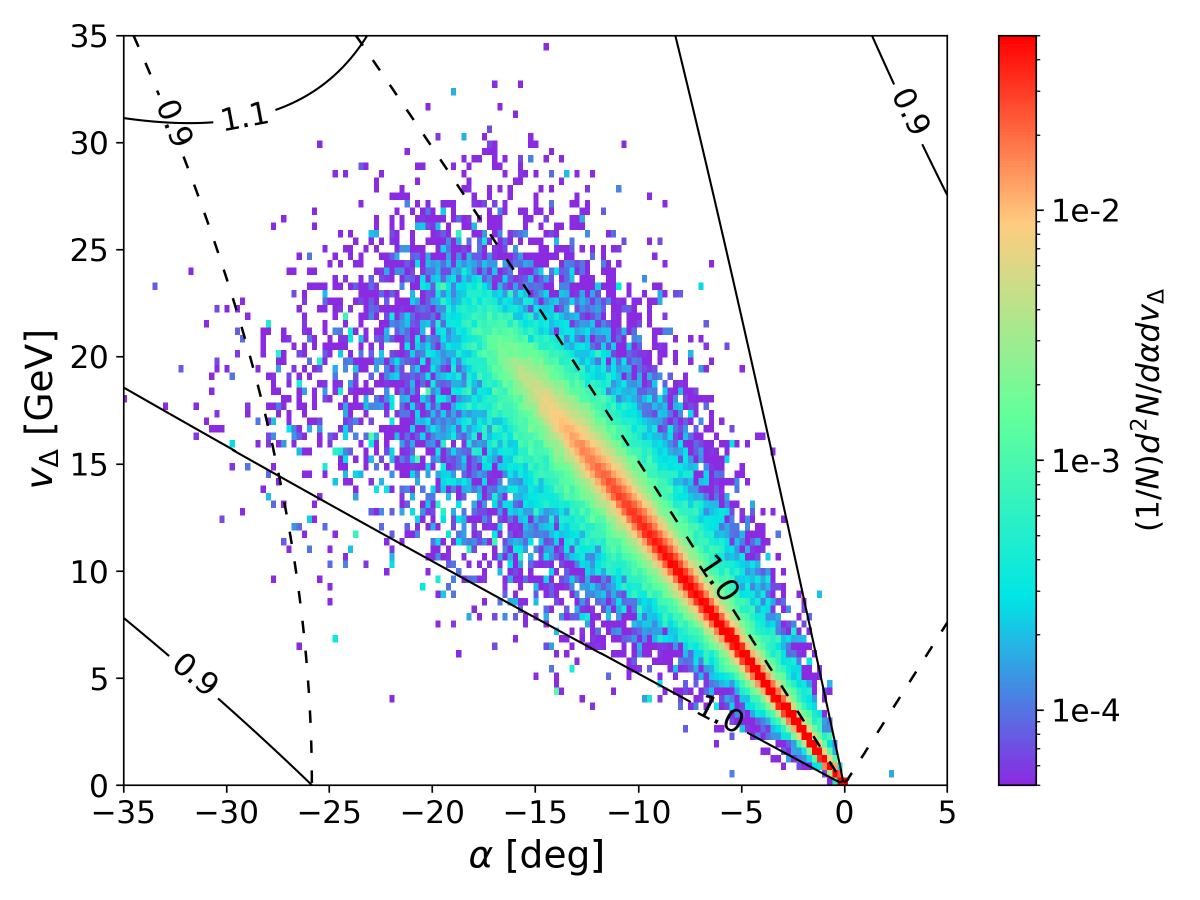}
\label{fig: ThC+HS}
}
\subfigure[\ The posterior distribution]{
\includegraphics[width=0.48\textwidth]{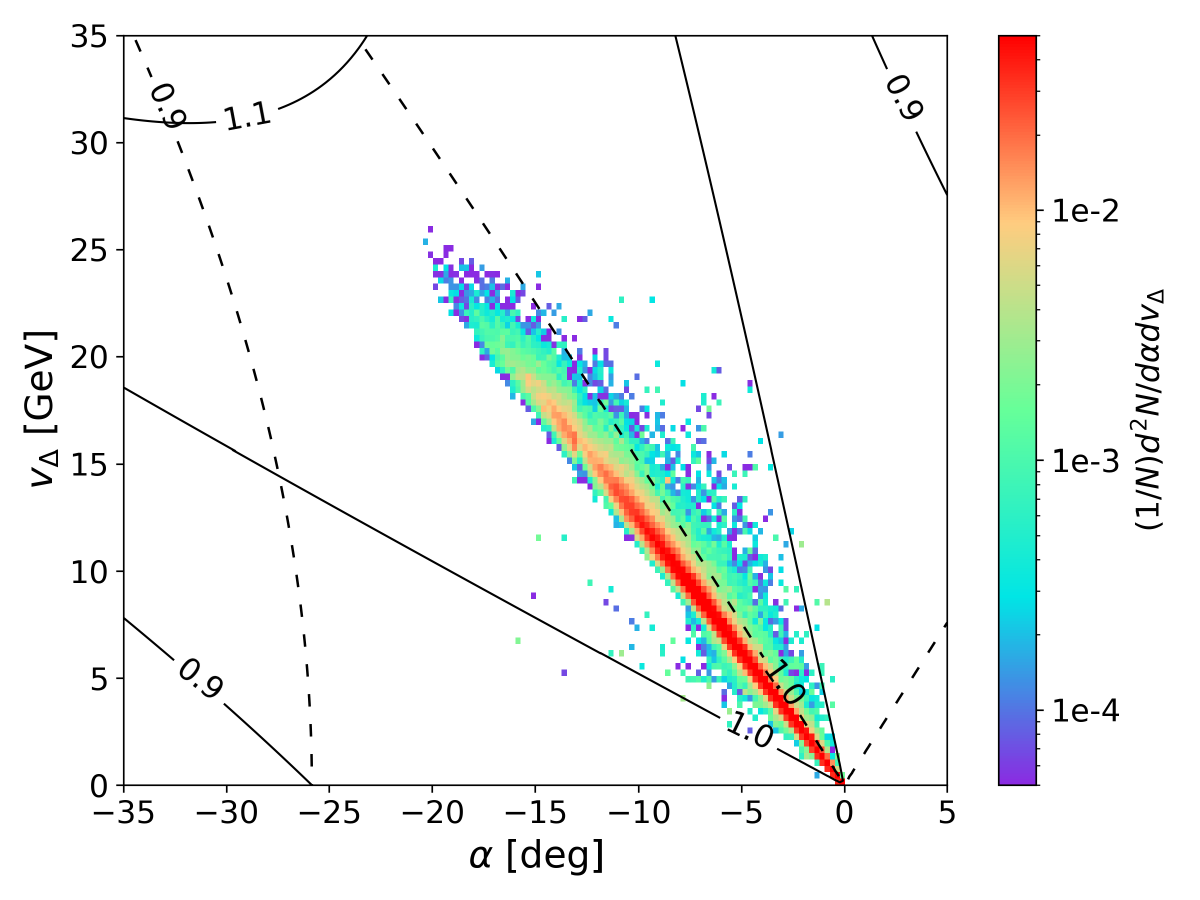}
\label{fig: global_fit}
}
\caption{Posterior distributions in the $\alpha$-$v_\Delta$ plane: (a) The prior distribution (b) The theoretical constraint and the auxiliary distribution are imposed (c) The Higgs coupling constraint is additionally imposed (d) All the constraints are imposed.  The solid and dashed contours represent $\kappa_V$ and $\kappa_f$ for the 125 GeV Higgs boson, respectively.  }
\label{fig: posteriors}
\end{center}
\end{figure}

We have generated $1.05 \times 10^6$ points from a single initial point, where $5 \times 10^4$ points are discarded in the burn-in. 
The obtained posterior distributions in the $\alpha$-$v_\Delta$ plane are shown in Fig.~\ref{fig: posteriors}.
The solid and dashed black lines on the figures are the contours for the effective vertex factors for the Higgs-vector couplings $\kappa_V$ and for the Higgs-fermion couplings $\kappa_f$, respectively, which are given by
\begin{align}
\kappa_V = \sin \beta \cos \alpha - \sqrt{\frac{8}{3}} \cos \beta \sin \alpha, \quad 
\kappa_f = \frac{ \cos \alpha }{ \sin \beta }.
\end{align}

Fig.~\ref{fig: prior} shows the prior distribution, which does not include the theoretical and experimental constraints and the auxiliary distributions. 
In Fig.~\ref{fig: ThC}, the theoretical constraint is imposed in the likelihood function in addition to the auxiliary distributions. 
The constraint from the $\chi^2$ for the 125 GeV Higgs couplings is also imposed in Fig.~\ref{fig: ThC+HS}. 
Finally, all the constraints are imposed in Fig.~\ref{fig: global_fit}, which is the final result of the global fit. 
We can observe that the final posterior distribution is concentrated around the line $\kappa_f \simeq 1$.

We have also made the same figures in the case of multiple MCMC chains and have confirmed that the shape of the distribution does not largely change. 
Thus, there is little dependence on the choice of the initial point of the MCMC. 

\section{Numerical results for the electroweak phase transition and phenomenological predictions}
\label{sec: numerical_studies}
\setcounter{equation}{0}

\subsection{Phase transition analysis}

Here, we perform the phase transition analysis by using \texttt{CosmoTransitions}~\cite{Wainwright:2011kj} for the parameter points found by the global fit in the previous section. 
In Fig.~\ref{fig: global_fit}, there are about $2.5\times 10^5$ independent parameter points, which are $10^6$ points if we include the multiplicity in the MCMC. 
Since performing a detailed phase transition analysis for every point would take an enormous amount of time, we perform the preselection via the following procedure to reduce the computational cost. 

\begin{enumerate}

\item We examine the theoretical constraints in Sec.~\ref{sec: theoretical_constraint} by using the RGE. 
We identify the scale $\Lambda$ at which at least one of the constraints is violated as the cutoff scale, above which new physics is expected to appear instead of the GM model. 
We require $\Lambda$ to satisfy
\begin{align}
\Lambda > \mathrm{max}\Bigl[ 1~\mathrm{TeV}, \ M_\mathrm{new} \Bigr], 
\end{align}
where $M_\mathrm{new}$ is the heaviest of the masses of new particles: $m_{H_1}$, $m_{H_3}$, and $m_{H_5}$. 

As a threshold effect due to the decoupling of the new particles, we only consider the running of the SM couplings with the SM beta functions at $\mu < \sqrt{|m_2^2|}$, where the right-hand side is the typical mass scale of new particles. 
The full running including all the scalar couplings starts from $\mu = \sqrt{|m_2^2|}$ after the matching with the SM running.

\item For the consistency of the one-loop effective potential, we impose that all the running quartic scalar couplings are in the perturbative regions at the scales relevant to the EWPT analysis. We note that this is different from the previous requirement because we do not consider the threshold effect of the RGE in evaluating the effective potential to completely cancel the one-loop $\mu$-dependence.\footnote{If we use the SM beta function at some scales, the one-loop $\mu$-dependence remains there. To incorporate the decoupling effect of new physics correctly, we also need to switch the effective potential formula depending on scales with the matching conditions.}

\item To select data preferred by the strong first-order EWPT, we require that the Higgs triple coupling $\lambda_{hhh}$ is enhanced by at least $5~\%$ from the SM prediction at the one-loop level~\cite{Kanemura:2004ch}. 
We evaluate $\lambda_{hhh}$ by using the third derivative of the one-loop effective potential (without the RGE improvement). 
Then, we require that 
\begin{align}
\label{eq: def_Rhhh}
R_{hhh} = \frac{ \lambda_{hhh} - \lambda_{hhh}^\mathrm{SM} }{ \lambda_{hhh}^\mathrm{SM} } > 0.05, 
\end{align}
where $\lambda_{hhh}^\mathrm{SM}$ is the SM prediction. 
Formulas to evaluate $\lambda_{hhh}$ and $\lambda_{hhh}^\mathrm{SM}$ are shown in Appendix~\ref{app: hhh}.

\end{enumerate}

\begin{table}[b]
\begin{center}
\caption{The classification of the phase transition behavior at the preselected parameter points: strong first-order (Strong 1st), weak first-order (Weak 1st), second-order (2nd), and no phase transition (NoPT). The percentage in the rightmost column indicates the proportion within the preselected data.} 
\label{table: data_classification}
\begin{tabular}{|l|c|c|}  \hline
    & Number of points & Percentage (\%) \\ \hline
Strong 1st & 1869 & 4.0 \\
Weak 1st & 38 & $8 \times 10^{-2}$ \\
2nd & 44122 & 93.7 \\
NoPT & 1063 & 2.3 \\ \hline
\end{tabular}
\end{center}
\end{table}

After the preselection, the number of independent parameter points is 47092 (about 19~\% of the original data). 
We use the \texttt{CalcTcTrans} command in \texttt{CosmoTransitions} to investigate the transition behavior at these points. 
As shown in Table~\ref{table: data_classification}, the EWPT behavior of the preselected data is classified into four groups: strong first-order (Strong 1st), weak first-order (Weak 1st), second-order (2nd), and no phase transition (NoPT). 
In the first three groups, the EWPT occurs in a single step, and no multistep EWPT occurs at any of the points. 
This fact does not rule out the possibility of multistep phase transitions in the GM model because the current analysis focuses on the preselected parameter points. 
At the strong (weak) first-order EWPT, $\varphi_c/T_c$ is larger (smaller) than one, where $\varphi_c$ is defined as $\sqrt{\varphi_1^2 + 4 \varphi_2^2 + 4 \varphi_3^2}$ at the critical temperature $T_c$.
The NoPT data points include cases such as smooth crossover, the non-restoration of the EW symmetry~\cite{Weinberg:1974hy}, and an unstable EW vacuum due to radiative corrections.

\begin{figure}[b]
\begin{center} 
\includegraphics[width=0.6\textwidth]{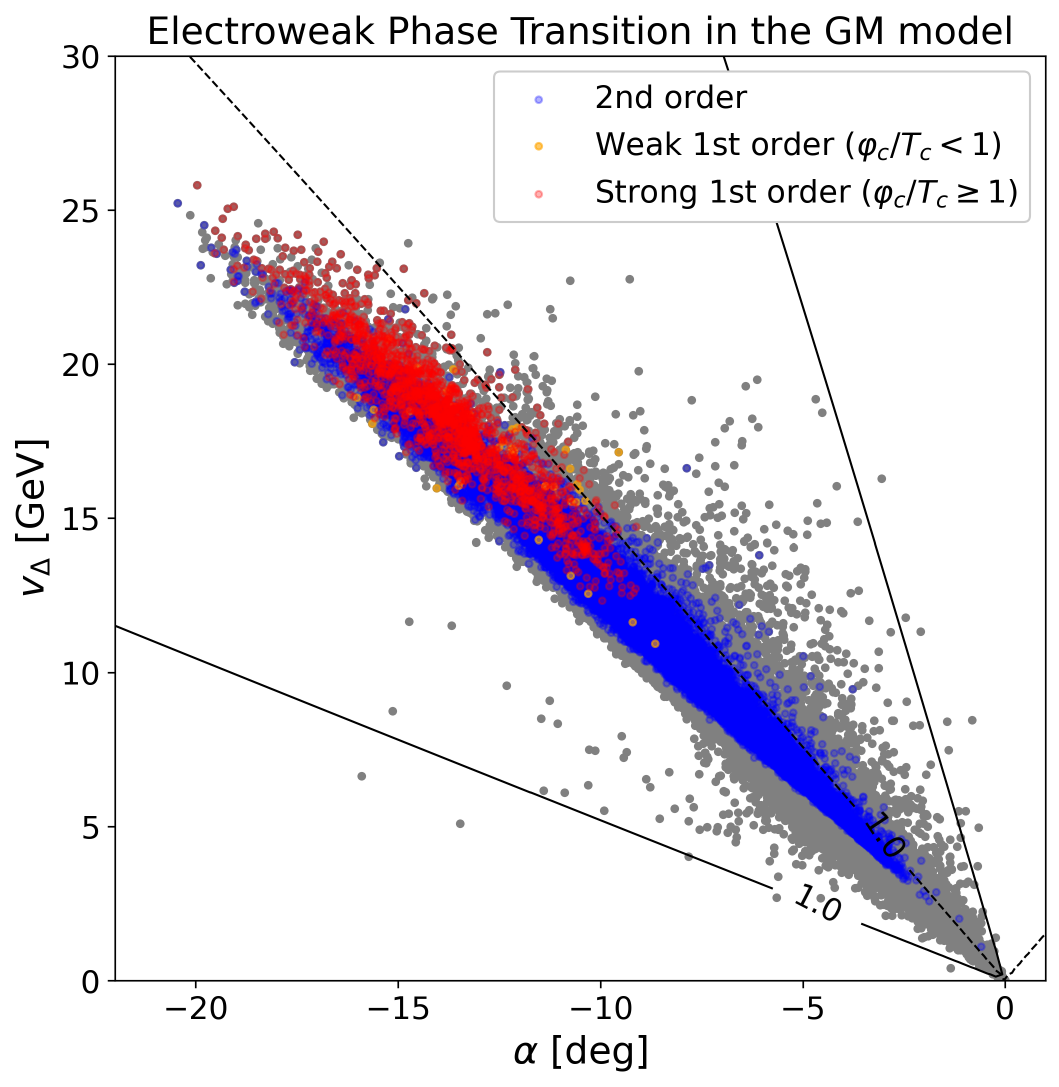}
\caption{The behavior of the electroweak phase transition in the GM model. Gray points are those excluded by the preselection or those at which no phase transition occurs.}
\label{fig: EWPT_distribution}
\end{center}
\end{figure}

In Table~\ref{table: data_classification}, we can observe that the second-order EWPT is expected at almost all the parameter points. 
The strong first-order EWPT, which is the phenomenologically most interesting case, occurs at about 4~\% of the preselected parameter points, which is 0.76~\% of the original parameter points given by the global fit.

The preselected parameter points and the EWPT behaviors are shown on the $\alpha$-$v_\Delta$ plane in Fig.~\ref{fig: EWPT_distribution}. 
The red, orange, and blue points represent the preselected points for the strong first-order, weak first-order, and second-order EWPT, respectively. 
The gray points are the points for no phase transition and those excluded by the preselection. 
We can observe that large $v_\Delta$ and $|\alpha|$ are preferred for the strong first-order EWPT. 

\begin{figure}[t]
\begin{center} 
\includegraphics[width=0.7\textwidth]{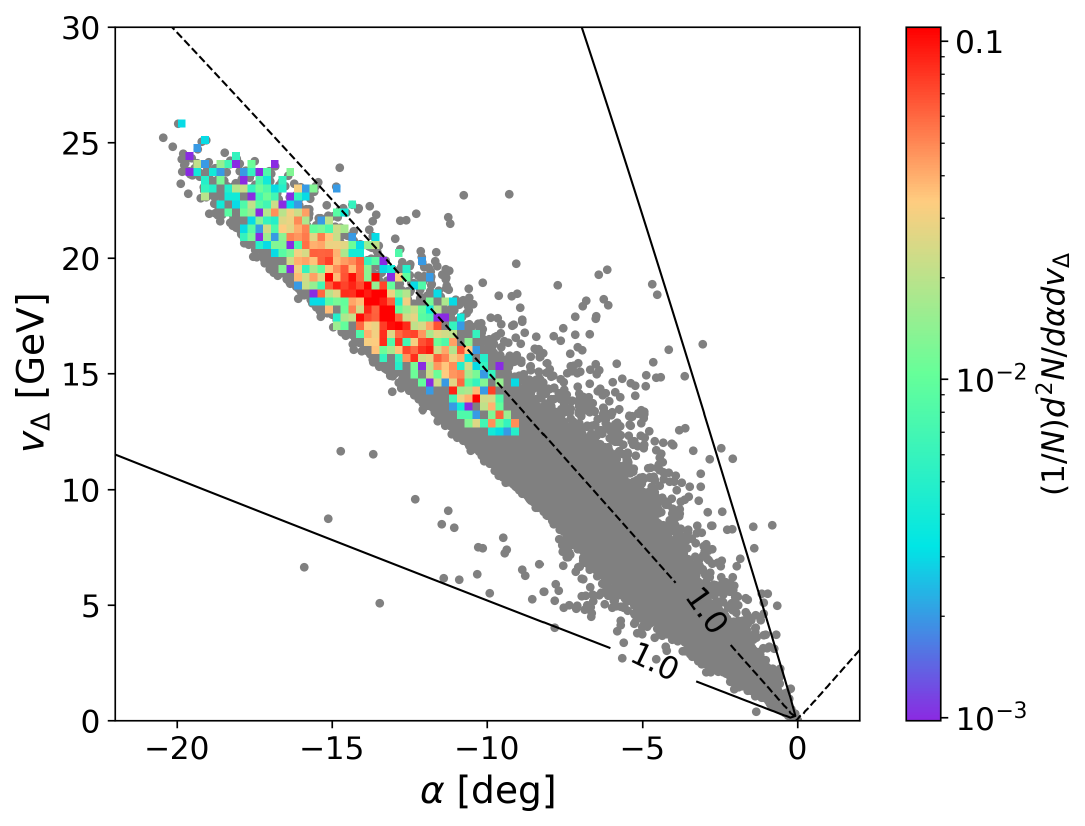}
\caption{The distribution of the parameter points for the strong first-order EWPT. The gray points are the other points generated in the global fit.}
\label{fig: 1stOPT_distribution}
\end{center}
\end{figure}

Furthermore, we show the distribution of the parameter points for the strong first-order EWPT in Fig.~\ref{fig: 1stOPT_distribution}, where the colors represent the density of the points. 
Most of the points are concentrated in the region $15~\mathrm{GeV} \lesssim v_\Delta \lesssim 20~\mathrm{GeV}$ and $-15~\mathrm{deg} \lesssim \alpha \lesssim -12~\mathrm{deg}$, which would provide phenomenological predictions at various experiments, for example, the Higgs precision measurements and direct searches for new scalar bosons. 

\begin{figure}[t]
\begin{center} 
\includegraphics[width=0.85\textwidth]{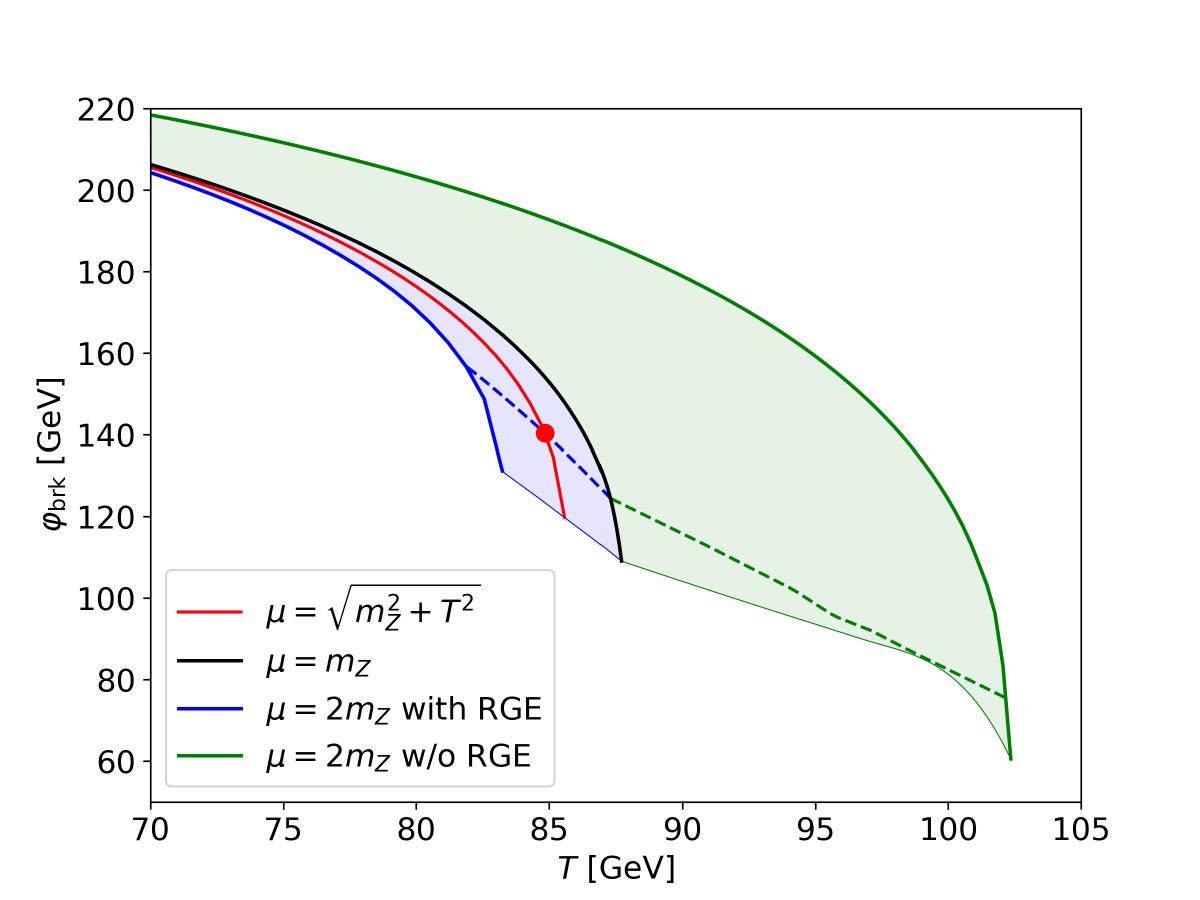}
\caption{The scale dependence of the thermal history of the broken phase on the $T$-$\varphi_\mathrm{brk}$ plane, where $\varphi_\mathrm{brk} = \sqrt{\varphi_1^2 + 4\varphi_2^2 + 4\varphi_3^2}$ at the symmetry-breaking local minimum. The black line is the result using the effective potential with $\mu = m_Z$. 
Objects colored in blue and green are given by using the effective potential with and without the RGE improvement, respectively. 
The blue and green solid lines are contours for $\varphi_\mathrm{brk}(T)$ with $\mu = 2 m_Z$, and each colored region represents possible $\varphi_\mathrm{brk}(T)$ for $m_Z < \mu < 2m_Z$. 
The dashed lines are contours of ($T_c$, $\varphi_c$) for each renormalization scale. 
The solid red line is the result with the RGE improvement using $\mu = \sqrt{ m_Z^2 + T^2 }$, and the red point represents the critical point: $(T_c, \varphi_c) \simeq$ (84.8 GeV, 140 GeV).}
\label{fig: Effect_of_RGE}
\end{center}
\end{figure}

In Fig.~\ref{fig: Effect_of_RGE}, effects of the RGE running on the scale dependence are shown at one of the parameter points for the strong EWPT. 
The input parameters are set to $m_2^2 \simeq 9.49\times 10^5~\mathrm{GeV}^2$, $\mu_1 \simeq -1.42\times 10^3$ GeV, $\mu_2 \simeq -2.98 \times 10^2$ GeV, $\lambda_2 \simeq 0.554$, $\lambda_3 \simeq -0.360$, $\lambda_4 \simeq 0.173$, and $\lambda_5 \simeq 1.31$. 
We show thermal histories of the broken phase VEV: $\varphi_\mathrm{brk} = \sqrt{ \varphi_1^2 + 4\varphi_2^2 + 4\varphi_3^2}$ with several renormalization scales. 
We compare results using the effective potential with and without the RGE improvement. 

The solid black line is a result with a fixed renormalization scale $\mu = m_Z$. 
This is common to the effective potentials with and without the RGE improvement because $\mu = m_Z$ is the initial scale of the RGE running. 
The green solid line is a result at a fixed scale $\mu = 2m_Z$ without the RGE running. 
The green region between the black and green solid lines is possible contours of $\varphi_\mathrm{brk}(T)$ at intermediate scales $m_Z < \mu < 2m_Z$. 
Thus, the size of the green region represents the scale dependence of the phase transition analysis when we use the one-loop effective potential without the RGE improvement. 
This scale dependence arises at the one-loop level. 

On the other hand, the blue solid line is a result using the RGE-improved effective potential with a fixed scale $\mu = 2m_Z$, and the blue region represents possible $\varphi_\mathrm{brk}(T)$ at intermediate scales $m_Z < \mu < 2m_Z$.  
We can observe that the blue region is much smaller than the green one. 
This is because the one-loop scale dependence is canceled in the RGE-improved effective potential as discussed in Sec.~\ref{sec: RGE}. 
The finite blue region is caused by higher-order effects of the running couplings, which will be canceled by including contributions from higher-loop diagrams. 
Consequently, the RGE improvement significantly reduces the scale dependence, {\it i.e.}, a theoretical uncertainty in the phase transition analysis using the effective potential. 
It is also remarkable that the direction of changes in the contours resulting from variations in $\mu$ is opposite between the one-loop and higher-loop effects. 
This may not be a common feature of all the parameter points, but a distinctive attribute of the current input parameters. 

The blue and green dashed lines are contours for $(T_c, \varphi_c)$ with and without the RGE improvement, respectively. 
The red solid line corresponds to the prescription employed in our analysis, where the renormalization scale is dynamically chosen: $\mu = \sqrt{m_Z^2 + T^2}$. 
The critical temperature $T_c \simeq 84.8$ GeV is located at a crossing point with the blue dashed line and is shown by the red point in the figure.
The VEV and the renormalization scale at the critical temperature are given by $\varphi_c \simeq 140$~GeV and $\mu(T_c) \simeq 125$~GeV, respectively.

\subsection{Phenomenological predictions}

Next, we discuss phenomenological predictions at the parameter points for the strong first-order EWPT. 
We consider observables closely related to the EWPT: $h \to \gamma \gamma$, $h \to Z \gamma$, the Higgs triple coupling $\lambda_{hhh}$, which plays an important role in the di-Higgs production at high-energy colliders, and the expected gravitational wave (GW) spectrum at future space-based GW observatories. 

\subsubsection{Higgs observables}

First, we discuss the Higgs observables: $h \to \gamma \gamma$, $h \to Z \gamma$, and $\lambda_{hhh}$. 
The branching ratios of $h \to \gamma \gamma$ and $h \to Z \gamma$ are derived from the analytic decay rates and the total decay width of $h$ evaluated by \texttt{HiggsTools}. 
See Sec.~\ref{subsec: experimental_constraint} for the details. 
The Higgs triple coupling $\lambda_{hhh}$ and its SM value are evaluated as explained in Appendix~\ref{app: hhh}. 

\begin{figure}[b]
\begin{center} 
\includegraphics[width=0.8\textwidth]{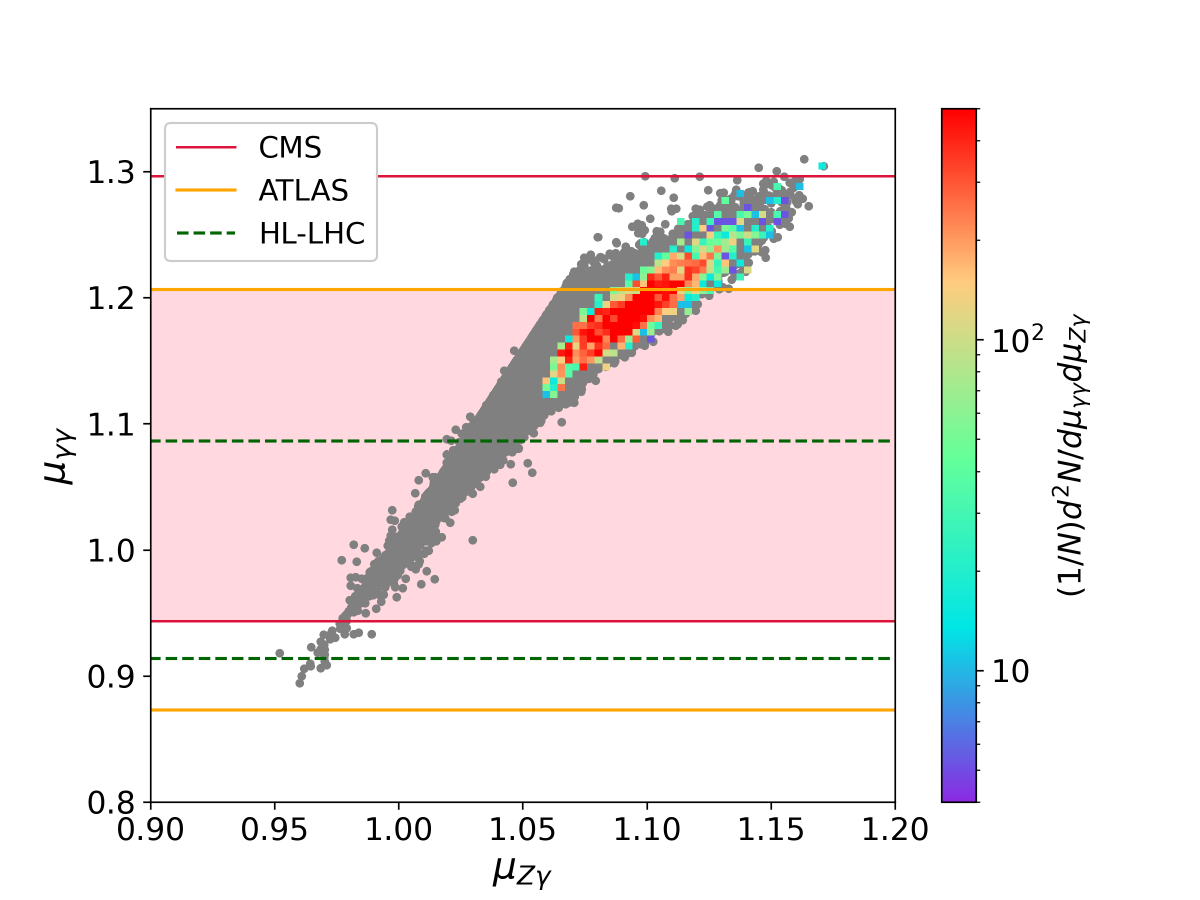}
\caption{The predictions of the signal strengths $\mu_{\gamma\gamma}$ and $\mu_{Z \gamma}$ at the parameter points for the strong first-order EWPT, whose density is represented by the colors. 
The red and orange solid lines are the latest $95~\%$ confidence level (C.L.) bounds on $\mu_{\gamma\gamma}$ reported by CMS and ATLAS, respectively. 
The pink region is consistent with both of these bounds. 
The green dashed lines are the expected sensitivity at $95~\%$ C.L. at the High-Luminosity LHC (HL-LHC). 
The gray points are the same as in Fig.~\ref{fig: 1stOPT_distribution}.}
\label{fig: gamgam_vs_Zgam}
\end{center}
\end{figure}

In Fig.~\ref{fig: gamgam_vs_Zgam}, we show the correlation between the predictions of the signal strength of $h\to \gamma \gamma$ ($\mu_{\gamma \gamma}$) and that of $h \to Z \gamma$ ($\mu_{Z\gamma}$).  
At the colored points, the strong first-order EWPT is expected, and the colors represent the density of the points. 
The gray points represent the other parameter points given by the MCMC in the global fit (before the preselection). 
We can observe a positive correlation between the signal strengths because they are induced by the same loop diagrams~\cite{Degrande:2017naf}. 

The red (orange) solid lines are the latest 95~\% C.L. bounds on $\mu_{\gamma\gamma}$ given by the CMS~\cite{CMS:2021kom} (ATLAS~\cite{ATLAS:2026pdi}) experiment from the Run 2 result. 
The pink region is consistent with both of these bounds. 
The $95~\%$ C.L. expected sensitivity at the CMS experiment at the HL-LHC is represented by the green dashed lines while assuming the central value $\mu_{\gamma\gamma} = 1$~\cite{Cepeda:2019klc}. 
The expectation in ATLAS is more conservative~\cite{Cepeda:2019klc}.

We can observe that most of the colored points are consistent with the current $h \to \gamma \gamma$ observation at $95~\%$ C.L.
At the HL-LHC, the uncertainty is reduced to about half, and a large part of the colored points can be tested more precisely. 

The current experimental uncertainty of the $h \to Z \gamma$ observation is large, which is $\mu_{Z\gamma} = 2.03^{+0.96}_{-0.93}$ by ATLAS~\cite{ATLAS:2026pdi} and $2.4 \pm 0.9$ by CMS~\cite{CMS:2022ahq} at the $1\sigma$ level. 
Thus, all the range shown in Fig.~\ref{fig: gamgam_vs_Zgam} is consistent with both observations at $95~\%$ C.L.
The future expected uncertainty at the HL-LHC is estimated as $24.2~\%$ at the $1~\sigma$ level~\cite{Cepeda:2019klc}; the corresponding region also covers the whole range of the figure if we assume the central value $\mu_{Z \gamma} = 1$. 

\begin{figure}[t]
\begin{center} 
\includegraphics[width=0.8\textwidth]{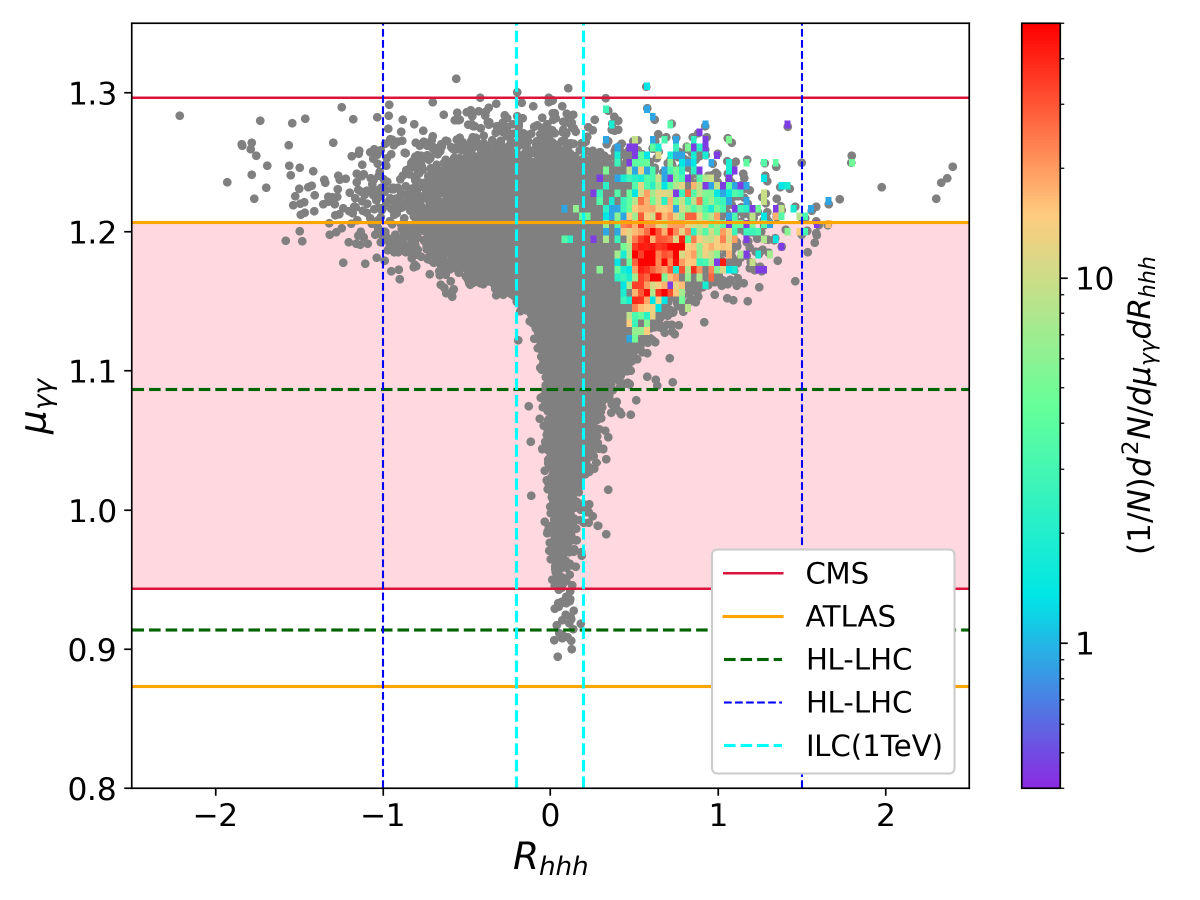}
\caption{The predictions of $\mu_{\gamma\gamma}$ and the deviation in the Higgs triple coupling $R_{hhh}$ at the parameter points for the strong first-order EWPT. 
The solid red line, solid orange line, dashed green line, and pink region are the same as in Fig.~\ref{fig: gamgam_vs_Zgam}. 
The dashed blue and dashed light blue lines are the expected sensitivities at the HL-LHC and the International Linear Collider (ILC) with $\sqrt{s}=1$ TeV, respectively. 
For the latter, we assume the central value is $R_{hhh} = 0$.}
\label{fig: gamgam_vs_hhh}
\end{center}
\end{figure}

In Fig.~\ref{fig: gamgam_vs_hhh}, we show the prediction of $\mu_{\gamma \gamma}$ and $R_{hhh}$ in Eq.~(\ref{eq: def_Rhhh}).   
The colored points represent the distribution of the parameter points for the strong first-order EWPT, and the gray points are the same as in Fig.~\ref{fig: gamgam_vs_Zgam}. 
At the colored points, the minimum of $R_{hhh}$ is about $6.7~\%$, which is larger than the criterion in the preselection: $5~\%$. 
Also, the distribution is concentrated around $R_{hhh} \simeq 0.6$. 
Thus, we expect that the preselection does not largely change the parameter region for the strong first-order EWPT. 

The solid red line, solid orange line, dashed green line, and pink region are the same as in Fig.~\ref{fig: gamgam_vs_Zgam}.  
The current observed limit on $\kappa_{hhh} = \lambda_{hhh}/\lambda_{hhh}^\mathrm{SM} = R_{hhh} + 1$ is given by $\kappa_{hhh} = 1.3^{+3.1}_{-1.6}$ at the $1~\sigma$ level~\cite{ATLAS:2026pdi}.
Thus, all the regions in Fig.~\ref{fig: gamgam_vs_hhh} are consistent with the current $\lambda_{hhh}$ observation. 
The future sensitivity at the HL-LHC is estimated as $0.5<\kappa_{hhh}<1.6$ and $0.0<\kappa_{hhh}<2.5$ at the 68~\% and 95~\% C.L., respectively~\cite{ATLAS:2022faz}.
The latter is shown by the dashed blue lines in Fig.~\ref{fig: gamgam_vs_hhh}. 
On the other hand, at the ILC with $\sqrt{s} = 1~\mathrm{TeV}$, $10~\%$ precision can be ensured for any value of $\lambda_{hhh}$, according to Ref.~\cite{Torndal:2023fky}.  
The dashed light blue lines in Fig.~\ref{fig: gamgam_vs_hhh} represent the $95~\%$ C.L. expected sensitivity at the ILC assuming a central value $R_{hhh}=0$. 
We can observe that the future precise $\lambda_{hhh}$ measurement at the ILC is especially effective in testing the strong first-order EWPT in the GM model. 

\begin{figure}[t]
\begin{center} 
\includegraphics[width=0.8\textwidth]{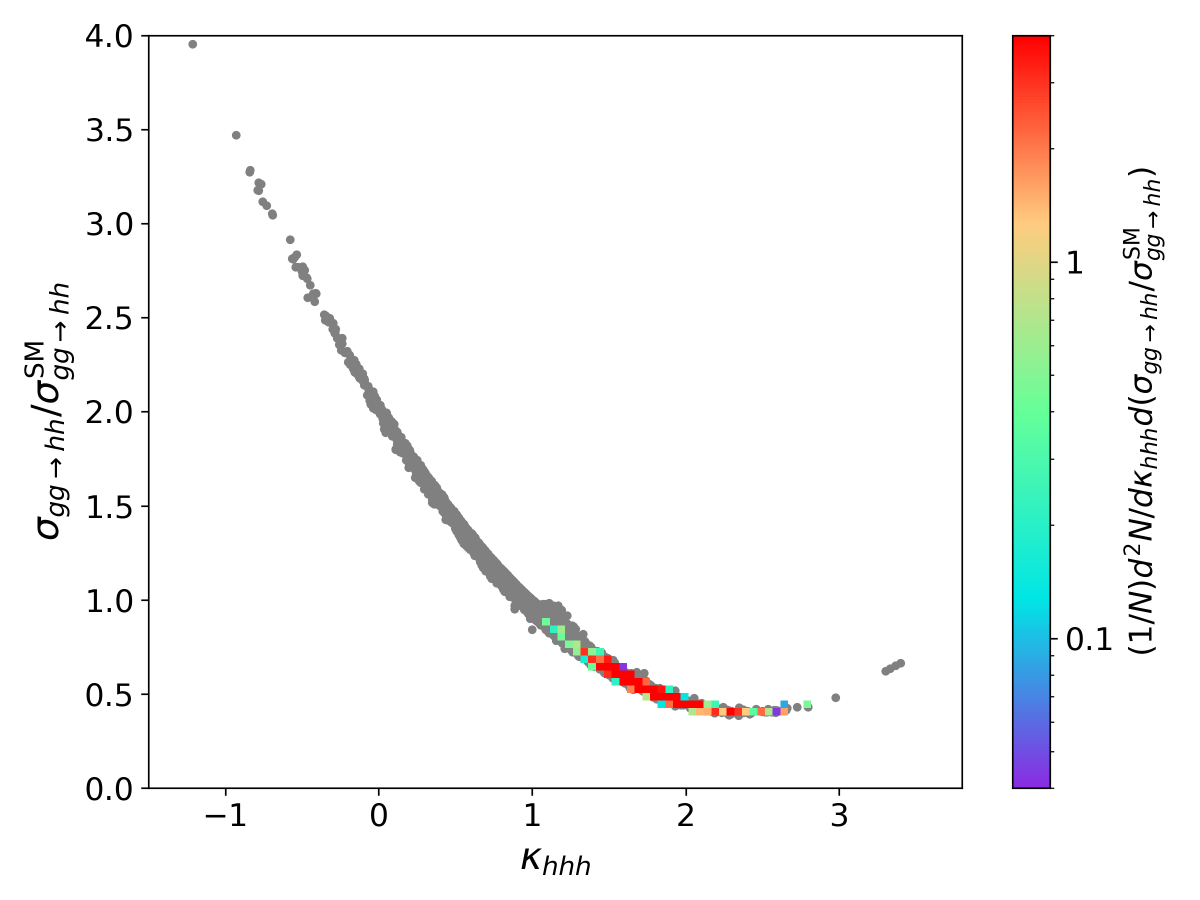}
\caption{The ratio of the gluon fusion cross section for the di-Higgs production to the SM prediction and the deviation in the Higgs triple coupling: $\kappa_{hhh} = R_{hhh} + 1$. At the colored points, the EWPT is strongly first-order. The gray points are the same as in Fig.~\ref{fig: 1stOPT_distribution}. }
\label{fig: kap3_vs_diHiggs}
\end{center}
\end{figure}

Nonzero $R_{hhh}$ results in the deviation in the di-Higgs production cross section from the SM prediction. 
In Fig.~\ref{fig: kap3_vs_diHiggs}, the prediction of the cross section of the gluon fusion $gg \to hh$ ($\sigma_{gg\to hh}$) is shown. 
The colored points represent the distribution of the parameter points, and the gray points are the same as in Figs.~\ref{fig: gamgam_vs_Zgam} and \ref{fig: gamgam_vs_hhh}. 
The horizontal axis is $\kappa_{hhh} = R_{hhh} + 1 = \lambda_{hhh}/\lambda_{hhh}^\mathrm{SM}$, and the vertical axis is the ratio of the cross section to the SM value ($\sigma_{gg \to hh}^\mathrm{SM}$). 
The cross section has been evaluated by using \texttt{HPAIR}~\cite{ref:HPAIR}. 
The gluon fusion process is also affected by $\kappa_f$, the deviation in the Yukawa coupling. 
However, as shown in Fig.~\ref{fig: global_fit}, the distribution of the parameter points is clustered around the contour $\kappa_f = 1$. 
Therefore, the distribution exhibits a tight correlation, and the cross section resembles a single-valued function of $\kappa_{hhh}$. 
The points for the strong first-order EWPT predict $\kappa_{hhh} > 1$, which results in a smaller cross section than the SM one. 
In addition, they are concentrated around $\sigma_{gg \to hh}/\sigma_{gg\to hh}^\mathrm{SM} = 0.5$.

\subsubsection{Gravitational waves observation}

At the first-order EWPT, bubbles of the true vacuum are nucleated. 
If the nucleation rate is sufficiently large, the nucleated bubbles expand and finally cover the entire Universe.
As bubbles grow, they collide with other bubbles, and GWs are generated in association with the collisions. 
These GWs can be detected as a stochastic background at future space-based observatories such as LISA~\cite{LISA:2017pwj}, DECIGO~\cite{Seto:2001qf}, and BBO~\cite{Corbin:2005ny} if the phase transition is sufficiently strong. 
See Refs.~\cite{Caprini:2015zlo, Athron:2023xlk} and references therein for details. 
The observation of such GWs is a powerful test of the strong first-order EWPT~\cite{Nicolis:2003tg, Grojean:2006bp}. 

The GW spectra can be estimated by using three parameters determined by the nature of the nucleation: phase transition strength parameter $\alpha_\theta$, inverse time duration $\beta_t$ ($\tilde{\beta}_t$), and the bubble wall velocity $v_w$. 
$\alpha_\theta$ and $\beta_t$ ($\tilde{\beta}_t$) are determined by the effective potential and the bounce action $S_3$: 
\begin{align}
& \alpha_\theta = \frac{ 1 }{ \rho_\ast } 
    \biggl( \Delta V - \frac{T_\ast}{4} \frac{ \mathrm{d} \Delta V }{ \mathrm{d} T } \biggr)_{T=T_\ast}, 
\quad \tilde{\beta}_t = \frac{ \beta_t }{ H_\ast } = T_\ast \biggl(\frac{ \mathrm{d} }{ \mathrm{d} T } \frac{S_3}{T} \biggr)_{T=T_\ast}, 
\end{align}
where $\Delta V$ is the potential difference between the false vacuum and the true vacuum, $\rho_\ast$ and $H_\ast$ are the radiation energy density of the Universe and the Hubble parameter at $T=T_\ast$, and $T_\ast$ is the temperature at which the bubble collision occurs. 
In the following, we assume that the phase transition proceeds promptly, {\it i.e.}, the bubbles collide soon after the nucleation. 
Then, $T_\ast$ is approximated by the nucleation temperature $T_n$, where one bubble is expected to be nucleated in each Hubble volume. 
We use the standard criterion $S_3 / T = 140$ to estimate $T_n$~\cite{Anderson:1991zb, Wainwright:2011kj}. 

To evaluate the bubble wall velocity, we need to examine the dynamics of the bubble expansion, taking into account the friction with non-equilibrium background plasma~\cite{Moore:1995si}. 
Despite recent progress in these calculations, there is still a large theoretical uncertainty in the estimation of $v_w$~\cite{vandeVis:2025plm}. 
Here, we thus treat $v_w$ as a free parameter and use $v_w = 0.1$ and 0.6 as typical input values. 

There are three sources of the GW production at the EWPT: (i) bubble collisions~\cite{Kosowsky:1991ua, Kosowsky:1992rz, Kosowsky:1992vn, Kamionkowski:1993fg, Caprini:2007xq, Huber:2008hg}, (ii) sound waves~\cite{Hindmarsh:2013xza, Giblin:2013kea, Giblin:2014qia, Hindmarsh:2015qta}, and (iii) magnetohydrodynamic turbulence~\cite{Caprini:2006jb, Kahniashvili:2008pf, Kahniashvili:2008pe, Kahniashvili:2009mf, Caprini:2009yp, Kisslinger:2015hua}.  
By using $\alpha_\theta$, $\tilde{\beta}_t$, and $v_w$, the GW spectra as function of the frequency $f$ due to each source are given by~\cite{Caprini:2015zlo, Ellis:2020awk, Guo:2020grp} 
\begin{align}
& h^2 \Omega_\mathrm{col} = 1.67 \times 10^{-5} \frac{1}{\tilde{\beta}^{2}_t} 
    \biggl( \frac{ \kappa_\mathrm{col} \alpha_\theta }{ 1 + \alpha_\theta } \biggr)^2
    \biggl( \frac{ 100 }{ g_\ast } \biggr)^{1/3}
    \biggl( \frac{ 0.11 v_w^3 }{ 0.42 + v_w^2 } \biggr)
    \frac{ 3.8(f/f_\mathrm{col})^{2.8} }{ 1 + 2.8(f/f_\mathrm{col})^{3.8} }, \\[10pt]
\label{eq: GW_sw}
& h^2 \Omega_\mathrm{sw} = 1.21\times 10^{-6} \Upsilon \frac{ \mathrm{max}(v_w, c_s) }{ \tilde{\beta}_t }
    \biggl( \frac{ \kappa_\mathrm{sw} \alpha_\theta }{ 1 + \alpha_\theta } \biggr)^2
    \biggl( \frac{ 100 }{ g_\ast } \biggr)^{1/3} 
    \biggl( \frac{ f }{ f_\mathrm{sw} } \biggr)^3 
    \biggl( \frac{ 7 }{ 4 + 3 (f/f_\mathrm{sw})^2 } \biggr)^{7/2}, \\[10pt]
& h^2 \Omega_\mathrm{turb} = 3.35 \times 10^{-4} \frac{\mathrm{max}(v_w, c_s)}{ \tilde{\beta}_t }
    \biggl( \frac{ \kappa_\mathrm{turb} \alpha_\theta }{ 1 + \alpha_\theta } \biggr)^{3/2}
    \biggl( \frac{ 100 }{ g_\ast } \biggr)^{1/3}
    \frac{ (f/f_\mathrm{turb})^3 }{ (1 + f/f_\mathrm{turb})^{11/3}(1 + 8\pi f/h_\ast) }, 
\end{align}
respectively, where $c_s$ is the sound speed, $g_\ast$ is the effective number of relativistic degrees of freedom at the collision, and $h_\ast$ is the inverse Hubble time at GW production, redshifted to today:
\begin{align}
h_\ast = 16.5\times10^{-3}\mathrm{mHz}
    \biggl( \frac{ T_n }{ 100~\mathrm{GeV} } \biggr)
    \biggl( \frac{ g_\ast }{ 100 } \biggr)^{1/6}.
\end{align}
In Eq.~(\ref{eq: GW_sw}), we include a suppression factor $\Upsilon$ due to the finite lifetime of
the sound waves $\tau_\mathrm{sw}$~\cite{Ellis:2020awk, Guo:2020grp}:
\begin{align}
\Upsilon = 1 - \frac{ 1 }{ \sqrt{ 1 + 2 \tau_\mathrm{sw} H_\ast} }, 
\end{align}
where $\tau_\mathrm{sw} H_\ast = H_\ast R_\ast / \bar{U}_f$, and 
\begin{align}
H_\ast R_\ast = (8\pi)^{1/3}\frac{ \mathrm{max}(v_w, c_s) }{ \tilde{\beta}_t }, \quad
\bar{U}_f = \sqrt{\frac{3}{4} \frac{\kappa_\mathrm{sw} \alpha_\theta }{ 1 + \alpha_\theta} }.
\end{align}
Peak frequencies $f_\mathrm{col}$, $f_\mathrm{sw}$, and $f_\mathrm{turb}$ are given by 
\begin{align}
& f_\mathrm{col} = 1.7\times 10^{-2}\mathrm{mHz} \times 
    \tilde{\beta}_t
    \biggl( \frac{ 0.62 }{ 1.8 - 0.1 v_w + v_w^2 } \biggr) 
    \biggl( \frac{ T_n }{ 100~\mathrm{GeV} } \biggr)
    \biggl( \frac{ g_\ast }{ 100 } \biggr)^{1/6}, \\[10pt]
& f_\mathrm{sw} = 8.9\times 10^{-3} \mathrm{mHz}
    \times \frac{ \tilde{\beta}_t }{ \mathrm{max}(v_w, c_s) }
    \biggl( \frac{ T_n }{ 100~\mathrm{GeV} } \biggr)
    \biggl( \frac{ g_\ast }{ 100 } \biggr)^{1/6}, \\[10pt]
& f_\mathrm{turb} = 2.7\times 10^{-2} \mathrm{mHz}
    \times \frac{ \tilde{\beta}_t }{ \mathrm{max}(v_w, c_s) }
    \biggl( \frac{ T_n }{ 100~\mathrm{GeV} } \biggr)
    \biggl( \frac{ g_\ast }{ 100 } \biggr)^{1/6}.
\end{align}
$\kappa_\mathrm{col}$, $\kappa_\mathrm{sw}$, and $\kappa_\mathrm{turb}$ are the conversion efficiencies of the phase transition energy to kinetic energy, bulk motion of the fluid, and turbulence, respectively~\cite{Espinosa:2010hh, Huber:2008hg}: 
\begin{align}
& \kappa_\mathrm{col} = \frac{ 1 }{ 1+ 0.715 \alpha_\theta } \biggl[ 0.715\alpha_\theta + \frac{4}{27}\sqrt{\frac{ 3 \alpha_\theta }{ 2 } } \biggr], \\[10pt]
& \kappa_\mathrm{sw} = \left\{
\begin{array}{ll}
\displaystyle{ 
    \kappa_A = \frac{ 6.9 v_w^{6/5} \alpha_\theta}{ 1.36 - 0.037\sqrt{\alpha_\theta} + \alpha_\theta }
}& (v_w < c_s), \\[10pt]
\displaystyle{ 
    \kappa_B = \frac{ \alpha_\theta^{2/5} }{ 0.017 + (0.997 + \alpha_\theta)^{2/5} }
}& (v_w = c_s), 
\end{array}
\right. \\
& \kappa_\mathrm{turb} = \xi_\mathrm{turb}\kappa_\mathrm{sw}, 
\end{align}
where $\xi_\mathrm{turb}$ is the fraction of turbulent bulk motion. 
A typical choice of $\kappa_\mathrm{turb}$ is a few percent of $\kappa_\mathrm{sw}$: $\xi_\mathrm{turb} = 5$--10~\%~\cite{Caprini:2015zlo}. 
On the other hand, in Ref.~\cite{Ellis:2020awk}, the finite lifetime of the sound waves is taken into account, and the fraction is estimated as $\xi_\mathrm{turb} = \bigl\{ 1 - \mathrm{min}(\tau_\mathrm{sw} H_\ast, 1) \bigr\}^{2/3}$, assuming that all the energy in the bulk fluid motion remaining at the end of the sound wave period is transferred into turbulence. 
As discussed in Ref.~\cite{Ellis:2020awk}, this conversion efficiency may result in an upper bound on the GW spectrum generated by the turbulence. 
In the following analysis, we employ $\xi_\mathrm{turb} = 0.1$, according to Ref.~\cite{Caprini:2015zlo}. 
As the conversion efficiency for the sound waves, we use $\kappa_A$ for $v_w = 0.1$ and $\kappa_B$ for $v_w = 0.6$. 
The latter is a reasonable approximation because $c_s \simeq 0.577$. 

In the above estimation of the GW spectra, it is assumed that $T_n$ is the typical temperature at the bubble collisions, and the Universe is radiation-dominated at $T=T_n$. 
For the consistency of these assumptions, we require that the following two conditions are satisfied at the parameter points for the strong first-order EWPT:\footnote{At a strongly supercooled EWPT in which these conditions are violated, we should use the percolation temperature alternatively~\cite{Leitao:2015fmj, Athron:2022mmm}.}  
\begin{align}
\label{eq: moderate_supercooling}
\frac{ T_n }{ T_c } > 0.3, \quad \alpha_\theta^{} < 1. 
\end{align}
Since $\alpha_\theta^{} \simeq \Delta V/\rho_\ast$ in the strong supercooling case, the latter is a good approximation for the condition such that the radiation energy dominates the Universe~\cite{Ellis:2018mja}. 
After this cut, 1455 parameter points remain, which is about 78~\% of the points for the strong first-order EWPT. 

\begin{figure}[t]
\includegraphics[width=0.49\textwidth]{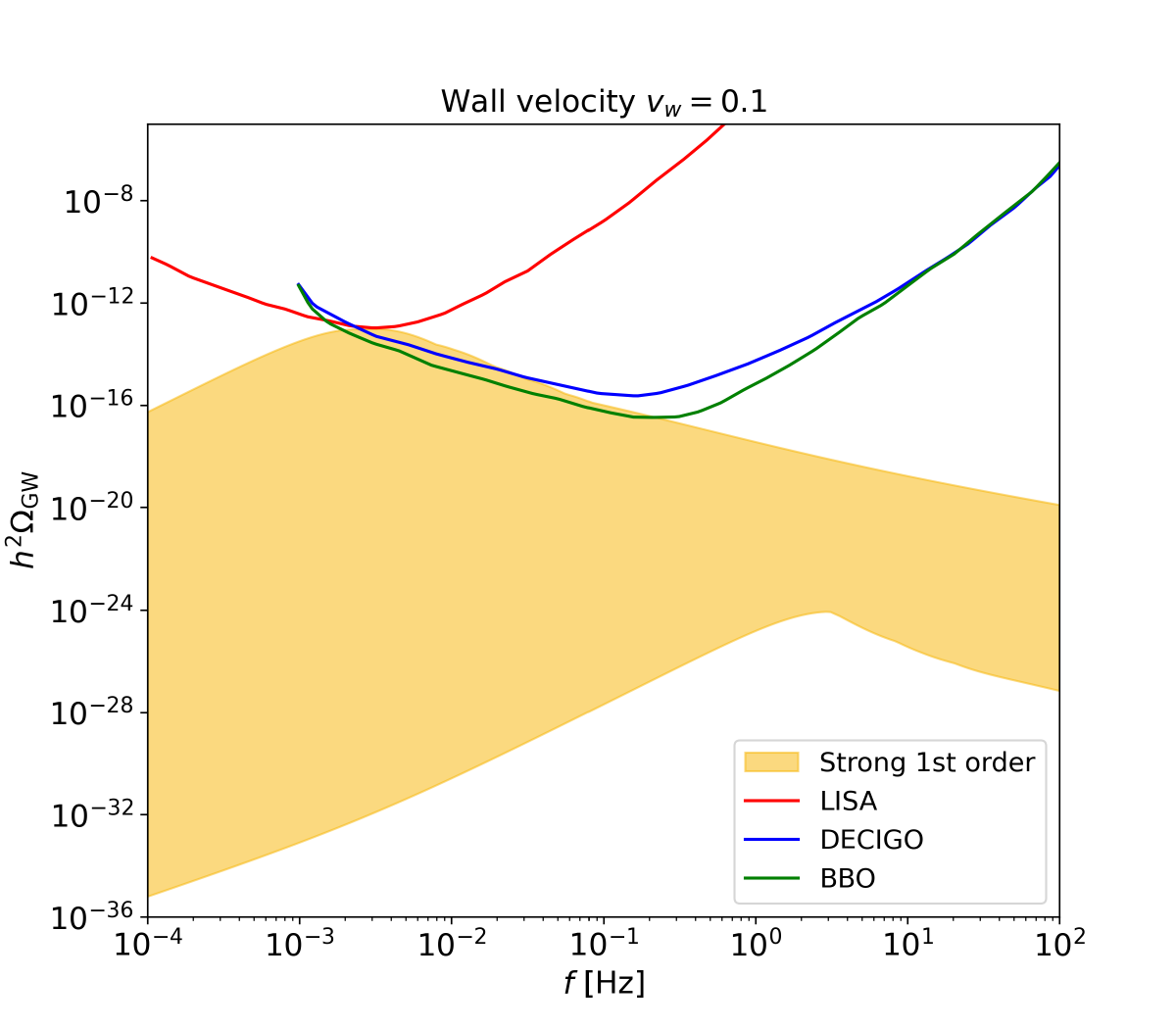}
\includegraphics[width=0.49\textwidth]{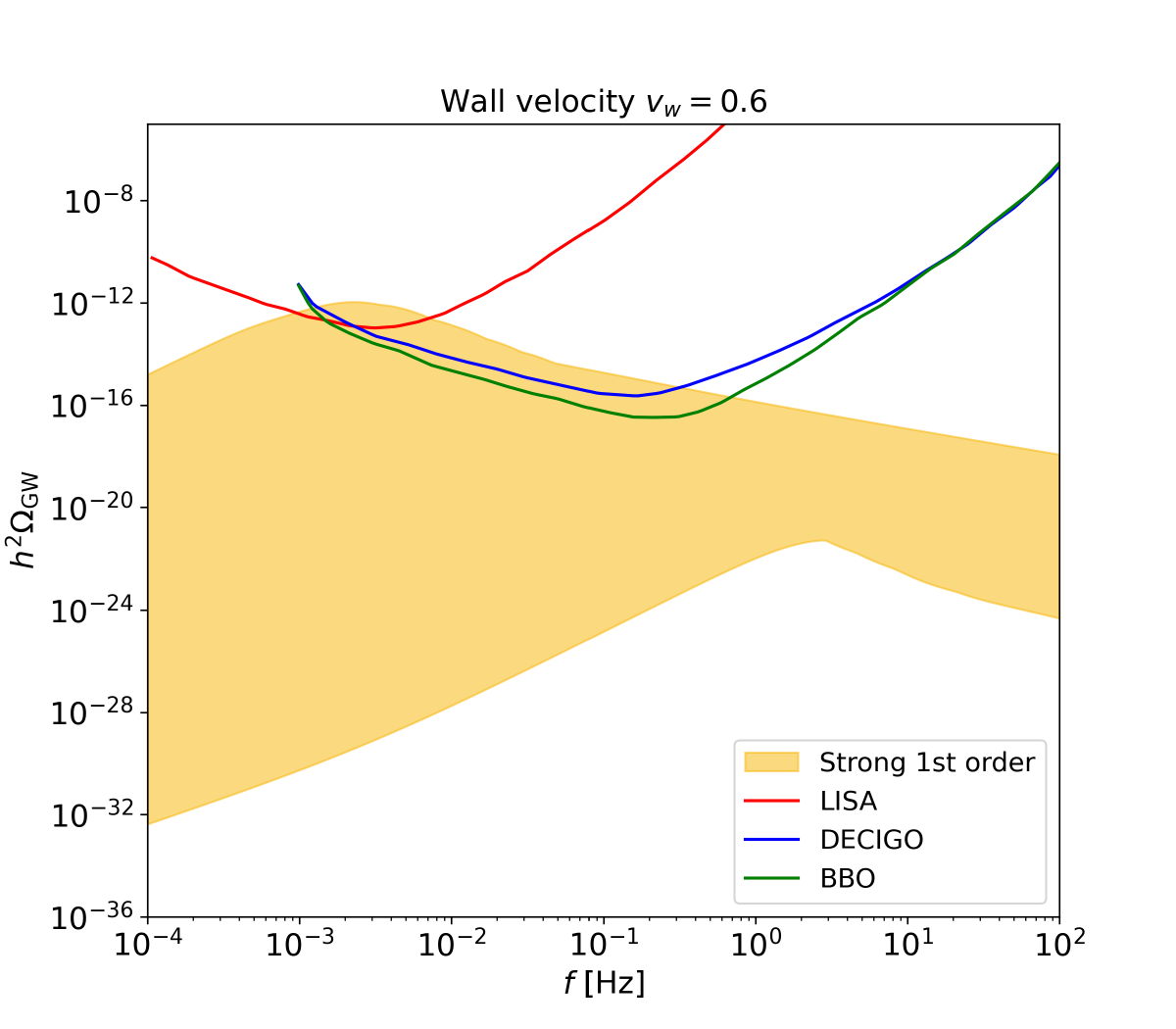}
\caption{The GW spectra at the parameter points for the strong first-order EWPT with $v_w=0.1$ (left) and $v_w=0.6$ (right). The colored curves are the sensitivity curves at LISA (red), DECIGO (blue), and BBO (green).}
\label{fig: GW_spectrum}
\end{figure}

In Fig.~\ref{fig: GW_spectrum}, a region swept by the possible GW spectra at the parameter points satisfying Eq.~(\ref{eq: moderate_supercooling}) is shown by the yellow band. 
The colored lines are the sensitivity curves at LISA (red), DECIGO (blue), and BBO (green), which are from Ref.~\cite{Chen:2022zsh}.  
If the spectra are beyond the sensitivity curves, they are expected to be detected at each observatory. 
The detectability is hierarchical for the points considered here, {\it i.e.}, the spectra that exceed the sensitivity of LISA also exceed the sensitivity of the other two experiments; similarly, spectra that exceed the sensitivity of DECIGO also exceed that of BBO.

\begin{table}[t]
\begin{center}
\caption{The number of parameter points testable at each observatory.}
\label{table: GW_testability}
\begin{tabular}{|c|c|c|c|c|} \hline
    & \ LISA \& DECIGO \& BBO \quad & \ DECIGO \& BBO \quad & \ BBO \quad & \ undetectable \quad \\ \hline
$v_w=0.1$ & 0 (0~\%) & 32 (2.2~\%) & 79 (5.4~\%) & 1344 (92.4~\%) \\ \hline
$v_w=0.6$ & 27 (1.9~\%) & 254 (17.5~\%) & 237 (16.3~\%) & 937 (64.4~\%) \\ \hline
\end{tabular}
\end{center}
\end{table}

The number of testable parameter points at each observatory is summarized in Table~\ref{table: GW_testability}. 
We can see that only about $7.6$~\% of the points satisfying Eq.~(\ref{eq: moderate_supercooling}) are testable at the observatories in the case of $v_w = 0.1$. 
The testability increases for the larger velocity $v_w=0.6$, and GWs are expected to be detected at about 36~\% of the points. 

We note that the suppression factor $\Upsilon$ in $h^2 \Omega_\mathrm{sw}$ changes the $v_w$ dependence of the spectrum. 
The velocity dependence of $h^2\Omega_\mathrm{sw}$ at the peak frequency is given by
\begin{align}
h^2\Omega_\mathrm{sw} (f = f_\mathrm{sw}) \propto
\mathrm{max}(v_w, c_s) \Upsilon \kappa_\mathrm{sw}^2
= \left\{ 
\begin{array}{ll}
\mathrm{max}(v_w,c_s)^2 \kappa_\mathrm{sw}^{3/2} & (\Upsilon \simeq \tau_\mathrm{sw} H_\ast \ll 1) \\[5pt]
\mathrm{max}(v_w,c_s) \kappa_\mathrm{sw}^2 & (\Upsilon = 1), 
\end{array}
\right.
\end{align}
where the latter corresponds to the case neglecting the suppression.
Since $\Upsilon \ll 1$ is valid in most realistic models~\cite{Ellis:2018mja}, the inclusion of $\Upsilon$ significantly changes not only the spectrum but also the velocity dependence. 
For example, in the present analysis, the velocity dependence is reduced by $\Upsilon$.

\begin{figure}[t]
\includegraphics[width=0.49\textwidth]{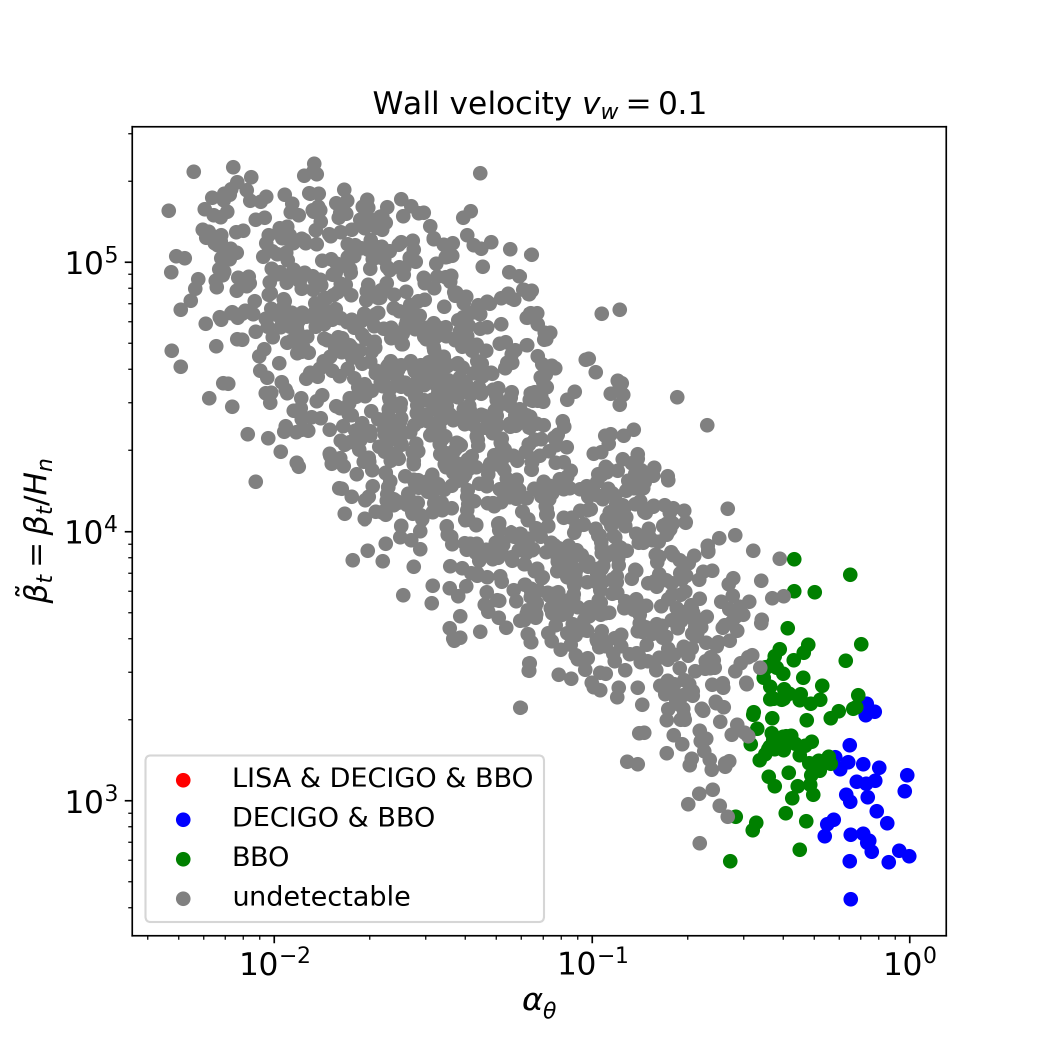}
\includegraphics[width=0.49\textwidth]{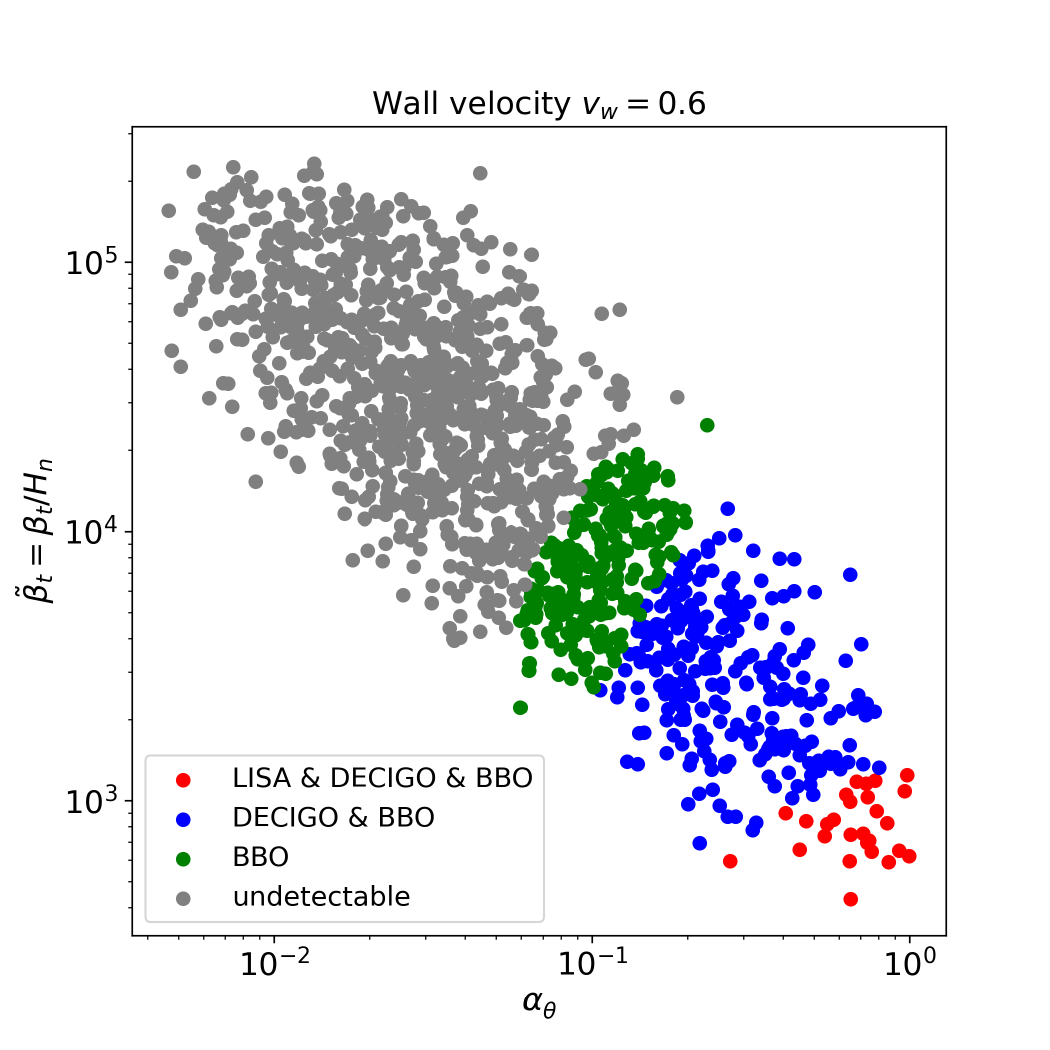}
\caption{Detectability of the GWs at the future space-based observatories on the $\alpha_\theta$-$\tilde{\beta}_t$ plane.
Green points represent parameter points where GWs are expected to be detected by BBO; blue points, by both BBO and DECIGO; and red points, by LISA, BBO, and DECIGO.
At gray points, the energy density spectrum is too small to be detected at these experiments.
In the left (right) figure, the bubble wall velocity is assumed to be $0.1$ ($0.6$).} 
\label{fig: Gwplot1}
\end{figure}

\begin{figure}[t]
\includegraphics[width=0.49\textwidth]{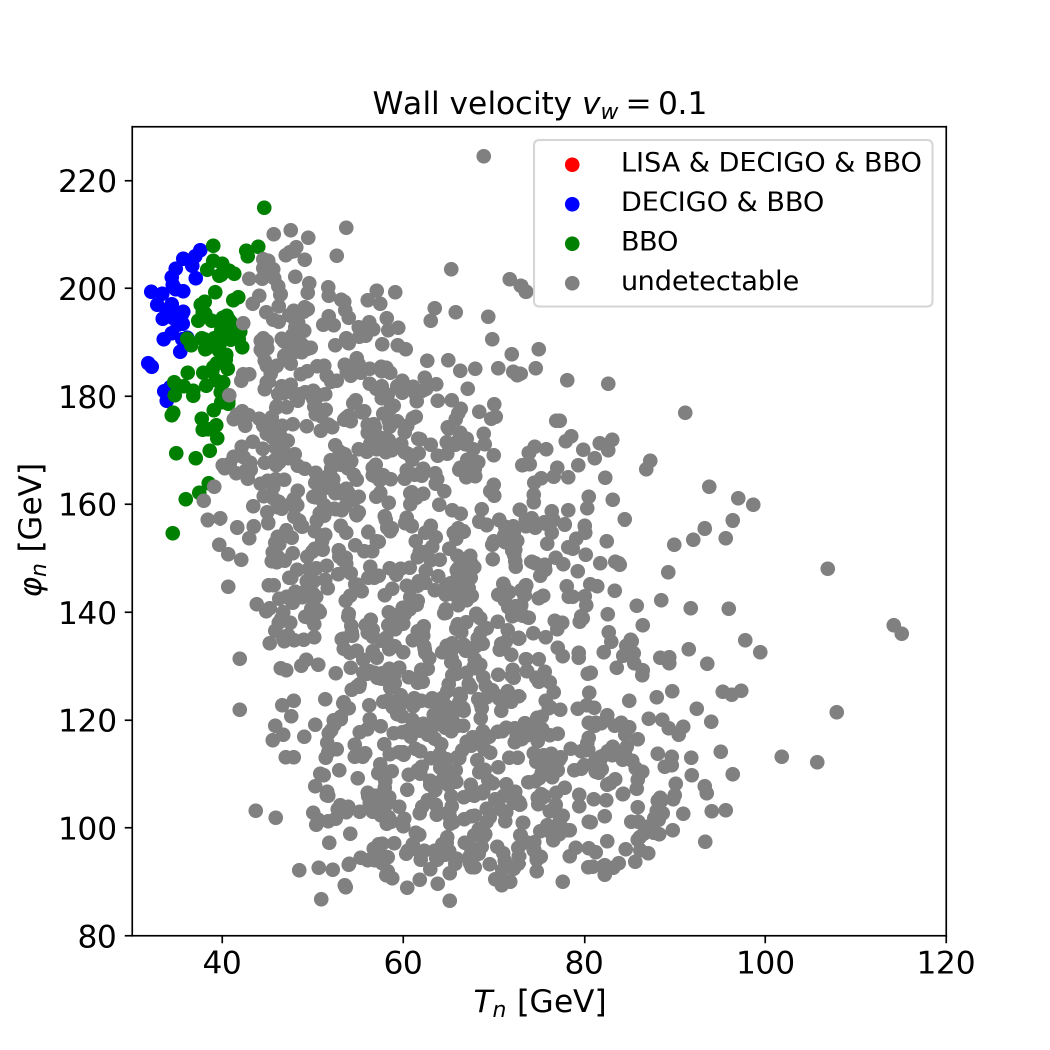}
\includegraphics[width=0.49\textwidth]{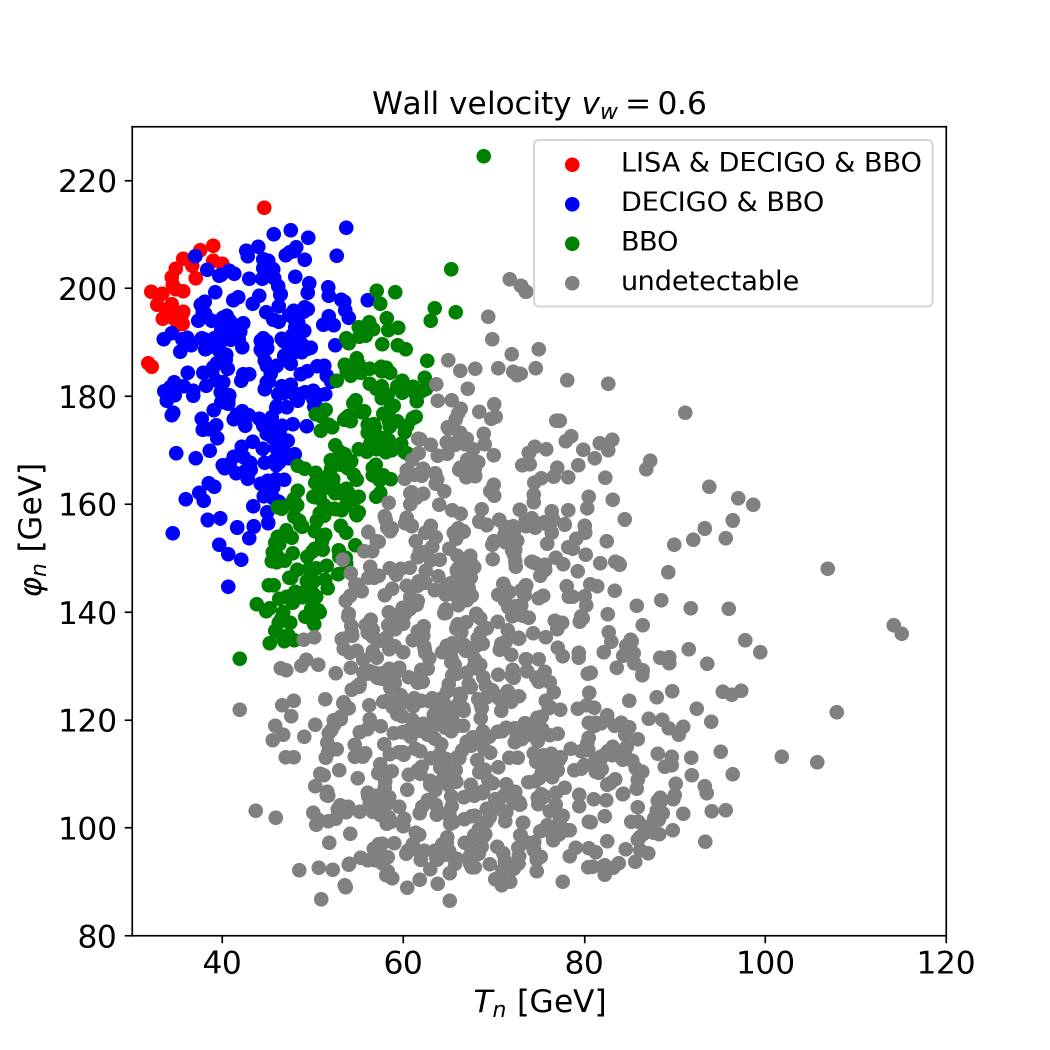}
\caption{The same as Fig.~\ref{fig: Gwplot1} but on the $T_n$-$\varphi_n$ plane.}
\label{fig: Gwplot2}
\end{figure}

In Figs.~\ref{fig: Gwplot1} and \ref{fig: Gwplot2}, we show the testability of the points in the $\alpha_\theta^{}$-$\tilde{\beta}_t$ plane and $T_n$-$\varphi_n$ plane, respectively, where $\varphi_n = \sqrt{\varphi_1^2 + 4 \varphi_2^2 + 4 \varphi_3^2}$ at $T=T_n$. 
We can observe the strong correlation between these parameters and the detectability of the GWs. 
In particular, $\log_{10} \tilde{\beta}_t/ \log_{10} \alpha_\theta$ and $\varphi_n/T_n$ seem to be good reference quantities of the detectability. 

\section{Conclusions}
\label{sec: conclusion}

In this paper, we have improved the EWPT study in the GM model by using the thermally resummed one-loop effective potential with the running couplings and running classical fields according to the renormalization group equations. 
We have used the general Higgs potential, which includes the custodial-symmetry-breaking terms, to construct the one-loop effective potential. 
General configurations for $CP$-even classical fields have been considered, and the one-loop effective potential has been derived. 
The renormalization conditions have been imposed on the first and second derivatives of the potential. 
Thermal resummation has been performed in the Parwani scheme. 
We have used the renormalization group equations to evolve the couplings and the classical fields from the initial scale $m_Z$ to the temperature-dependent scale $\mu = \sqrt{m_Z^2 + T^2}$. 
This prescription has eliminated the scale dependence of the effective potential at the one-loop level. 

We have also performed the Bayesian analysis to find the posterior distributions of the model parameters by using the theoretical requirements and the latest experimental bounds. 
According to the posterior distribution, we have generated about $2.5\times 10^5$ independent parameter points and have investigated the EWPT at the preselected parameter points. 
We have classified the behavior of the phase transition into four cases: strong first-order, weak first-order, second-order, and no transition. 
No multistep EWPT has been observed at any of the parameter points in our analysis. 

We have discussed the phenomenological predictions at the points for the strong first-order EWPT. 
We have examined the Higgs boson decays to neutral gauge bosons ($h \to \gamma \gamma$ and $h \to Z \gamma$), the deviation in the Higgs triple coupling, which modifies the di-Higgs production at high-energy colliders, and the GWs associated with the bubble collisions at the EWPT. 
We have compared their predictions with expected sensitivities at future experiments. 
It has been shown that the parameter points for the strong first-order EWPT can be verified at the future high-energy colliders, such as the HL-LHC and ILC, and the future GW observatories, such as LISA, DECIGO, and BBO. 
In particular, at all the parameter points for the strong first-order EWPT in the present analysis, it is expected that predicted deviations in $h \to \gamma \gamma$ are beyond the $95~\%$ C.L. expected sensitivity at the HL-LHC with the central value $\mu_{\gamma\gamma} = 1$. 
Therefore, given the time scale of the future experiments, precision measurement of $h \to \gamma \gamma$ at the HL-LHC would provide the most effective and earliest validation. 
The other observables are also essential for the cross-examination of these parameter points.

\section*{Acknowledgments}

This work was supported by the National Science and Technology Council under Grant Nos. NSTC-111-2112-M002-018-MY3, NSTC-113-2811-M-002-043, and NSTC-114-2112-M002-020-MY3.

\appendix

\section{Notation for the general Higgs potential}
\label{app: notation}

In this appendix, we compare the notation of the Higgs potential in Eq.~(\ref{eq: general_Higgs_potential}) with notations in related previous works~\cite{Blasi:2017xmc, Keeshan:2018ypw, Chen:2023ins, Du:2024mry}.  
In the following equations in this appendix, the left-hand sides are the couplings in this paper, and the right-hand sides are those in the other works. 

First, relations with the scalar couplings in Ref.~\cite{Keeshan:2018ypw} are given by
\begin{align}
\label{eq: relation_with_Keeshan2018}
\begin{array}{l}
\displaystyle{
    m_\phi^2 = \tilde{\mu}_2^2, \quad
	m_\chi^2 = \tilde{\mu}_3^{\prime2}, \quad
	  m_\xi^2 = \tilde{\mu}_3^2, \quad
	\sigma_1 = - \frac{ 1 }{ 4 } \tilde{M}_1, \quad
	\sigma_2 = - \frac{ 1 }{ 2 } \tilde{M}_1^\prime, \quad
	\sigma_3 = - 6 \tilde{M}_2
}, \\[10pt]
\displaystyle{
    \lambda = \frac{ 1 }{ 4 } \tilde{\lambda}, \quad
	\rho_1 = \tilde{\lambda}_7 + 2 \tilde{\lambda}_2, \quad
	\rho_2 = - 2 \tilde{\lambda}_2, \quad
	\rho_3 = \tilde{\lambda}_8, \quad
	\rho_4 = \tilde{\lambda}_{10}, \quad
	\rho_5 = \tilde{\lambda}_9
}, \\[10pt]
\displaystyle{
    \kappa_1 = \frac{ 1 }{ 2 } \tilde{\lambda}_5, \quad
	\kappa_2 = \frac{ 1 }{ 4 } \tilde{\lambda}_3, \quad
	\kappa_3 = \frac{ 1 }{ 2 } \tilde{\lambda}_6, \quad
	\kappa_4 = \frac{ 1 }{ \sqrt{2} } \tilde{\lambda}_4
}.
\end{array}
\end{align}

Next, relations with couplings in Refs.~\cite{Blasi:2017xmc, Du:2024mry} are given by
\begin{align}
\label{eq: relation_Blasi2017}
\begin{array}{l}
\displaystyle{
    m_\phi^2 = m_\phi^{\prime 2}, \quad 
	m_\chi^2 = m_\chi^{\prime 2}, \quad 
	m_\xi^2 = 2 m_\xi^{\prime 2}, \quad 
	\sigma_1 = - \frac{ 1 }{ 2 \sqrt{2} } \mu_1^\prime , \quad 
	\sigma_2 = - \mu_2^\prime, \quad 
	\sigma_3 = \frac{ 1 }{ \sqrt{2} } \mu_3^\prime
}, \\[10pt]
\displaystyle{
    \lambda = \frac{ 1 }{ 4 } \lambda^\prime, \quad 
	\rho_1 = \rho_1^\prime, \quad 
	\rho_2 = \rho_2^\prime, \quad 
	\rho_3 = \frac{ 1 }{ 2 } \rho_3^\prime, \quad 
	\rho_4 = \rho_4^\prime, \quad 
	\rho_5 = \rho_5^\prime
},  \\[10pt]
\displaystyle{
    \kappa_1 = \frac{ 1 }{ 2 } \sigma_1^\prime + \frac{ 1 }{ 4 } \sigma_2^\prime, \quad 
	\kappa_2 = \frac{ 1 }{ 4 } \sigma_2^\prime, \quad 
	\kappa_3 = \frac{ 1 }{ 2 } \sigma_3^\prime, \quad 
	\kappa_4 = \frac{ 1 }{ \sqrt{2} } \sigma_4^\prime 
}. 
\end{array}
\end{align}
Although the couplings used in Refs.~\cite{Blasi:2017xmc, Du:2024mry} are without the superscript ${}^\prime$, we represent them with ${}^\prime$ in the above relations to avoid confusion of notation. 

The coupling relations with Ref.~\cite{Chen:2023ins} are almost the same as Eq.~(\ref{eq: relation_Blasi2017}). 
Only the relation for $\rho_3$ is different and is given by $\rho_3 = \rho_3^\prime$, where we use ${}^\prime$ for the coupling in Ref.~\cite{Chen:2023ins} for the same reason as in Eq.~(\ref{eq: relation_Blasi2017}).

\section{Field-dependent masses for the effective potential}
\label{app: field_dependent_mass}

In this appendix, we show the field-dependent masses necessary for computing the loop corrections to the effective potential. 
As discussed in Sec.~\ref{sec: effective_potential}, the SM fermions, the EW gauge bosons, and the scalar bosons play a part in the one-loop correction $V_1$. 

The field-dependent mass of the SM fermions $f$ is given by
\begin{align}
\tilde{m}_f^2 = \frac{ y_f^2 }{ 2 } \varphi_1^2, 
\end{align}
where $y_f = \sqrt{2} m_f/(vs_\beta)$ is the Yukawa coupling, and $m_f$ is the mass of $f$. 
The field-dependent masses of $W$ and $Z$ bosons are given by
\begin{align}
\tilde{m}_W^2 = \frac{ g_L^2 }{ 4 } \bigl( \varphi_1^2 + 4 \varphi_2^2 + 4 \varphi_3^2 \bigr), \quad 
\tilde{m}_Z^2 = \frac{ g_Z^2 }{ 4 } \bigl( \varphi_1^2 + 8 \varphi_2^2 \bigr),  
\end{align}
where $g_Z^2 = g_L^2 + g_Y^2$, and $g_L$ and $g_Y$ are the gauge couplings for $SU(2)_L$ and $U(1)_Y$, respectively. 

The field-dependent mass for the doubly charged scalar bosons $\chi^{\pm\pm}$ is given by
\begin{align}
\tilde{m}_{cc}^2 = m_\chi^2
	- \sigma_3 \varphi_3
	+ 2 \rho_1 \varphi_2^2
	+ \rho_4 \varphi_3^2
	+ (\kappa_1 - \kappa_2) \varphi_1^2.
\end{align}
The masses for the other scalar bosons are more complicated due to the mixing, and they are given by the eigenvalues of each mass matrix: 
\begin{align}
\mathcal{L} = - \omega_i^- (\tilde{M}_c^2)_{ij} \omega_j^+ 
    - \frac{ 1 }{ 2 } h_i (\tilde{M}_h^2)_{ij} h_j 
    - \frac{ 1 }{ 2 } a_i (\tilde{M}_a^2)_{ij} a_j. 
\end{align}
The components of each mass matrix are given as follows. 
\begin{itemize}
\item Charged Higgs 
\begin{align}
& (\tilde{M}_c^2)_{11} = m_\phi^2
	- 2 \sigma_1 \varphi_3
	+ 4 \lambda \varphi_1^2
	+ 2 (\kappa_1-\kappa_2) \varphi_2^2
	+ 2 \kappa_3 \varphi_3^2, \\[5pt]
& (\tilde{M}_c^2)_{22} 
	= m_\chi^2
	+ 2 (\rho_1 + \rho_2) \varphi_2^2
	+ (\rho_4 + \rho_5) \varphi_3^2
	+ \kappa_1 \varphi_1^2, \\[5pt]
& (\tilde{M}_c^2)_{33} 
	= m_\xi^2 
	+ 4 \rho_3 \varphi_3^2
	+ (2 \rho_4 + \rho_5) \varphi_2^2
	+ 2 \kappa_3 \varphi_1^2, \\[5pt]
& (\tilde{M}_c^2)_{12} = (\tilde{M}_c^2)_{21}
	= \sigma_2 \varphi_1 
	+ 2 \kappa_2 \varphi_1 \varphi_2, \\[5pt]
& (\tilde{M}_c^2)_{13} = (\tilde{M}_c^2)_{31}
	= 2 \sigma_1 \varphi_1 
	+ \kappa_4 \varphi_1 \varphi_2, \\[5pt]
& (\tilde{M}_c^2)_{23} = (\tilde{M}_c^2)_{32}
	= \sigma_3 \varphi_2 
	- \rho_5 \varphi_2 \varphi_3
	+ \frac{ \kappa_4 }{ 2 } \varphi_1^2.
\end{align}

\item CP-even Higgs 
\begin{align}
& (\tilde{M}_h^2)_{11} = m_\phi^2 
	+ 2 \sigma_1 \varphi_3 
	+ 2 \sigma_2 \varphi_2 
	+ 12 \lambda \varphi_1^2
	+ 2 (\kappa_1 + \kappa_2) \varphi_2^2 
	+ 2 \kappa_3 \varphi_3^2 
	+ 2 \kappa_4 \varphi_2 \varphi_3, \\[5pt]
& (\tilde{M}_h^2)_{22} = m_\chi^2 
	+ \sigma_3 \varphi_3
	+ 6 (\rho_1+\rho_2) \varphi_2^2 
	+ \rho_4 \varphi_3^2
	+ (\kappa_1 + \kappa_2) \varphi_1^2, \\[5pt]
& (\tilde{M}_h^2)_{33} = m_\xi^2 
	+ 12 \rho_3 \varphi_3^2 
	+ 2 \rho_4 \varphi_2^2
	+ 2 \kappa_3 \varphi_1^2, \\[5pt]
& (\tilde{M}_h^2)_{12} = (\tilde{M}_h^2)_{21}
	= \sqrt{2} \sigma_2 \varphi_1
	+ 2 \sqrt{2} (\kappa_1 + \kappa_2) \varphi_1\varphi_2
	+ \sqrt{2} \kappa_4 \varphi_1 \varphi_3, \\[5pt]
& (\tilde{M}_h^2)_{13} = (\tilde{M}_h^2)_{31}
	= 2 \sigma_1 \varphi_1 
	+ 4 \kappa_3 \varphi_1 \varphi_3
	+ 2 \kappa_4 \varphi_1 \varphi_2, \\[5pt]
& (\tilde{M}_h^2)_{23} = (\tilde{M}_h^2)_{32}
	= \sqrt{2} \sigma_3 \varphi_2
	+ 2 \sqrt{2} \rho_4 \varphi_2 \varphi_3
	+ \frac{ \kappa_4 }{ \sqrt{2} } \varphi_1^2. 
\end{align}

\item CP-odd Higgs 
\begin{align}
& (\tilde{M}_a^2)_{11} 
	= m_\phi^2 
	+ 2 \sigma_1 \varphi_3 
	- 2 \sigma_2 \varphi_2 
	+ 4 \lambda \varphi_1^2 
	+ 2 (\kappa_1 + \kappa_2) \varphi_2^2
	+ 2 \kappa_3 \varphi_3^2
	- 2 \kappa_4 \varphi_2 \varphi_3, \\[5pt]
& (\tilde{M}_a^2)_{22} 
	= m_\chi^2
	+ \sigma_3 \varphi_3
	+ 2 (\rho_1 + \rho_2) \varphi_2^2
	+ (\kappa_1 + \kappa_2) \varphi_1^2
	+ \rho_4 \varphi_3^2, \\[5pt]
& (\tilde{M}_a^2)_{12} = (\tilde{M}_a^2)_{21}
	= \sqrt{2}\sigma_2 \varphi_1 + \sqrt{2} \kappa_4 \varphi_1 \varphi_3. 
\end{align}
\end{itemize}

Next, we present the formulas for thermal mass corrections, which are necessary for the thermal resummation in the Parwani scheme. 
First, the thermal masses for the EW gauge bosons are given by
\begin{align}
\Delta m_{B_\mu B_\nu}^2 = \frac{17}{6} g_Y^2 T^2 \delta_{\mu\nu} \delta_{\mu \parallel}, \quad 
\Delta m_{W^a_\mu W^b_\nu}^2 = \frac{17}{6} g_L^2 T^2 \delta_{ab} \delta_{\mu\nu} \delta_{\mu \parallel}, \quad 
\Delta m_{W^a_\mu B_\nu}^2 = 0, 
\end{align}
where $\delta_{\mu \parallel}$ represents the fact that only the longitudinal mode of the gauge bosons feels the thermal mass due to the interaction with the background plasma at the perturbative level~\cite{Espinosa:1992kf}.  

Second, thermal corrections to the scalar masses are common to fields in the same isospin multiplets. 
They are given by
\begin{align}
\Delta (M_c^2)_{11} = & \Delta (M_h^2)_{11} = \Delta (M_a^2)_{11} 
    \nonumber \\[5pt]
    = & \frac{T^2}{24} \Bigl( 48\lambda + 12 \kappa_1 + 12 \kappa_3 
        + \frac{ 9 }{ 2 } g_L^2 + \frac{ 3 }{ 2 } g_Y^2 
        + 2 \sum_f N_c^f y_f^2 \Bigr), \\[5pt]
\Delta (M_c^2)_{22} = & \Delta (M_h^2)_{22} = \Delta (M_a^2)_{22} = \Delta m_{cc}^2 
    \nonumber \\[5pt]
    = & \frac{ T^2 }{ 24 } \Bigl( 16 \rho_1 + 12 \rho_2 + 6 \rho_4 + 2 \rho_5 + 8 \kappa_1
    + 12 g_L^2 + 6 g_Y^2 \Bigr), \\[5pt]
\Delta (M_c^2)_{33} = & \Delta (M_h^2)_{33} 
    \nonumber \\[5pt]
    = & \frac{ T^2 }{ 24 } \Bigl( 40\rho_3 + 12 \rho_4 + 4 \rho_5 + 16 \kappa_3 + 12 g_L^2 \Bigr), 
\end{align}
where $N_c^f$ is the number of color degrees of freedom of the SM fermions $f$. 

The fermion mass does not receive thermal corrections at this order. 
Thus, $\hat{m}^2_f$ in the loop corrections are the same as the zero-temperature field-dependent masses $\tilde{m}^2_f$.

\section{Beta functions}
\label{app: beta_function}

In this appendix, we present the one-loop beta functions necessary for eliminating the one-loop $\mu$-dependence of the effective potential. 
As discussed in Sec.~\ref{sec: RGE}, the beta functions can be separated into the universal part $\bar{\beta}_g$ due to the divergent part of the counterterms and the renormalization-scheme-dependent part $\delta \beta_g$ due to their explicit $\mu$-dependence. 

First, the universal parts for the scalar couplings and the vacuum energy $\Omega$ are given by 
\begin{align}
& \bar{\beta}_\Omega = \frac{ 1 }{ 16\pi^2 }
	\biggl( 2 m_\phi^4 + 3 m_\chi^4 + \frac{ 3 }{ 2 } m_\xi^4 \biggr), \\[5pt]
& \bar{\beta}_{m_\phi^2} = \frac{ 1 }{ 16 \pi^2 }
	\Bigl( 
		12 \kappa_{1} m_{\chi}^{2} 
		+ 12 \kappa_{3} m_{\xi}^{2} 
		+ 48 \lambda m_{\phi}^{2} 
		+ 24 \sigma_{1}^{2} 
		+ 12 \sigma_2^2 \Bigr)
    + 2 m_\phi^2 \gamma_\phi,  \\[5pt]
& \bar{\beta}_{m_\chi^2} = \frac{ 1 }{ 16 \pi^2 }
	\Bigl( 
		8 \kappa_{1} m_{\phi}^{2} 
		+ 16 \rho_{1} m_{\chi}^{2} 
		+ 12 \rho_{2} m_{\chi}^{2} 
		+ 6 \rho_{4} m_{\xi}^{2} 
		+ 2 \rho_{5} m_{\xi}^{2} 
		+ 4 \sigma_{2}^{2} 
		+ 4 \sigma_3^2 \Bigr)
    + 2 m_\chi^2 \gamma_\chi,  \\[5pt]
& \bar{\beta}_{m_\xi^2} = \frac{ 1 }{ 16\pi^2 }
	\Bigl( 
		16 \kappa_{3} m_{\phi}^{2} 
		+ 40 \rho_{3} m_{\xi}^{2} 
		+ 12 \rho_{4} m_{\chi}^{2} 
		+ 4 \rho_{5} m_{\chi}^{2} 
		+ 16 \sigma_{1}^{2} 
		+ 4 \sigma_3^2\Bigr)
    + 2 m_\xi^2 \gamma_\xi,  \\[5pt]
& \bar{\beta}_{\sigma_1} = \frac{ 1 }{ 16 \pi^2 } 
	\Bigl(
		4 \kappa_{2} \sigma_{3} 
		+ 16 \kappa_{3} \sigma_{1} 
		+ 8 \kappa_{4} \sigma_{2} 
		+ 16 \lambda \sigma_{1} \Bigr)
    + \sigma_1 \bigl( 2 \gamma_\phi + \gamma_\xi \bigr),  \\[5pt]
& \bar{\beta}_{\sigma_2} = \frac{ 1 }{ 16 \pi^2 } 
	\Bigl(
		8 \kappa_{1} \sigma_{2} 
		+ 16 \kappa_{2} \sigma_{2} 
		+ 16 \kappa_{4} \sigma_{1} 
		+ 4 \kappa_4 \sigma_3
		+ 16 \lambda \sigma_{2} \Bigr),
    + \sigma_2 \bigl( 2 \gamma_\phi + \gamma_\chi \bigr), \\[5pt]
& \bar{\beta}_{\sigma_3} = \frac{ 1 }{ 16 \pi^2 } 
	\Bigl( 
		16 \kappa_{2} \sigma_{1} 
		+ 8 \kappa_{4} \sigma_{2} 
		+ 4 \rho_{1} \sigma_{3} 
		+ 8 \rho_{2} \sigma_{3} 
		+ 8 \rho_{4} \sigma_{3} 
		- 4 \rho_{5} \sigma_{3} \Bigr)
    + \sigma_3 \bigl( 2 \gamma_\chi + \gamma_\xi \bigr),  \\[5pt]
& \bar{\beta}_\lambda = \frac{ 1 }{ 16 \pi^2 } 
	\Biggl( 
		3 \kappa_{1}^{2} 
		+ 2 \kappa_{2}^{2} 
		+ 6 \kappa_{3}^{2} 
		+ \kappa_{4}^{2} 
		+ 96 \lambda^{2} 
		-\frac{3}{2} (y_t^4 + y_b^4)
		+ \frac{9}{32} g_L^4 
		+ \frac{3 }{ 16 } g_L^2 g_Y^2
		+ \frac{3}{32} g_Y^4
		\Biggr)
    + 4 \lambda \gamma_\phi, \\[5pt]
& \bar{\beta}_{\rho_1^{}} = \frac{ 1 }{ 16 \pi^2 } 
	\Bigl( 
		8 \kappa_1^2 
		- 8 \kappa_2^2
		+ 28 \rho_1^2 
		+ 24 \rho_1 \rho_2 
		+ 6 \rho_2^2 
		+ 6 \rho_4^2 
		+ 4 \rho_4 \rho_5 
		+ 3 \rho_5^2
        \nonumber \\
		& \hspace{60pt} + 15 g_L^4
		- 12 g_L^2 g_Y^2 
		+ 6 g_Y^4 \Bigr)
    + 4 \rho_1 \gamma_\chi, \\[5pt]
& \bar{\beta}_{\rho_2^{}} = \frac{ 1 }{ 16\pi^2 } 
	\Bigl( 
		16 \kappa_2^2
		+ 24 \rho_1 \rho_2 
		+ 18 \rho_2^2 
		- 2 \rho_5^2 
		-6 g_L^4 + 24 g_L^2g_Y^2 \Bigr)
    + 4 \rho_2 \gamma_\chi, \\[5pt]
& \bar{\beta}_{\rho_3} = \frac{ 1 }{ 16 \pi^2 } 
	\biggl( 
		8 \kappa_{3}^{2} 
		+ 88 \rho_{3}^{2} 
		+ 3 \rho_{4}^{2} 
		+ 2 \rho_{4} \rho_{5} 
		+ \rho_{5}^{2} 
		+ 3  g_L^4 \biggr)
    + 4 \rho_3 \gamma_\xi, \\[5pt]
& \bar{\beta}_{\rho_4} = \frac{ 1 }{ 16 \pi^2 }
	\Bigl( 
		16 \kappa_{1} \kappa_{3} 
		+ 4 \kappa_{4}^{2} 
		+ 16 \rho_{1} \rho_{4} 
		+ 4 \rho_{1} \rho_{5} 
		+ 12 \rho_{2} \rho_{4} 
		+ 4 \rho_{2} \rho_{5} 
		+ 40 \rho_{3} \rho_{4} 
		\nonumber \\
		& \hspace{60pt} 
		+ 8 \rho_{3} \rho_{5} 
		+ 8 \rho_{4}^{2} 
		+ 2 \rho_{5}^{2} 
		+ 6 g_L^4 \Bigr)
    + 2 \rho_4 \bigl( \gamma_\chi + \gamma_\xi \bigr),  \\[5pt]
& \bar{\beta}_{\rho_5} = \frac{ 1 }{ 16\pi^2 } 
	\Bigl( 
		- 4 \kappa_4^2
		+ 4 \rho_1 \rho_5 
		+ 16 \rho_3 \rho_5 
		+ 16 \rho_4 \rho_5 
		+ 10 \rho_5^2 
		+ 6 g_L^4 \Bigr)
    + 2 \rho_5 \bigl( \gamma_\chi + \gamma_\xi \bigr), \\[5pt]
& \bar{\beta}_{\kappa_1} = \frac{ 1 }{ 16\pi^2 } 
	\biggl( 
		8 \kappa_1^2 
		+ 16 \kappa_2^2 
		+ 48 \kappa_1 \lambda
		+ 16 \kappa_1 \rho_1 
		+ 12 \kappa_1 \rho_2 
		+ 12 \kappa_3 \rho_4 
		+ 4 \kappa_3 \rho_5 
		+ 4 \kappa_4^2
        \nonumber \\
		& \hspace{60pt} + 3 g_L^4 
		+ \frac{ 3 }{ 2 } g_Y^4 \biggr)
    + 2 \kappa_1 \bigl( \gamma_\chi + \gamma_\phi \bigr), \\[5pt]
& \bar{\beta}_{\kappa_2} = \frac{ 1 }{ 16\pi^2 } 
	\Bigl( 
		16 \kappa_1 \kappa_2
		+ 16 \kappa_2 \lambda
		+ 4 \kappa_2 \rho_1 
		+ 8 \kappa_2 \rho_2 
		+ 2 \kappa_4^2 
		+ 3 g_L^2 g_Y^2 \Bigr)
    + 2 \kappa_2 \bigl( \gamma_\phi + \gamma_\chi \bigr), \\[5pt]
& \bar{\beta}_{\kappa_3} = \frac{ 1 }{ 16\pi^2 }
	\biggl( 
		6 \kappa_{1} \rho_{4} 
		+ 2 \kappa_{1} \rho_{5} 
		+ 16 \kappa_{3}^{2} 
		+ 48 \kappa_{3} \lambda 
		+ 40 \kappa_{3} \rho_{3} 
		+ 4 \kappa_{4}^{2} 
		+ \frac{3}{2} g_L^4
		\biggr)
    + 2 \kappa_3 \bigl( \gamma_\phi + \gamma_\xi \bigr),  \\[5pt]
& \bar{\beta}_{\kappa_4} = \frac{ 1 }{ 16\pi^2 }
	\Bigl( 
		8 \kappa_{1} \kappa_{4} 
		+ 8 \kappa_{2} \kappa_{4} 
		+ 16 \kappa_{3} \kappa_{4} 
		+ 16 \kappa_{4} \lambda 
		+ 4 \kappa_{4} \rho_{4} 
		- 2 \kappa_{4} \rho_{5} \Bigr)
    + \kappa_4 \bigl(2\gamma_\phi + \gamma_\chi + \gamma_\xi \bigr), 
\end{align}
where the anomalous dimensions are given by
\begin{align}
\gamma_\phi = \frac{ 1 }{ 16\pi^2 } 
	\Bigl\{ \sum_f N_c^f y_f^2 
		- \frac{ 3 }{ 4 } (3 g_L^2 + g_Y^2 ) \Bigr\}, \quad
	\gamma_\chi = - \frac{3}{16\pi^2} (2 g_L^2 + g_Y^2 ), \quad
	\gamma_\xi = - \frac{6}{16\pi^2} g_L^2. 
\end{align}
These results are consistent with Ref.~\cite{Blasi:2017xmc}. 

The divergent terms in the VEV counterterms $\delta v_\chi$ and $\delta v_\xi$ are determined to cancel the divergence in the linear terms of $\delta V$ due to the wave function renormalizations. 
By using them, the universal parts of their beta functions are given by
\begin{align}
\bar{\beta}_{v_\phi} = - v_\phi \gamma_\phi, \quad
\bar{\beta}_{v_\chi} = - v_\chi \gamma_\chi, \quad 
\bar{\beta}_{v_\xi} = - v_\xi \gamma_\xi.
\end{align}

Next, we discuss $\delta \beta_g$ for the scalar couplings, which are given by 
\begin{align}
\delta \beta_g = - \mu \frac{ \partial }{ \partial \mu} \delta g. 
\end{align}
As discussed in Sec.~\ref{sec: renormalization}, only the finite part of the counterterms contributes to $\delta \beta_g$. 
Their finite parts are given by the derivatives of $V_1$ in the present renormalization scheme. 
Thus, $\delta \beta_g$ can be evaluated by using the $\mu$-dependence of the derivatives of $V_1$, which are given as follows. 
\begin{align}
\mu \frac{ \partial }{ \partial \mu }V_1^\prime 
	= \frac{ 1 }{ 16\pi^2 } 
	\Bigl[
		& - 16 \kappa_{1}^{2} v_{\chi}^{2} v_{\phi} 
            - 12 \kappa_{1}^{2} v_{\phi}^{3} 
            - 32 \kappa_{1} \kappa_{2} v_{\chi}^{2} v_{\phi} 
            - 16 \kappa_{1} \kappa_{4} v_{\chi} v_{\phi} v_{\xi} 
            - 96 \kappa_{1} \lambda v_{\chi}^{2} v_{\phi} 
\nonumber \\
            & - 32 \kappa_{1} \rho_{1} v_{\chi}^{2} v_{\phi} 
            - 24 \kappa_{1} \rho_{2} v_{\chi}^{2} v_{\phi} 
            - 12 \kappa_{1} \rho_{4} v_{\phi} v_{\xi}^{2} 
            - 4 \kappa_{1} \rho_{5} v_{\phi} v_{\xi}^{2} 
\nonumber \\[5pt]
            & - 16 \kappa_{1} \sigma_{2} v_{\chi} v_{\phi} 
            - 12 \kappa_{1} m^{2}_{\chi} v_{\phi} 
            - 32 \kappa_{2}^{2} v_{\chi}^{2} v_{\phi} 
            - 8 \kappa_{2}^{2} v_{\phi}^{3} 
            - 16 \kappa_{2} \kappa_{4} v_{\chi} v_{\phi} v_{\xi} 
\nonumber \\[5pt]
            & - 32 \kappa_{2} \lambda v_{\chi}^{2} v_{\phi} 
            - 8 \kappa_{2} \rho_{1} v_{\chi}^{2} v_{\phi} 
            - 16 \kappa_{2} \rho_{2} v_{\chi}^{2} v_{\phi} 
            - 32 \kappa_{2} \sigma_{2} v_{\chi} v_{\phi} 
            - 8 \kappa_{2} \sigma_{3} v_{\phi} v_{\xi}  
\nonumber \\[5pt]
            & - 24 \kappa_{3}^{2} v_{\phi}^{3} 
            - 32 \kappa_{3}^{2} v_{\phi} v_{\xi}^{2} 
            - 32 \kappa_{3} \kappa_{4} v_{\chi} v_{\phi} v_{\xi}
            - 96 \kappa_{3} \lambda v_{\phi} v_{\xi}^{2} 
            - 80 \kappa_{3} \rho_{3} v_{\phi} v_{\xi}^{2} 
\nonumber \\[5pt]
            & - 24 \kappa_{3} \rho_{4} v_{\chi}^{2} v_{\phi} 
            - 8 \kappa_{3} \rho_{5} v_{\chi}^{2} v_{\phi} 
            - 32 \kappa_{3} \sigma_{1} v_{\phi} v_{\xi} 
            - 12 \kappa_{3} m^{2}_{\xi} v_{\phi} 
            - 12 \kappa_{4}^{2} v_{\chi}^{2} v_{\phi} 
\nonumber \\[5pt]
           & - 4 \kappa_{4}^{2} v_{\phi}^{3} 
            - 8 \kappa_{4}^{2} v_{\phi} v_{\xi}^{2} 
            - 32 \kappa_{4} \lambda v_{\chi} v_{\phi} v_{\xi} 
            - 8 \kappa_{4} \rho_{4} v_{\chi} v_{\phi} v_{\xi} 
            + 4 \kappa_{4} \rho_{5} v_{\chi} v_{\phi} v_{\xi}  
\nonumber \\[5pt]
            & - 32 \kappa_{4} \sigma_{1} v_{\chi} v_{\phi} 
            - 16 \kappa_{4} \sigma_{2} v_{\phi} v_{\xi} 
            - 8 \kappa_{4} \sigma_{3} v_{\chi} v_{\phi} 
            - 384 \lambda^{2} v_{\phi}^{3} 
            - 32 \lambda \sigma_{1} v_{\phi} v_{\xi} 
\nonumber \\
            & - 32 \lambda \sigma_{2} v_{\chi} v_{\phi}
            - 48 \lambda m^{2}_{\phi} v_{\phi} 
            - 24 \sigma_{1}^{2} v_{\phi} 
            - 12 \sigma_{2}^{2} v_{\phi} 
            - 3 g_{L}^{4} v_{\chi}^{2} v_{\phi} 
            - \frac{3 g_{L}^{4} v_{\phi}^{3}}{4} 
\nonumber \\
            & - 3 g_{L}^{4} v_{\phi} v_{\xi}^{2} 
            - 3 g_{Z}^{4} v_{\chi}^{2} v_{\phi} 
            - \frac{3 g_{Z}^{4} v_{\phi}^{3}}{8} 
            + 2 v_{\phi}^{3} \sum_f N_c^f y_f^{4}
	\Bigr], \\[10pt]
\mu \frac{ \partial }{ \partial \mu }V_2^\prime 
	= \frac{ 1 }{ 16\pi^2 } 
	\Bigl[ 
		& - 32 \kappa_{1}^{2} v_{\chi}^{3} 
                - 16 \kappa_{1}^{2} v_{\chi} v_{\phi}^{2} 
                - 32 \kappa_{1} \kappa_{2} v_{\chi} v_{\phi}^{2} 
                - 32 \kappa_{1} \kappa_{3} v_{\chi} v_{\xi}^{2} 
                - 8 \kappa_{1} \kappa_{4} v_{\phi}^{2} v_{\xi} 
\nonumber \\
                & - 96 \kappa_{1} \lambda v_{\chi} v_{\phi}^{2} 
                - 32 \kappa_{1} \rho_{1} v_{\chi} v_{\phi}^{2} 
                - 24 \kappa_{1} \rho_{2} v_{\chi} v_{\phi}^{2} 
                - 8 \kappa_{1} \sigma_{2} v_{\phi}^{2} 
                - 16 \kappa_{1} m^{2}_{\phi} v_{\chi} 
\nonumber \\[5pt]
                & - 32 \kappa_{2}^{2} v_{\chi}^{3} 
                - 32 \kappa_{2}^{2} v_{\chi} v_{\phi}^{2} 
                - 8 \kappa_{2} \kappa_{4} v_{\phi}^{2} v_{\xi} 
                - 32 \kappa_{2} \lambda v_{\chi} v_{\phi}^{2} 
                - 8 \kappa_{2} \rho_{1} v_{\chi} v_{\phi}^{2}  
\nonumber \\[5pt]
                & - 16 \kappa_{2} \rho_{2} v_{\chi} v_{\phi}^{2} 
                - 32 \kappa_{2} \sigma_{1} v_{\chi} v_{\xi} 
                - 16 \kappa_{2} \sigma_{2} v_{\phi}^{2}
                - 16 \kappa_{3} \kappa_{4} v_{\phi}^{2} v_{\xi} 
                - 24 \kappa_{3} \rho_{4} v_{\chi} v_{\phi}^{2} 
\nonumber \\[5pt]
                & - 8 \kappa_{3} \rho_{5} v_{\chi} v_{\phi}^{2} 
                - 12 \kappa_{4}^{2} v_{\chi} v_{\phi}^{2} 
                - 8 \kappa_{4}^{2} v_{\chi} v_{\xi}^{2} 
                - 16 \kappa_{4} \lambda v_{\phi}^{2} v_{\xi} 
                - 4 \kappa_{4} \rho_{4} v_{\phi}^{2} v_{\xi} 
\nonumber \\[5pt]
                & + 2 \kappa_{4} \rho_{5} v_{\phi}^{2} v_{\xi} 
                - 16 \kappa_{4} \sigma_{1} v_{\phi}^{2} 
                - 16 \kappa_{4} \sigma_{2} v_{\chi} v_{\xi} 
                - 4 \kappa_{4} \sigma_{3} v_{\phi}^{2} 
                - 16 \lambda \sigma_{2} v_{\phi}^{2}  
\nonumber \\[5pt]
                & - 112 \rho_{1}^{2} v_{\chi}^{3} 
                - 192 \rho_{1} \rho_{2} v_{\chi}^{3} 
                - 32 \rho_{1} \rho_{4} v_{\chi} v_{\xi}^{2} 
                - 8 \rho_{1} \rho_{5} v_{\chi} v_{\xi}^{2} 
                - 8 \rho_{1} \sigma_{3} v_{\chi} v_{\xi} 
\nonumber \\[5pt]
                & - 32 \rho_{1} m^{2}_{\chi} v_{\chi}
                - 96 \rho_{2}^{2} v_{\chi}^{3} 
                - 24 \rho_{2} \rho_{4} v_{\chi} v_{\xi}^{2} 
                - 8 \rho_{2} \rho_{5} v_{\chi} v_{\xi}^{2} 
                - 16 \rho_{2} \sigma_{3} v_{\chi} v_{\xi}  
\nonumber \\[5pt]
                &  - 24 \rho_{2} m^{2}_{\chi} v_{\chi} 
                - 80 \rho_{3} \rho_{4} v_{\chi} v_{\xi}^{2} 
                - 16 \rho_{3} \rho_{5} v_{\chi} v_{\xi}^{2} 
                - 24 \rho_{4}^{2} v_{\chi}^{3} 
                - 16 \rho_{4}^{2} v_{\chi} v_{\xi}^{2} 
\nonumber \\
                & - 16 \rho_{4} \rho_{5} v_{\chi}^{3} 
                - 16 \rho_{4} \sigma_{3} v_{\chi} v_{\xi} 
                - 12 \rho_{4} m^{2}_{\xi} v_{\chi}
                - 4 \rho_{5}^{2} v_{\chi}^{3} 
                - 4 \rho_{5}^{2} v_{\chi} v_{\xi}^{2} 
                + 8 \rho_{5} \sigma_{3} v_{\chi} v_{\xi} 
\nonumber \\
                & - 4 \rho_{5} m^{2}_{\xi} v_{\chi} 
                - 8 \sigma_{2}^{2} v_{\chi} 
                - 8 \sigma_{3}^{2} v_{\chi} 
                - 12 g_{L}^{4} v_{\chi}^{3} 
                - 3 g_{L}^{4} v_{\chi} v_{\phi}^{2} 
                - 12 g_{L}^{4} v_{\chi} v_{\xi}^{2} 
\nonumber \\
                & - 24 g_{Z}^{4} v_{\chi}^{3} 
                - 3 g_{Z}^{4} v_{\chi} v_{\phi}^{2}
	\Bigr], \\[10pt]
\mu \frac{ \partial }{ \partial \mu }V_3^\prime 
	= \frac{ 1 }{ 16\pi^2 } 
	\Bigl[
		& - 32 \kappa_{1} \kappa_{3} v_{\chi}^{2} v_{\xi} 
                - 8 \kappa_{1} \kappa_{4} v_{\chi} v_{\phi}^{2} 
                - 12 \kappa_{1} \rho_{4} v_{\phi}^{2} v_{\xi} 
                - 4 \kappa_{1} \rho_{5} v_{\phi}^{2} v_{\xi} 
                - 8 \kappa_{2} \kappa_{4} v_{\chi} v_{\phi}^{2}  
\nonumber \\
               & - 16 \kappa_{2} \sigma_{1} v_{\chi}^{2}
               - 4 \kappa_{2} \sigma_{3} v_{\phi}^{2} 
                - 32 \kappa_{3}^{2} v_{\phi}^{2} v_{\xi} 
                - 32 \kappa_{3}^{2} v_{\xi}^{3} 
                - 16 \kappa_{3} \kappa_{4} v_{\chi} v_{\phi}^{2} 
\nonumber \\[5pt]
                & - 96 \kappa_{3} \lambda v_{\phi}^{2} v_{\xi} 
                - 80 \kappa_{3} \rho_{3} v_{\phi}^{2} v_{\xi} 
                - 16 \kappa_{3} \sigma_{1} v_{\phi}^{2} 
                - 16 \kappa_{3} m^{2}_{\phi} v_{\xi} 
                - 8 \kappa_{4}^{2} v_{\chi}^{2} v_{\xi}
\nonumber \\[5pt]
                &  - 8 \kappa_{4}^{2} v_{\phi}^{2} v_{\xi} 
                - 16 \kappa_{4} \lambda v_{\chi} v_{\phi}^{2} 
                - 4 \kappa_{4} \rho_{4} v_{\chi} v_{\phi}^{2} 
                + 2 \kappa_{4} \rho_{5} v_{\chi} v_{\phi}^{2} 
                - 8 \kappa_{4} \sigma_{2} v_{\chi}^{2} 
\nonumber \\[5pt]
                & - 8 \kappa_{4} \sigma_{2} v_{\phi}^{2} 
                - 16 \lambda \sigma_{1} v_{\phi}^{2} 
                - 32 \rho_{1} \rho_{4} v_{\chi}^{2} v_{\xi} 
                - 8 \rho_{1} \rho_{5} v_{\chi}^{2} v_{\xi} 
                - 4 \rho_{1} \sigma_{3} v_{\chi}^{2} 
\nonumber \\[5pt]
                & - 24 \rho_{2} \rho_{4} v_{\chi}^{2} v_{\xi} 
                - 8 \rho_{2} \rho_{5} v_{\chi}^{2} v_{\xi} 
                - 8 \rho_{2} \sigma_{3} v_{\chi}^{2} 
                - 352 \rho_{3}^{2} v_{\xi}^{3} 
                - 80 \rho_{3} \rho_{4} v_{\chi}^{2} v_{\xi}  
\nonumber \\[5pt]
                &  - 16 \rho_{3} \rho_{5} v_{\chi}^{2} v_{\xi} 
                - 40 \rho_{3} m^{2}_{\xi} v_{\xi} 
                - 16 \rho_{4}^{2} v_{\chi}^{2} v_{\xi} 
                - 12 \rho_{4}^{2} v_{\xi}^{3} 
                - 8 \rho_{4} \rho_{5} v_{\xi}^{3} 
                - 8 \rho_{4} \sigma_{3} v_{\chi}^{2}
\nonumber \\[5pt]
                & - 12 \rho_{4} m^{2}_{\chi} v_{\xi} 
                - 4 \rho_{5}^{2} v_{\chi}^{2} v_{\xi} 
                - 4 \rho_{5}^{2} v_{\xi}^{3} 
                + 4 \rho_{5} \sigma_{3} v_{\chi}^{2} 
                - 4 \rho_{5} m^{2}_{\chi} v_{\xi} 
                - 16 \sigma_{1}^{2} v_{\xi} 
\nonumber \\[5pt]
                & - 4 \sigma_{3}^{2} v_{\xi} 
                - 12 g_{L}^{4} v_{\chi}^{2} v_{\xi} 
                - 3 g_{L}^{4} v_{\phi}^{2} v_{\xi} 
                - 12 g_{L}^{4} v_{\xi}^{3}
	\Bigr], \\[10pt]
\mu \frac{ \partial }{ \partial \mu }V_{11}^{\prime \prime}
	= \frac{ 1 }{ 16\pi^2 } 
	\Bigl[
		& - 16 \kappa_{1}^{2} v_{\chi}^{2} 
                - 36 \kappa_{1}^{2} v_{\phi}^{2} 
                - 32 \kappa_{1} \kappa_{2} v_{\chi}^{2} 
                - 16 \kappa_{1} \kappa_{4} v_{\chi} v_{\xi} 
                - 96 \kappa_{1} \lambda v_{\chi}^{2} 
                - 32 \kappa_{1} \rho_{1} v_{\chi}^{2} 
\nonumber \\
                & - 24 \kappa_{1} \rho_{2} v_{\chi}^{2} 
                - 12 \kappa_{1} \rho_{4} v_{\xi}^{2} 
                - 4 \kappa_{1} \rho_{5} v_{\xi}^{2} 
                - 16 \kappa_{1} \sigma_{2} v_{\chi} 
                - 12 \kappa_{1} m^{2}_{\chi} 
                - 32 \kappa_{2}^{2} v_{\chi}^{2}  
\nonumber \\[5pt]
                & - 24 \kappa_{2}^{2} v_{\phi}^{2}
                - 16 \kappa_{2} \kappa_{4} v_{\chi} v_{\xi} 
                - 32 \kappa_{2} \lambda v_{\chi}^{2} 
                - 8 \kappa_{2} \rho_{1} v_{\chi}^{2} 
                - 16 \kappa_{2} \rho_{2} v_{\chi}^{2} 
                - 32 \kappa_{2} \sigma_{2} v_{\chi} 
\nonumber \\[5pt]
                & - 8 \kappa_{2} \sigma_{3} v_{\xi} 
                - 72 \kappa_{3}^{2} v_{\phi}^{2} 
                - 32 \kappa_{3}^{2} v_{\xi}^{2} 
                - 32 \kappa_{3} \kappa_{4} v_{\chi} v_{\xi} 
                - 96 \kappa_{3} \lambda v_{\xi}^{2} 
                - 80 \kappa_{3} \rho_{3} v_{\xi}^{2} 
\nonumber \\[5pt]
                & - 24 \kappa_{3} \rho_{4} v_{\chi}^{2} 
                - 8 \kappa_{3} \rho_{5} v_{\chi}^{2} 
                - 32 \kappa_{3} \sigma_{1} v_{\xi} 
                - 12 \kappa_{3} m^{2}_{\xi} 
                - 12 \kappa_{4}^{2} v_{\chi}^{2} 
                - 12 \kappa_{4}^{2} v_{\phi}^{2} 
\nonumber \\[5pt]
                &  - 8 \kappa_{4}^{2} v_{\xi}^{2} 
                - 32 \kappa_{4} \lambda v_{\chi} v_{\xi} 
                - 8 \kappa_{4} \rho_{4} v_{\chi} v_{\xi} 
                + 4 \kappa_{4} \rho_{5} v_{\chi} v_{\xi} 
                - 32 \kappa_{4} \sigma_{1} v_{\chi} 
                - 16 \kappa_{4} \sigma_{2} v_{\xi}  
\nonumber \\[5pt]
                & - 8 \kappa_{4} \sigma_{3} v_{\chi} 
                - 1152 \lambda^{2} v_{\phi}^{2} 
                - 32 \lambda \sigma_{1} v_{\xi} 
                - 32 \lambda \sigma_{2} v_{\chi} 
                - 48 \lambda m^{2}_{\phi}
                - 24 \sigma_{1}^{2} 
                - 12 \sigma_{2}^{2} 
\nonumber \\[5pt]
                & - 3 g_{L}^{4} v_{\chi}^{2} 
                - \frac{9 g_{L}^{4} v_{\phi}^{2}}{4} 
                - 3 g_{L}^{4} v_{\xi}^{2} 
                - 3 g_{Z}^{4} v_{\chi}^{2} 
                - \frac{9 g_{Z}^{4} v_{\phi}^{2}}{8} 
                + 6 v_\phi^2 \sum_f N_c^f y_f^4 
	\Bigr], \\[10pt]
\mu \frac{ \partial }{ \partial \mu }V_{12}^{\prime \prime}
	= \frac{ 1 }{ 16\pi^2 } 
	\Bigl[
		& - 32 \kappa_{1}^{2} v_{\chi} v_{\phi} 
                - 64 \kappa_{1} \kappa_{2} v_{\chi} v_{\phi} 
                - 16 \kappa_{1} \kappa_{4} v_{\phi} v_{\xi} 
                - 192 \kappa_{1} \lambda v_{\chi} v_{\phi} 
                - 64 \kappa_{1} \rho_{1} v_{\chi} v_{\phi} 
\nonumber \\
                & - 48 \kappa_{1} \rho_{2} v_{\chi} v_{\phi} 
                - 16 \kappa_{1} \sigma_{2} v_{\phi} 
                - 64 \kappa_{2}^{2} v_{\chi} v_{\phi} 
                - 16 \kappa_{2} \kappa_{4} v_{\phi} v_{\xi} 
                - 64 \kappa_{2} \lambda v_{\chi} v_{\phi} 
\nonumber \\[5pt]
                & - 16 \kappa_{2} \rho_{1} v_{\chi} v_{\phi} 
                - 32 \kappa_{2} \rho_{2} v_{\chi} v_{\phi} 
                - 32 \kappa_{2} \sigma_{2} v_{\phi} 
                - 32 \kappa_{3} \kappa_{4} v_{\phi} v_{\xi} 
                - 48 \kappa_{3} \rho_{4} v_{\chi} v_{\phi}  
\nonumber \\[5pt]
               & - 16 \kappa_{3} \rho_{5} v_{\chi} v_{\phi} 
                - 24 \kappa_{4}^{2} v_{\chi} v_{\phi}
                - 32 \kappa_{4} \lambda v_{\phi} v_{\xi} 
                - 8 \kappa_{4} \rho_{4} v_{\phi} v_{\xi} 
                + 4 \kappa_{4} \rho_{5} v_{\phi} v_{\xi} 
\nonumber \\[5pt]
               & - 32 \kappa_{4} \sigma_{1} v_{\phi} 
                - 8 \kappa_{4} \sigma_{3} v_{\phi} 
                - 32 \lambda \sigma_{2} v_{\phi} 
                - 6 g_{L}^{4} v_{\chi} v_{\phi} 
                - 6 g_{Z}^{4} v_{\chi} v_{\phi}
	\Bigr], \\[10pt]
\mu \frac{ \partial }{ \partial \mu }V_{13}^{\prime \prime}
	= \frac{ 1 }{ 16\pi^2 } 
	\Bigl[
		& - 16 \kappa_{1} \kappa_{4} v_{\chi} v_{\phi} 
                - 24 \kappa_{1} \rho_{4} v_{\phi} v_{\xi} 
                - 8 \kappa_{1} \rho_{5} v_{\phi} v_{\xi} 
                - 16 \kappa_{2} \kappa_{4} v_{\chi} v_{\phi} 
                - 8 \kappa_{2} \sigma_{3} v_{\phi}  
\nonumber \\
                & - 64 \kappa_{3}^{2} v_{\phi} v_{\xi}
                - 32 \kappa_{3} \kappa_{4} v_{\chi} v_{\phi} 
                - 192 \kappa_{3} \lambda v_{\phi} v_{\xi} 
                - 160 \kappa_{3} \rho_{3} v_{\phi} v_{\xi} 
                - 32 \kappa_{3} \sigma_{1} v_{\phi}  
\nonumber \\[5pt]
               & - 16 \kappa_{4}^{2} v_{\phi} v_{\xi} 
                - 32 \kappa_{4} \lambda v_{\chi} v_{\phi}
                - 8 \kappa_{4} \rho_{4} v_{\chi} v_{\phi} 
                + 4 \kappa_{4} \rho_{5} v_{\chi} v_{\phi} 
                - 16 \kappa_{4} \sigma_{2} v_{\phi} 
\nonumber \\[5pt]
                & - 32 \lambda \sigma_{1} v_{\phi} 
                - 6 g_{L}^{4} v_{\phi} v_{\xi}
	\Bigr], \\[10pt]
\mu \frac{ \partial }{ \partial \mu }V_{22}^{\prime \prime}
	= \frac{ 1 }{ 16\pi^2 } 
	\Bigl[
		& - 96 \kappa_{1}^{2} v_{\chi}^{2} 
                - 16 \kappa_{1}^{2} v_{\phi}^{2} 
                - 32 \kappa_{1} \kappa_{2} v_{\phi}^{2} 
                - 32 \kappa_{1} \kappa_{3} v_{\xi}^{2} 
                - 96 \kappa_{1} \lambda v_{\phi}^{2} 
                - 32 \kappa_{1} \rho_{1} v_{\phi}^{2} 
\nonumber \\
                & - 24 \kappa_{1} \rho_{2} v_{\phi}^{2} 
                - 16 \kappa_{1} m^{2}_{\phi} 
                - 96 \kappa_{2}^{2} v_{\chi}^{2} 
                - 32 \kappa_{2}^{2} v_{\phi}^{2} 
                - 32 \kappa_{2} \lambda v_{\phi}^{2} 
                - 8 \kappa_{2} \rho_{1} v_{\phi}^{2}  
\nonumber \\[5pt]
                & - 16 \kappa_{2} \rho_{2} v_{\phi}^{2}
                - 32 \kappa_{2} \sigma_{1} v_{\xi} 
                - 24 \kappa_{3} \rho_{4} v_{\phi}^{2} 
                - 8 \kappa_{3} \rho_{5} v_{\phi}^{2} 
                - 12 \kappa_{4}^{2} v_{\phi}^{2} 
                - 8 \kappa_{4}^{2} v_{\xi}^{2}  
\nonumber \\[5pt]
                & - 16 \kappa_{4} \sigma_{2} v_{\xi} 
                - 336 \rho_{1}^{2} v_{\chi}^{2} 
                - 576 \rho_{1} \rho_{2} v_{\chi}^{2}
                - 32 \rho_{1} \rho_{4} v_{\xi}^{2} 
                - 8 \rho_{1} \rho_{5} v_{\xi}^{2} 
                - 8 \rho_{1} \sigma_{3} v_{\xi}  
\nonumber \\[5pt]
                & - 32 \rho_{1} m^{2}_{\chi} 
                - 288 \rho_{2}^{2} v_{\chi}^{2} 
                - 24 \rho_{2} \rho_{4} v_{\xi}^{2} 
                - 8 \rho_{2} \rho_{5} v_{\xi}^{2}
                - 16 \rho_{2} \sigma_{3} v_{\xi} 
                - 24 \rho_{2} m^{2}_{\chi}  
\nonumber \\[5pt]
                & - 80 \rho_{3} \rho_{4} v_{\xi}^{2} 
                - 16 \rho_{3} \rho_{5} v_{\xi}^{2} 
                - 72 \rho_{4}^{2} v_{\chi}^{2} 
                - 16 \rho_{4}^{2} v_{\xi}^{2} 
                - 48 \rho_{4} \rho_{5} v_{\chi}^{2}
                - 16 \rho_{4} \sigma_{3} v_{\xi} 
\nonumber \\[5pt]
                & - 12 \rho_{4} m^{2}_{\xi} 
                - 12 \rho_{5}^{2} v_{\chi}^{2} 
                - 4 \rho_{5}^{2} v_{\xi}^{2} 
                + 8 \rho_{5} \sigma_{3} v_{\xi} 
                - 4 \rho_{5} m^{2}_{\xi} 
                - 8 \sigma_{2}^{2} 
                - 8 \sigma_{3}^{2} 
\nonumber \\[5pt]
                & - 36 g_{L}^{4} v_{\chi}^{2} 
                - 3 g_{L}^{4} v_{\phi}^{2} 
                - 12 g_{L}^{4} v_{\xi}^{2} 
                - 72 g_{Z}^{4} v_{\chi}^{2} 
                - 3 g_{Z}^{4} v_{\phi}^{2}
	\Bigr], \\[10pt]
\mu \frac{ \partial }{ \partial \mu }V_{23}^{\prime \prime}
	= \frac{ 1 }{ 16\pi^2 } 
	\Bigl[
		& - 64 \kappa_{1} \kappa_{3} v_{\chi} v_{\xi} 
                - 8 \kappa_{1} \kappa_{4} v_{\phi}^{2} 
                - 8 \kappa_{2} \kappa_{4} v_{\phi}^{2} 
                - 32 \kappa_{2} \sigma_{1} v_{\chi} 
                - 16 \kappa_{3} \kappa_{4} v_{\phi}^{2} 
\nonumber \\
                & - 16 \kappa_{4}^{2} v_{\chi} v_{\xi} 
                - 16 \kappa_{4} \lambda v_{\phi}^{2} 
                - 4 \kappa_{4} \rho_{4} v_{\phi}^{2} 
                + 2 \kappa_{4} \rho_{5} v_{\phi}^{2} 
                - 16 \kappa_{4} \sigma_{2} v_{\chi} 
                - 64 \rho_{1} \rho_{4} v_{\chi} v_{\xi} 
\nonumber \\[5pt]
                &  - 16 \rho_{1} \rho_{5} v_{\chi} v_{\xi} 
                - 8 \rho_{1} \sigma_{3} v_{\chi} 
                - 48 \rho_{2} \rho_{4} v_{\chi} v_{\xi} 
                - 16 \rho_{2} \rho_{5} v_{\chi} v_{\xi} 
                - 16 \rho_{2} \sigma_{3} v_{\chi} 
\nonumber \\[5pt]
                & - 160 \rho_{3} \rho_{4} v_{\chi} v_{\xi} 
                - 32 \rho_{3} \rho_{5} v_{\chi} v_{\xi} 
                - 32 \rho_{4}^{2} v_{\chi} v_{\xi} 
                - 16 \rho_{4} \sigma_{3} v_{\chi} 
                - 8 \rho_{5}^{2} v_{\chi} v_{\xi} 
\nonumber \\[5pt]
                &  + 8 \rho_{5} \sigma_{3} v_{\chi} 
                - 24 g_{L}^{4} v_{\chi} v_{\xi}
	\Bigr], \\[10pt]
\mu \frac{ \partial }{ \partial \mu }V_{33}^{\prime \prime} 
	= \frac{ 1 }{ 16\pi^2 } 
	\Bigl[
		& - 32 \kappa_{1} \kappa_{3} v_{\chi}^{2} 
                - 12 \kappa_{1} \rho_{4} v_{\phi}^{2} 
                - 4 \kappa_{1} \rho_{5} v_{\phi}^{2} 
                - 32 \kappa_{3}^{2} v_{\phi}^{2} 
                - 96 \kappa_{3}^{2} v_{\xi}^{2} 
                - 96 \kappa_{3} \lambda v_{\phi}^{2}  
\nonumber \\
                & - 80 \kappa_{3} \rho_{3} v_{\phi}^{2}
                - 16 \kappa_{3} m^{2}_{\phi} 
                - 8 \kappa_{4}^{2} v_{\chi}^{2} 
                - 8 \kappa_{4}^{2} v_{\phi}^{2} 
                - 32 \rho_{1} \rho_{4} v_{\chi}^{2} 
                - 8 \rho_{1} \rho_{5} v_{\chi}^{2} 
\nonumber \\[5pt]
                & - 24 \rho_{2} \rho_{4} v_{\chi}^{2} 
                - 8 \rho_{2} \rho_{5} v_{\chi}^{2} 
                - 1056 \rho_{3}^{2} v_{\xi}^{2} 
                - 80 \rho_{3} \rho_{4} v_{\chi}^{2} 
                - 16 \rho_{3} \rho_{5} v_{\chi}^{2} 
                - 40 \rho_{3} m^{2}_{\xi}  
\nonumber \\[5pt]
                & - 16 \rho_{4}^{2} v_{\chi}^{2} 
                - 36 \rho_{4}^{2} v_{\xi}^{2} 
                - 24 \rho_{4} \rho_{5} v_{\xi}^{2}
                - 12 \rho_{4} m^{2}_{\chi} 
                - 4 \rho_{5}^{2} v_{\chi}^{2} 
                - 12 \rho_{5}^{2} v_{\xi}^{2} 
                - 4 \rho_{5} m^{2}_{\chi}  
\nonumber \\[5pt]
                & - 16 \sigma_{1}^{2} 
                - 4 \sigma_{3}^{2} 
                - 12 g_{L}^{4} v_{\chi}^{2} 
                - 3 g_{L}^{4} v_{\phi}^{2}
                - 36 g_{L}^{4} v_{\xi}^{2}
	\Bigr].
\end{align}
By using these formulas and the definition of the finite counterterms in Eqs.~(\ref{eq: CT1})-(\ref{eq: CT9}), we can find $\delta \beta_g$ for the triplet VEVs and the scalar couplings: $v_\chi$, $v_\xi$, $m_\phi^2$, $m_\chi^2$, $m_\xi^2$, $\sigma_1$, $\sigma_2$, $\lambda$, and $\kappa_4$. 
For $v_\phi$ and the other scalar couplings, $\delta \beta_g$ is zero, and their beta functions are given by their universal part: $\beta_g = \bar{\beta}_g$. 

Finally, we present the beta functions for the Yukawa interactions and the gauge couplings. 
They are evaluated in the mass-independent scheme and are given by~\cite{Keeshan:2018ypw,Machacek:1983fi,Hamada:2015bra}
\begin{align}
& \beta_{g_C} = - \frac{ 7 }{ 16\pi^2 } g_C^3, \\[5pt]
& \beta_{g_L} = - \frac{ 1 }{ 16\pi^2 } \biggl( \frac{ 13 }{ 6 } \biggr) g_L^3, \\[5pt]
& \beta_{g_Y} = \frac{ 1 }{ 16\pi^2 } \biggl( \frac{ 47 }{ 6 } \biggr) g_Y^3, \\[5pt]
& \beta_{y_t} = \frac{ y_t }{ 16\pi^2 } \biggl(
    \frac{ 9 }{ 2 } y_t^2 
    + \frac{ 3 }{ 2 } y_b^2 
    - 8 g_C^2 
    - \frac{ 9 }{ 4 } g_L^2 
    - \frac{ 17 }{ 12 } g_Y^2 
    \biggr), \\[5pt]
& \beta_{y_b} = \frac{ y_b }{ 16\pi^2 } \biggl(
     \frac{ 9 }{ 2 } y_b^2 
     + \frac{ 3 }{ 2 } y_t^2 
     - 8 g_C^2 
     - \frac{ 9 }{ 4 } g_L^2
     - \frac{ 5 }{ 12 } g_Y^2
    \biggr), 
\end{align}
where $g_C$ is the QCD coupling. 

\section{The calculation of the Higgs triple coupling}
\label{app: hhh}
 
In this appendix, we describe how to calculate the Higgs triple coupling at the one-loop level. 
We use the third derivative of the effective potential, which has been derived in Sec.~\ref{sec: effective_potential}. 
We note that the resulting $hhh$ coupling is a vertex with zero external momenta, not on the mass shell. 
Also, the one-loop formula depends on the renormalization scheme used to derive the effective potential. 

Let $\varphi_h$, $\varphi_{H_1}$, and $\varphi_{H_5^0}$ be the classical fields along the direction of the scalar bosons $h$, $H_1$, and $H_5^0$, respectively. 
Then, the classical fields $\varphi_1$, $\varphi_2$, and $\varphi_3$ are given by
\begin{align}
\left\{
\begin{array}{l}
\varphi_1 = c_\alpha \varphi_h + s_\alpha \varphi_{H_1}, \\[10pt]
\displaystyle{
	\sqrt{2}\varphi_2 = \sqrt{\frac{2}{3}}(-s_\alpha \varphi_h + c_\alpha \varphi_{H_1} ) + \sqrt{ \frac{ 1 }{ 3 } } \varphi_{H_5^0}
}, \\[10pt]
\displaystyle{
	\varphi_3 = \sqrt{\frac{1}{3}}(-s_\alpha \varphi_h + c_\alpha \varphi_{H_1} ) - \sqrt{ \frac{ 2 }{ 3 } } \varphi_{H_5^0}
}. 
\end{array}
\right.  
\end{align}
Hence, the potential for $\varphi_h$ around the EW vacuum is given by the replacements: 
\begin{align}
\varphi_1 \to c_\alpha \varphi_h + s_\alpha v_{H_1}, \quad 
\sqrt{3}\varphi_2, \sqrt{3}\varphi_3 \to - s_\alpha \varphi_h + c_\alpha v_{H_1}, 
\end{align}
where $v_{H_1}= s_\alpha v_\phi + \sqrt{3} c_\alpha v_\Delta$ is the VEV along the $H_1$ direction. 
We note that $\varphi_{H_5^0}$ does not acquire the VEV at the EW vacuum. 

The derivative of the effective potential along $\varphi_h$ is given by (i) representing the potential on the plane $\varphi_2 = \varphi_3$ as a function of $\varphi_1$ and $\varphi^\prime \equiv \sqrt{3}\varphi_2 = \sqrt{3} \varphi_3$ and (ii) applying the derivative operator: 
\begin{align}
\frac{ \partial }{ \partial \varphi_h } = c_\alpha \frac{ \partial }{ \partial \varphi_1 } - s_\alpha \frac{ \partial }{ \partial \varphi^\prime }.
\end{align}
The Higgs triple coupling is given by 
\begin{align}
\label{eq: third_derivative_V}
\lambda_{hhh} = \left. \frac{ \partial^3 V_{T=0} }{ \partial \varphi_h^3 } \right|_0, 
\end{align}
where $V_{T=0} = V_0 + V_1 + \delta V$ is the zero-temperature effective potential, and the subscript 0 means that the derivative is evaluated at the EW vacuum: $\varphi_1 = v_\Phi, \ \varphi^\prime = \sqrt{3} v_\Delta$. 

At the tree level, $\lambda_{hhh}$ is given by using Eq.~(\ref{eq: third_derivative_V}) for the tree-level potential $V_0$:
\begin{align}
\lambda_{hhh}^\mathrm{tree} 
    = & \ 24 c_\alpha^3 \lambda_1 v_\phi
	- \frac{ 3 \sqrt{3} }{ 2 }s_\alpha c_\alpha^2 
	\Bigl\{ \mu_1 + 4 \bigl( 2 \lambda_4  + \lambda_5 \bigr) v_\Delta \Bigr\}
	\nonumber \\[5pt]
	& + 6 s_\alpha^2 c_\alpha \Bigl( 2 \lambda_4 + \lambda_5 \Bigr) v_\phi
	- 4 \sqrt{3} s_\alpha^3 
	\Bigl\{ \mu_2 + 2 \Bigl(3\lambda_2 + \lambda_3 \Bigr) v_\Delta \Bigr\}. 
\end{align}
This is consistent with Ref.~\cite{Chen:2022zsh}. 

At the one-loop level, we need to evaluate the derivatives of the Coleman--Weinberg type potential. 
First, we consider the scalar boson contribution without the mixing:
\begin{align}
V_{\text{no-mix}} = \frac{ n \tilde{m}^4 }{ 64 \pi^2 }
    \biggl( \log \frac{ \tilde{m}^2 }{ \mu^2 } - \frac{ 3 }{ 2 } - D \biggr), 
\end{align}
where $n$ is the number of degrees of freedom, $\tilde{m}$ is the field-dependent mass, and $\mu^2$ and $D$ are the same as those in Sec.~\ref{sec: effective_potential}. 
Since $\tilde{m}^2$ is at most a quadratic function of the classical fields, the third derivative is given by
\begin{align}
\label{eq: third_derivative_single_scalar}
V_{\text{no-mix}}^{\prime \prime \prime}
	 = \frac{ n }{ 32\pi^2 }
	\Biggl\{ \frac{ \bigl(m^{2\prime} \bigr)^3 }{ m^2 }
		+ 3 m^{2\prime} m^{2\prime \prime}
		\biggl( \log \frac{ m^2 }{ \mu^2 } - D \biggr)
	\Biggr\}, 
\end{align}
where ${}^\prime$ means the derivative with respect to $\varphi_h$ at the EW vacuum like
\begin{align}
m^{2\prime} = \left. \frac{ \partial \tilde{m}^2 }{ \partial \varphi_h } \right|_0, \quad 
m^{2\prime\prime} = \left. \frac{ \partial^2 \tilde{m}^2 }{ \partial \varphi_h^2 } \right|_0. 
\end{align}

Next, we consider the contributions from mixed scalar bosons, where the field-dependent mass is given by a matrix $\tilde{M}$:
\begin{align}
V_\mathrm{mix} = - n \times \frac{ i }{ 2 } \int \frac{ \mathrm{d}^4 p }{ (2\pi)^4 } \log \Bigl(\mathrm{det}[p^2 \cdot I - \tilde{M}^2] \Bigr), 
\end{align}
where $I$ is an identity matrix. 
Then, the third derivative is given by
\begin{align}
\label{eq: third_derivative_mixed_scalar}
V_\mathrm{mix}^{\prime \prime \prime}
	 = \frac{ n }{ 16\pi^2 }
	\Biggl[ & \sum_{a,b,c}
		\mathcal{A}_{ab} \mathcal{A}_{bc} \mathcal{A}_{ca} \mathcal{J}_{abc}^{(1)} 
	+ \frac{3}{2} 
		\sum_{a,b} \mathcal{A}_{ab} \mathcal{B}_{ba} \Bigl( \mathcal{J}_{ab}^{(2)} - D \Bigr)
	\Biggr], 
\end{align}
where $\mathcal{J}_{abc}^{(1)}$ and $\mathcal{J}_{ab}^{(2)}$ are defined as 
\begin{align}
& \mathcal{J}_{abc}^{(1)} = \frac{ 1 }{ (m_a^2 - m_b^2)(m_b^2-m_c^2)(m_c^2-m_a^2) }
		\biggl( m_a^2 m_b^2 \log \frac{m_b^2}{m_a^2}
            \nonumber \\
			& \hspace{210pt} + m_a^2 m_c^2 \log \frac{ m_a^2 }{ m_c^2 }
			+ m_b^2 m_c^2 \log \frac{ m_c^2 }{ m_b^2 }
		\biggr), \\[10pt]
& \mathcal{J}_{ab}^{(2)} = \frac{ f(m_a^2) - f(m_b^2) }{ m_a^2 - m_b^2 },
\end{align}
where $m_a^2$, $m_b^2$, and $m_c^2$ are eigenvalues of $\tilde{M}^2$ at the vacuum, $f(x) = x \bigl\{ \log (x/\mu^2) - 1 \bigr\}$, and $\mathcal{A}$ and $\mathcal{B}$ are given by
\begin{align}
\mathcal{A} = U^\dagger \left. \frac{\partial \tilde{M}^2 }{ \partial \varphi_h } \right|_0 U, \quad 
\mathcal{B} = U^\dagger \left. \frac{\partial^2 \tilde{M}^2 }{ \partial \varphi_h^2 } \right|_0 U,  
\end{align}
where $U$ is a unitary matrix that diagonalizes $\tilde{M}^2$ at the vacuum: $U^\dagger \tilde{M}^2\bigr|_0 U = \mathrm{diag}(m_1^2, m_2^2, \dots)$.
To derive Eq.~(\ref{eq: third_derivative_mixed_scalar}), we have used 
\begin{equation}
\frac{ \partial }{ \partial x } \log \bigl( \mathrm{det} [N] \bigr) = \mathrm{tr}\biggl[ N^{-1} \frac{ \partial N }{ \partial x } \biggr], 
\end{equation}
where $N$ is a regular matrix depending on $x$.

Since the denominators of $\mathcal{J}_{abc}^{(1)}$ and $\mathcal{J}_{ab}^{(2)}$ become zero when they contain the same indices, they must be evaluated in the limit of degenerate masses, like $m^2_b \to m_a^2$, as follows:
\begin{align}
& \mathcal{J}_{abb}^{(1)} = \frac{ 1 }{ (m_a^2 -m_b^2)^2 } \biggl( m_b^2 - m_a^2 - m_a^2 \log \frac{ m_b^2 }{ m_a^2 } \biggr), \\[5pt]
& \mathcal{J}_{aaa}^{(1)} = \frac{ 1 }{ 2m_a^2 }, \\[5pt]
& \mathcal{J}_{aa}^{(2)} = \log \frac{ m_a^2 }{ \mu^2 }. 
\end{align}
By using these formulas, we can show that Eq.~(\ref{eq: third_derivative_mixed_scalar}) reproduces Eq.~(\ref{eq: third_derivative_single_scalar}) in the case that $\tilde{M}^2$ is given by a single component. 

The contribution from the SM fermions and gauge bosons can be derived in the same way except for the difference in the constant $c_i$ in the Coleman--Weinberg type potential for the gauge bosons. 
The total contribution to $\lambda_{hhh}$ from $V_1$ is given by the sum of the third derivatives of all the particles. 
Formulas for first and second derivatives of the field-dependent masses with respect to $\varphi_h$ are summarized in the following subsection. 

We note that the first term of Eq.~(\ref{eq: third_derivative_single_scalar}) and $\mathcal{J}^{(1)}_{aaa}$ diverge in the massless limit. 
Thus, $\lambda_{hhh}$ includes the IR divergence due to the Nambu--Goldstone (NG) boson contributions to $\mathcal{J}^{(1)}$ in the Landau gauge. 
Since we are interested in the deviation from the SM value, we simply neglect these contributions. 

Finally, the third derivative of $\delta V$ is given by
\begin{align}
\left. \frac{\partial^3 }{ \partial \varphi_h^3 } \delta V \right|_0
= & \ 24 c_\alpha^3 v_\Phi \delta \lambda 
	+ 4 s_\alpha^2 c_\alpha v_\Phi \bigl( \delta \kappa_{12} + \delta \kappa_3 + \delta \kappa_4 \bigr)
\nonumber \\[5pt]
& - 2 \sqrt{3} s_\alpha c_\alpha^2
	\Bigl\{ \delta \sigma_1 + \delta \sigma_2 
		+ \bigl(2\lambda_4 + \lambda_5 \bigr) 
			\bigl( 2 \delta v_\chi + \delta v_\xi \bigr) 
		+ 2 v_\Delta^{} \bigl( \delta \kappa_{12} + \delta \kappa_3 + \delta \kappa_4 \bigr)
	\Bigr\}
\nonumber \\[5pt]
& - \frac{ 2 }{ \sqrt{3} } s_\alpha^3 
	\Bigl\{ \delta \sigma_3 + 4 \bigl( 3 \lambda_2 + \lambda_3 \bigr) 
		\bigl( 2 \delta v_\chi + \delta v_\xi \bigr)
	+ 4 v_\Delta \bigl( \delta \rho_{12} + \delta \rho_3 + \delta \rho_4 \bigr)
	\Bigr\}. 
 \end{align}
We have numerically confirmed that divergences in this formula are canceled by those in the third derivative of $V_1$. 

The total one-loop $\lambda_{hhh}$ is given by the sum of the third derivatives of $V_0$, $V_1$, and $\delta V$. 
The SM value is given by taking $\alpha \to 0$ and $\beta \to \pi/2$ and neglecting the additional scalar contributions in $\lambda_{hhh}$.

\subsection{Derivatives of field-dependent masses with respect to $\varphi_h$}

Here, we present formulas for first and second derivatives of field-dependent masses with respect to $\varphi_h$ at the EW vacuum for all the particles in the GM model. 
As used above, ${}^\prime$ represents the derivative with respect to $\varphi_h$ in the following. 

The derivatives are given as follows. 
\begin{itemize}
\item The fermions 
\begin{align}
m_f^{2\prime} = 2 c_\alpha \frac{ m_f^2 }{ v_\Phi }, \quad 
m_f^{2\prime\prime} = 2 c_\alpha^2 \frac{m_f^2}{v_\Phi^2}
\end{align}

\item The gauge bosons 
\begin{align}
m_V^{2\prime} = \frac{ 2m_V^2}{v}
	\Bigl( c_\alpha s_\beta - \sqrt{\frac{8}{3}} s_\alpha c_\beta \Bigr), \quad 
m_V^{2\prime\prime} = \frac{ 2 m_V^2 }{ v^2 }
		\Bigl( c_\alpha^2 + \frac{8}{3} s_\alpha^2 \Bigr), \quad
(V = W, \text{ or } Z). 
\end{align}

\item The doubly charged scalar bosons
\begin{align}
& m_{cc}^{2\prime } = v 
	\Biggl[ \bigl(4\lambda_4 - \lambda_5 \bigr) c_\alpha s_\beta
		+ 2 \sqrt{3}s_\alpha 
		\biggl( \frac{ \mu_2 }{ v } - \sqrt{2} c_\beta (\lambda_2 + \lambda_3 ) \biggr)
	\Biggr], \\[5pt]
& m_{cc}^{2\prime\prime} 
	= c_\alpha^2 \bigl( 4 \lambda_4 - \lambda_5 \bigr)
	+ 8 s_\alpha^2 \bigl( \lambda_2 + \lambda_3 \bigr). 
\end{align}

\item The singly charged scalar bosons
\begin{align}
& (M^2_c)_{11}^\prime = v
	\Biggl[ 8 c_\alpha s_\beta \lambda_1
		+ \frac{ s_\alpha }{ 2 \sqrt{3} }
		\biggl( \frac{ \mu_1 }{ v } 
			- \sqrt{2} c_\beta \bigl( 6 \lambda_4 - \lambda_5 \bigr)
		\biggr)
	\Biggr], \\[5pt]
& (M_c^2)_{12}^\prime = \frac{v}{2}
	\Biggl[
		c_\alpha \biggl( \frac{\mu_1}{v} + \frac{\lambda_5 }{ \sqrt{2} } c_\beta \biggr)
		- \frac{ 2 }{ \sqrt{3} } s_\alpha s_\beta \lambda_5 
	\Biggr], \\[5pt]
& (M_c^2)_{13}^\prime = \frac{v}{2}
	\Biggl[
		c_\alpha \biggl( \frac{\mu_1}{v} +\frac{\lambda_5}{\sqrt{2}} c_\beta \biggr)
		- \frac{ 2 }{ \sqrt{3} } s_\alpha s_\beta \lambda_5
	\Biggr], \\[5pt]
& (M_c^2)_{22}^\prime = v
	\Biggl[
		4 c_\alpha s_\beta \lambda_4 
		- \sqrt{\frac{ 8 }{ 3 }} s_\alpha c_\beta \bigl(3\lambda_2 + 2 \lambda_3 \bigr) 
	\Biggr], \\[5pt]
& (M_c^2)_{23}^\prime = v
	\Biggl[
		c_\alpha s_\beta \lambda_5
		- 2\sqrt{3} s_\alpha 
		\biggl( \frac{\mu_2}{v} - \frac{ \sqrt{2} }{ 3 } c_\beta \lambda_3
		\biggr)
	\Biggr], \\[5pt]
& (M_c^2)_{33}^\prime = v
	\Biggl[
		4 c_\alpha s_\beta \lambda_4
		- \sqrt{\frac{8}{3}}s_\alpha c_\beta
		\bigl( 3 \lambda_2 + 2 \lambda_3 \bigr)
	\Biggr], \\[5pt]
& (M_c^2)_{11}^{\prime \prime}
	= 8 c_\alpha^2 \lambda_1 
		+ \frac{2}{3}s_\alpha^2 (6 \lambda_4 - \lambda_5), \\[5pt]
& (M_c^2)_{12}^{\prime \prime} = -\frac{2}{\sqrt{3}}s_\alpha c_\alpha \lambda_5, \\[5pt]
& (M_c^2)_{13}^{\prime \prime} = - \frac{2}{\sqrt{3}}s_\alpha c_\alpha \lambda_5, \\[5pt]
& (M_c^2)_{22}^{\prime \prime}
	 = 4 c_\alpha^2 \lambda_4
		+ \frac{8}{3}s_\alpha^2 \bigl(3\lambda_2 + 2 \lambda_3 \bigr), \\[5pt]
& (M_c^2)_{23}^{\prime \prime}
	 = c_\alpha^2 \lambda_5 - \frac{8}{3}s_\alpha^2 \lambda_3, \\[5pt]
& (M_c^2)_{33}^{\prime \prime}
	 = 4 c_\alpha^2 \lambda_4 
		+ \frac{8}{3}s_\alpha^2 \bigl(3\lambda_2 + 2\lambda_3 \bigr). 
\end{align}

\item The $CP$-even scalar bosons 

\begin{align}
& (M^2_h)_{11}^\prime = v
	\Biggl[ 24 c_\alpha s_\beta \lambda_1
		- \frac{\sqrt{3}}{2} s_\alpha 
		\biggl( \frac{ \mu_1 }{ v } 
			+ \sqrt{2} c_\beta \bigl( 2 \lambda_4 + \lambda_5 \bigr)
		\biggr)
	\Biggr], \\[5pt]
& (M^2_h)_{12}^\prime = \frac{v}{\sqrt{2}}
	\Biggl[ c_\alpha
		\biggl( \frac{\mu_1}{v} 
			+ \sqrt{2}c_\beta \bigl(2 \lambda_4 + \lambda_5 \bigr)
		\biggr)
		- \frac{4}{\sqrt{3}} s_\alpha s_\beta
		\bigl( 2 \lambda_4 + \lambda_5 \bigr)
	\Biggr], \\[5pt]
& (M^2_h)_{13}^\prime = \frac{v}{2}
	\Biggl[ c_\alpha 
		\biggl( \frac{\mu_1}{v} 
			+ \sqrt{2}c_\beta \bigl(2 \lambda_4 + \lambda_5 \bigr)
		\biggr)
		- \frac{4}{\sqrt{3} } s_\alpha s_\beta
		\bigl( 2 \lambda_4 + \lambda_5 \bigr)
	\Biggr], \\[5pt]
& (M^2_h)_{22}^\prime = v
	\Biggl[ c_\alpha s_\beta \bigl( 4\lambda_4 +\lambda_5 \bigr)
		- 2\sqrt{3} s_\alpha
		\biggl( \frac{\mu_2}{v}
			+ \frac{\sqrt{2}}{3} c_\beta 
			\bigl( 7\lambda_2 + 3 \lambda_3 \bigr)
		\biggr)
	\Biggr], \\[5pt]
& (M^2_h)_{23}^\prime = \sqrt{2} v
	\Biggl[ c_\alpha s_\beta \lambda_5
		- 2 \sqrt{3} s_\alpha
		\biggl( \frac{\mu_2}{v}
			+ \frac{2\sqrt{2}}{3} c_\beta  \lambda_2
		\biggr)
	\Biggr], \\[5pt]
& (M^2_h)_{33}^\prime =  v
	\Biggl[ 4 c_\alpha s_\beta \lambda_4
		- \sqrt{ \frac{8}{3} } s_\alpha c_\beta
		\bigl( 5 \lambda_2 + 3 \lambda_3 \bigr)
	\Biggr], \\[5pt]
& (M_h^2)_{11}^{\prime \prime} 
	= 24 c_\alpha^2 \lambda_1 
		+ 2 s_\alpha^2 \bigl(2\lambda_4 + \lambda_5 \bigr), \\[5pt]
& (M_h^2)_{12}^{\prime \prime} 
	= -4 \sqrt{ \frac{2}{3} } s_\alpha c_\alpha 
		\bigl(2\lambda_4 + \lambda_5 \bigr), \\[5pt]
& (M_h^2)_{13}^{\prime \prime} 
	= - \frac{4}{\sqrt{3}} s_\alpha c_\alpha 
		\bigl(2\lambda_4 + \lambda_5 \bigr), \\[5pt]
& (M_h^2)_{22}^{\prime \prime} 
	= c_\alpha^2 \bigl(4\lambda_4 + \lambda_5 \bigr)
		+ \frac{8}{3}s_\alpha^2 \bigl(7\lambda_2 + 3\lambda_3 \bigr), \\[5pt]
& (M_h^2)_{23}^{\prime \prime} 
	= \sqrt{2} c_\alpha^2 \lambda_5 
		+ \frac{16\sqrt{2}}{3} s_\alpha^2 \lambda_2 , \\[5pt]
& (M_h^2)_{33}^{\prime \prime} 
	= 4 c_\alpha^2 \lambda_4 
		+ \frac{8}{3} s_\alpha^2
		\bigl(5\lambda_2 + 3\lambda_3 \bigr). 
\end{align}

\item The $CP$-odd scalar bosons

\begin{align}
& (M^2_a)_{11}^\prime = v
	\Biggl[ 8 c_\alpha s_\beta \lambda_1
		+ \frac{s_\alpha}{ 2\sqrt{3} } 
		\biggl( \frac{ \mu_1 }{ v } 
			- \sqrt{2} c_\beta \bigl( 6\lambda_4 - \lambda_5 \bigr)
		\biggr)
	\Biggr], \\[5pt]
& (M_a^2)_{12}^{\prime} 
	= \frac{v}{\sqrt{2}} \Biggl[
		c_\alpha 
		\biggl( \frac{\mu_1}{v} + \frac{c_\beta}{\sqrt{2}} \lambda_5 \biggr)
		- \frac{2}{\sqrt{3} } s_\alpha s_\beta \lambda_5 
	\Biggr], \\[5pt]
& (M^2_a)_{22}^\prime = v
	\Biggl[ c_\alpha s_\beta \bigl( 4\lambda_4 +\lambda_5 \bigr)
		- 2\sqrt{3} s_\alpha
		\biggl( \frac{\mu_2}{v}
			+ \frac{\sqrt{2}}{3} c_\beta 
			\bigl( 3\lambda_2 + \lambda_3 \bigr)
		\biggr)
	\Biggr], \\[5pt]
& (M_a^2)_{11}^{\prime \prime} 
	= 8c_\alpha^2 \lambda_1 
		+ \frac{2}{3}s_\alpha^2 \bigl(6\lambda_4-\lambda_5 \bigr), \\[5pt]
& (M_a^2)_{12}^{\prime \prime} 
	= - \sqrt{\frac{8}{3} } s_\alpha c_\alpha \lambda_5, \\[5pt]
& (M_a^2)_{22}^{\prime \prime} 
	= c_\alpha^2 \bigl(4\lambda_4 + \lambda_5 \bigr)
		+ \frac{8}{3}s_\alpha^2 \bigl(3\lambda_2 + \lambda_3 \bigr). 
\end{align}

\end{itemize}

To derive the matrices $\mathcal{A}$ and $\mathcal{B}$ for the mixed scalar particles, we need unitary matrices $U_c$, $U_h$, and $U_a$, which diagonalize the mass matrices for the singly charged, $CP$-even, and $CP$-odd scalars, respectively. 
They are given by
\begin{align}
U_c = \frac{1}{\sqrt{2} }
	\begin{pmatrix}
	\sqrt{2} s_\beta & -\sqrt{2}c_\beta & 0 \\
	c_\beta & s_\beta & 1\\
	c_\beta & s_\beta & -1 \\
	\end{pmatrix}, 
\quad 
U_h = \frac{1 }{ \sqrt{3} }
\begin{pmatrix}
\sqrt{3} c_\alpha & \sqrt{3} s_\alpha & 0 \\
-\sqrt{2} s_\alpha & \sqrt{2} c_\alpha & 1 \\
- s_\alpha & c_\alpha & -\sqrt{2} \\
\end{pmatrix}, 
\quad 
U_a = 
\begin{pmatrix}
s_\beta & -c_\beta \\
c_\beta & s_\beta \\
\end{pmatrix}. 
\end{align}


\begin{thebibliography}{99}

\bibitem{PDG2026}
F.~Takahashi \textit{et al.} [Particle Data Group],
``Review of particle physics,''
Int. J. Mod. Phys. A \textbf{41}, no.22, 2630011 (2026)

\bibitem{type-II_seesaw}
W.~Konetschny and W.~Kummer,
``Nonconservation of Total Lepton Number with Scalar Bosons,''
Phys. Lett. B \textbf{70}, 433-435 (1977); 
R.~N.~Mohapatra and G.~Senjanovic,
``Neutrino Mass and Spontaneous Parity Nonconservation,''
Phys. Rev. Lett. \textbf{44}, 912 (1980);
J.~Schechter and J.~W.~F.~Valle,
``Neutrino Masses in SU(2) x U(1) Theories,''
Phys. Rev. D \textbf{22}, 2227 (1980);
M.~Magg and C.~Wetterich,
``Neutrino Mass Problem and Gauge Hierarchy,''
Phys. Lett. B \textbf{94}, 61-64 (1980);
G.~Lazarides, Q.~Shafi and C.~Wetterich,
``Proton Lifetime and Fermion Masses in an SO(10) Model,''
Nucl. Phys. B \textbf{181}, 287-300 (1981)

\bibitem{GM_model}
H.~Georgi and M.~Machacek,
``DOUBLY CHARGED HIGGS BOSONS,''
Nucl. Phys. B \textbf{262}, 463-477 (1985); 
M.~S.~Chanowitz and M.~Golden,
``Higgs Boson Triplets With M ($W$) = M ($Z$) $\cos \theta \omega$,''
Phys. Lett. B \textbf{165}, 105-108 (1985)



\bibitem{Chiang:2014hia}
C.~W.~Chiang and T.~Yamada,
``Electroweak phase transition in Georgi{\textendash}Machacek model,''
Phys. Lett. B \textbf{735}, 295-300 (2014)
[arXiv:1404.5182 [hep-ph]].

\bibitem{Zhou:2018zli}
R.~Zhou, W.~Cheng, X.~Deng, L.~Bian and Y.~Wu,
``Electroweak phase transition and Higgs phenomenology in the Georgi-Machacek model,''
JHEP \textbf{01}, 216 (2019)
[arXiv:1812.06217 [hep-ph]].

\bibitem{Bian:2019bsn}
L.~Bian, H.~K.~Guo, Y.~Wu and R.~Zhou,
``Gravitational wave and collider searches for electroweak symmetry breaking patterns,''
Phys. Rev. D \textbf{101}, no.3, 035011 (2020)
[arXiv:1906.11664 [hep-ph]].

\bibitem{Chen:2022zsh}
T.~K.~Chen, C.~W.~Chiang, C.~T.~Huang and B.~Q.~Lu,
``Updated constraints on the Georgi-Machacek model and its electroweak phase transition and associated gravitational waves,''
Phys. Rev. D \textbf{106}, no.5, 055019 (2022)
[arXiv:2205.02064 [hep-ph]].

\bibitem{Garcia-Pepin:2016hvs}
M.~Garcia-Pepin and M.~Quiros,
``Strong electroweak phase transition from Supersymmetric Custodial Triplets,''
JHEP \textbf{05}, 177 (2016)
[arXiv:1602.01351 [hep-ph]].

\bibitem{Lu:2025vif}
C.~T.~Lu, Y.~Wu and S.~Xu,
``Dark matter and electroweak phase transition in the Z$_{2}$ symmetric Georgi-Machacek model,''
JHEP \textbf{12}, 155 (2025)
[arXiv:2504.10930 [hep-ph]].

\bibitem{Dolan:1973qd}
L.~Dolan and R.~Jackiw,
``Symmetry Behavior at Finite Temperature,''
Phys. Rev. D \textbf{9}, 3320-3341 (1974)

\bibitem{Chiang:2017vvo}
C.~W.~Chiang, A.~L.~Kuo and K.~Yagyu,
``Radiative corrections to Higgs couplings with weak gauge bosons in custodial multi-Higgs models,''
Phys. Lett. B \textbf{774}, 119-122 (2017)
[arXiv:1707.04176 [hep-ph]].

\bibitem{Chiang:2018xpl}
C.~W.~Chiang, A.~L.~Kuo and K.~Yagyu,
``One-loop renormalized Higgs boson vertices in the Georgi-Machacek model,''
Phys. Rev. D \textbf{98}, no.1, 013008 (2018)
[arXiv:1804.02633 [hep-ph]].

\bibitem{Bando:1992np}
M.~Bando, T.~Kugo, N.~Maekawa and H.~Nakano,
``Improving the effective potential,''
Phys. Lett. B \textbf{301}, 83-89 (1993)
[arXiv:hep-ph/9210228 [hep-ph]];

\bibitem{Parwani:1991gq}
R.~R.~Parwani,
``Resummation in a hot scalar field theory,''
Phys. Rev. D \textbf{45}, 4695 (1992)
[erratum: Phys. Rev. D \textbf{48}, 5965 (1993)]
[arXiv:hep-ph/9204216 [hep-ph]].

\bibitem{Chiang:2018cgb}
C.~W.~Chiang, G.~Cottin and O.~Eberhardt,
``Global fits in the Georgi-Machacek model,''
Phys. Rev. D \textbf{99}, no.1, 015001 (2019)
[arXiv:1807.10660 [hep-ph]].

\bibitem{Bahl:2022igd}
H.~Bahl, T.~Biek{\"o}tter, S.~Heinemeyer, C.~Li, S.~Paasch, G.~Weiglein and J.~Wittbrodt,
``HiggsTools: BSM scalar phenomenology with new versions of HiggsBounds and HiggsSignals,''
Comput. Phys. Commun. \textbf{291}, 108803 (2023)
[arXiv:2210.09332 [hep-ph]].

\bibitem{Wainwright:2011kj}
C.~L.~Wainwright,
``CosmoTransitions: Computing Cosmological Phase Transition Temperatures and Bubble Profiles with Multiple Fields,''
Comput. Phys. Commun. \textbf{183}, 2006-2013 (2012)
[arXiv:1109.4189 [hep-ph]].

\bibitem{Blasi:2017xmc}
S.~Blasi, S.~De Curtis and K.~Yagyu,
``Effects of custodial symmetry breaking in the Georgi-Machacek model at high energies,''
Phys. Rev. D \textbf{96}, no.1, 015001 (2017)
[arXiv:1704.08512 [hep-ph]].

\bibitem{Keeshan:2018ypw}
B.~Keeshan, H.~E.~Logan and T.~Pilkington,
``Custodial symmetry violation in the Georgi-Machacek model,''
Phys. Rev. D \textbf{102}, no.1, 015001 (2020)
[arXiv:1807.11511 [hep-ph]].

\bibitem{Coleman:1973jx}
S.~R.~Coleman and E.~J.~Weinberg,
``Radiative Corrections as the Origin of Spontaneous Symmetry Breaking,''
Phys. Rev. D \textbf{7}, 1888-1910 (1973)

\bibitem{Nielsen:1975fs}
N.~K.~Nielsen,
``On the Gauge Dependence of Spontaneous Symmetry Breaking in Gauge Theories,''
Nucl. Phys. B \textbf{101}, 173-188 (1975)

\bibitem{Fukuda:1975di}
R.~Fukuda and T.~Kugo,
``Gauge Invariance in the Effective Action and Potential,''
Phys. Rev. D \textbf{13}, 3469 (1976)

\bibitem{ref:KE_preparation}
M.~Aoki, K.~Enomoto, S.~Kanemura, S.~Taniguchi,
``The effective potential in the the on-shell renormalization scheme,''
in preparation

\bibitem{Bando:1992wy}
M.~Bando, T.~Kugo, N.~Maekawa and H.~Nakano,
``Improving the effective potential: Multimass scale case,''
Prog. Theor. Phys. \textbf{90}, 405-418 (1993)
[arXiv:hep-ph/9210229 [hep-ph]].

\bibitem{Lee:1977yc}
B.~W.~Lee, C.~Quigg and H.~B.~Thacker,
``The Strength of Weak Interactions at Very High-Energies and the Higgs Boson Mass,''
Phys. Rev. Lett. \textbf{38}, 883-885 (1977);
``Weak Interactions at Very High-Energies: The Role of the Higgs Boson Mass,''
Phys. Rev. D \textbf{16}, 1519 (1977)

\bibitem{Luscher:1988gc}
M.~Luscher and P.~Weisz,
``Is There a Strong Interaction Sector in the Standard Lattice Higgs Model?,''
Phys. Lett. B \textbf{212}, 472-478 (1988)

\bibitem{Chen:2023ins}
T.~K.~Chen, C.~W.~Chiang and K.~Yagyu,
``CP violation in a model with Higgs triplets,''
JHEP \textbf{06}, 069 (2023)
[erratum: JHEP \textbf{07}, 169 (2023)]
[arXiv:2303.09294 [hep-ph]].

\bibitem{Aoki:2007ah}
M.~Aoki and S.~Kanemura,
``Unitarity bounds in the Higgs model including triplet fields with custodial symmetry,''
Phys. Rev. D \textbf{77}, no.9, 095009 (2008)
[erratum: Phys. Rev. D \textbf{89}, no.5, 059902 (2014)]
[arXiv:0712.4053 [hep-ph]].

\bibitem{Hartling:2014zca}
K.~Hartling, K.~Kumar and H.~E.~Logan,
``The decoupling limit in the Georgi-Machacek model,''
Phys. Rev. D \textbf{90}, no.1, 015007 (2014)
[arXiv:1404.2640 [hep-ph]].

\bibitem{Degrande:2017naf}
C.~Degrande, K.~Hartling and H.~E.~Logan,
``Scalar decays to $\gamma\gamma$, $Z\gamma$, and $W\gamma$ in the Georgi-Machacek model,''
Phys. Rev. D \textbf{96}, no.7, 075013 (2017)
[erratum: Phys. Rev. D \textbf{98}, no.1, 019901 (2018)]
[arXiv:1708.08753 [hep-ph]].

\bibitem{Djouadi:2005gi}
A.~Djouadi,
``The Anatomy of electro-weak symmetry breaking. I: The Higgs boson in the standard model,''
Phys. Rept. \textbf{457}, 1-216 (2008)
[arXiv:hep-ph/0503172 [hep-ph]].

\bibitem{Alwall:2014hca}
J.~Alwall, R.~Frederix, S.~Frixione, V.~Hirschi, F.~Maltoni, O.~Mattelaer, H.~S.~Shao, T.~Stelzer, P.~Torrielli and M.~Zaro,
``The automated computation of tree-level and next-to-leading order differential cross sections, and their matching to parton shower simulations,''
JHEP \textbf{07}, 079 (2014)
[arXiv:1405.0301 [hep-ph]].

\bibitem{Bechtle:2020pkv}
P.~Bechtle, D.~Dercks, S.~Heinemeyer, T.~Klingl, T.~Stefaniak, G.~Weiglein and J.~Wittbrodt,
``HiggsBounds-5: Testing Higgs Sectors in the LHC 13 TeV Era,''
Eur. Phys. J. C \textbf{80}, no.12, 1211 (2020)
[arXiv:2006.06007 [hep-ph]].

\bibitem{Numerical_Recipes}
W.~H.~Press, S.~A.~Teukolsky, W.~T.~Vetterling, B.~P.~Flannery,
``Numerical Recipes: The Art of Scientific Computing (3rd Edition),''
Cambridge University Press (2007)

\bibitem{Kanemura:2004ch}
S.~Kanemura, Y.~Okada and E.~Senaha,
``Electroweak baryogenesis and quantum corrections to the triple Higgs boson coupling,''
Phys. Lett. B \textbf{606}, 361-366 (2005)
[arXiv:hep-ph/0411354 [hep-ph]].

\bibitem{Weinberg:1974hy}
S.~Weinberg,
``Gauge and Global Symmetries at High Temperature,''
Phys. Rev. D \textbf{9}, 3357-3378 (1974)

\bibitem{CMS:2021kom}
A.~M.~Sirunyan \textit{et al.} [CMS],
``Measurements of Higgs boson production cross sections and couplings in the diphoton decay channel at $ \sqrt{\mathrm{s}} $ = 13 TeV,''
JHEP \textbf{07}, 027 (2021)
[arXiv:2103.06956 [hep-ex]].

\bibitem{ATLAS:2026pdi}
G.~Aad \textit{et al.} [ATLAS],
``Combination of Higgs boson measurements at $\sqrt{s} =$ 13 TeV and their interpretations by the ATLAS experiment,''
[arXiv:2608.07332 [hep-ex]].

\bibitem{Cepeda:2019klc}
M.~Cepeda, S.~Gori, P.~Ilten, M.~Kado, F.~Riva, R.~Abdul Khalek, A.~Aboubrahim, J.~Alimena, S.~Alioli and A.~Alves, \textit{et al.}
``Report from Working Group 2: Higgs Physics at the HL-LHC and HE-LHC,''
CERN Yellow Rep. Monogr. \textbf{7}, 221-584 (2019)
[arXiv:1902.00134 [hep-ph]].

\bibitem{CMS:2022ahq}
A.~Tumasyan \textit{et al.} [CMS],
``Search for Higgs boson decays to a Z boson and a photon in proton-proton collisions at $ \sqrt{s} $ = 13 TeV,''
JHEP \textbf{05}, 233 (2023)
doi:10.1007/JHEP05(2023)233
[arXiv:2204.12945 [hep-ex]].

\bibitem{ATLAS:2022faz}
 [ATLAS],
``HL-LHC prospects for the measurement of Higgs boson pair production in the $b\bar{b}b\bar{b}$ final state and combination with the $b\bar{b}\gamma\gamma$ and $b\bar{b}\tau^+\tau^-$ final states at the ATLAS experiment,''
ATL-PHYS-PUB-2022-053.

\bibitem{Torndal:2023fky}
J.~M.~Torndal and J.~List,
``Higgs self-coupling measurement at the International Linear Collider,''
[arXiv:2307.16515 [hep-ph]].

\bibitem{ref:HPAIR}
T.~Plehn, M.~Spira and P.~M.~Zerwas,
``Pair production of neutral Higgs particles in gluon-gluon collisions,''
Nucl. Phys. B \textbf{479}, 46-64 (1996)
[erratum: Nucl. Phys. B \textbf{531}, 655-655 (1998)]
[arXiv:hep-ph/9603205 [hep-ph]];
S.~Dawson, S.~Dittmaier and M.~Spira,
``Neutral Higgs boson pair production at hadron colliders: QCD corrections,''
Phys. Rev. D \textbf{58}, 115012 (1998)
[arXiv:hep-ph/9805244 [hep-ph]];
R.~Grober, M.~Muhlleitner, M.~Spira and J.~Streicher,
``NLO QCD Corrections to Higgs Pair Production including Dimension-6 Operators,''
JHEP \textbf{09}, 092 (2015)
[arXiv:1504.06577 [hep-ph]]; 
%
The program can be downloaded at \url{https://ltpth.pages.psi.ch/tiger/}.

\bibitem{LISA:2017pwj}
P.~Amaro-Seoane \textit{et al.} [LISA],
``Laser Interferometer Space Antenna,''
[arXiv:1702.00786 [astro-ph.IM]].

\bibitem{Seto:2001qf}
N.~Seto, S.~Kawamura and T.~Nakamura,
``Possibility of direct measurement of the acceleration of the universe using 0.1-Hz band laser interferometer gravitational wave antenna in space,''
Phys. Rev. Lett. \textbf{87}, 221103 (2001)
[arXiv:astro-ph/0108011 [astro-ph]].

\bibitem{Corbin:2005ny}
V.~Corbin and N.~J.~Cornish,
``Detecting the cosmic gravitational wave background with the big bang observer,''
Class. Quant. Grav. \textbf{23}, 2435-2446 (2006)
[arXiv:gr-qc/0512039 [gr-qc]].

\bibitem{Caprini:2015zlo}
C.~Caprini, M.~Hindmarsh, S.~Huber, T.~Konstandin, J.~Kozaczuk, G.~Nardini, J.~M.~No, A.~Petiteau, P.~Schwaller and G.~Servant, \textit{et al.}
``Science with the space-based interferometer eLISA. II: Gravitational waves from cosmological phase transitions,''
JCAP \textbf{04}, 001 (2016)
[arXiv:1512.06239 [astro-ph.CO]].

\bibitem{Athron:2023xlk}
P.~Athron, C.~Bal{\'a}zs, A.~Fowlie, L.~Morris and L.~Wu,
``Cosmological phase transitions: From perturbative particle physics to gravitational waves,''
Prog. Part. Nucl. Phys. \textbf{135}, 104094 (2024)
[arXiv:2305.02357 [hep-ph]].

\bibitem{Nicolis:2003tg}
A.~Nicolis,
``Relic gravitational waves from colliding bubbles and cosmic turbulence,''
Class. Quant. Grav. \textbf{21}, L27 (2004)
[arXiv:gr-qc/0303084 [gr-qc]].

\bibitem{Grojean:2006bp}
C.~Grojean and G.~Servant,
``Gravitational Waves from Phase Transitions at the Electroweak Scale and Beyond,''
Phys. Rev. D \textbf{75}, 043507 (2007)
[arXiv:hep-ph/0607107 [hep-ph]].

\bibitem{Anderson:1991zb}
G.~W.~Anderson and L.~J.~Hall,
``The Electroweak phase transition and baryogenesis,''
Phys. Rev. D \textbf{45}, 2685-2698 (1992)

\bibitem{Moore:1995si}
G.~D.~Moore and T.~Prokopec,
``How fast can the wall move? A Study of the electroweak phase transition dynamics,''
Phys. Rev. D \textbf{52}, 7182-7204 (1995)
[arXiv:hep-ph/9506475 [hep-ph]].

\bibitem{vandeVis:2025plm}
J.~van de Vis, P.~Schicho, L.~Niemi, B.~Laurent, J.~Hirvonen and O.~Gould,
``WallGo investigates: Theoretical uncertainties in the bubble wall velocity,''
JHEP \textbf{04}, 041 (2026)
[arXiv:2510.27691 [hep-ph]].

\bibitem{Kosowsky:1991ua}
A.~Kosowsky, M.~S.~Turner and R.~Watkins,
``Gravitational Radiation from Colliding Vacuum Bubbles,''
Phys. Rev. D \textbf{45}, 4514-4535 (1992)

\bibitem{Kosowsky:1992rz}
A.~Kosowsky, M.~S.~Turner and R.~Watkins,
``Gravitational Waves from First Order Cosmological Phase Transitions,''
Phys. Rev. Lett. \textbf{69}, 2026-2029 (1992)

\bibitem{Kosowsky:1992vn}
A.~Kosowsky and M.~S.~Turner,
``Gravitational radiation from colliding vacuum bubbles: envelope approximation to many bubble collisions,''
Phys. Rev. D \textbf{47}, 4372-4391 (1993)
[arXiv:astro-ph/9211004 [astro-ph]].

\bibitem{Kamionkowski:1993fg}
M.~Kamionkowski, A.~Kosowsky and M.~S.~Turner,
``Gravitational radiation from first order phase transitions,''
Phys. Rev. D \textbf{49}, 2837-2851 (1994)
[arXiv:astro-ph/9310044 [astro-ph]].

\bibitem{Caprini:2007xq}
C.~Caprini, R.~Durrer and G.~Servant,
``Gravitational wave generation from bubble collisions in first-order phase transitions: An analytic approach,''
Phys. Rev. D \textbf{77}, 124015 (2008)
[arXiv:0711.2593 [astro-ph]].

\bibitem{Huber:2008hg}
S.~J.~Huber and T.~Konstandin,
``Gravitational Wave Production by Collisions: More Bubbles,''
JCAP \textbf{09}, 022 (2008)
[arXiv:0806.1828 [hep-ph]].

\bibitem{Hindmarsh:2013xza}
M.~Hindmarsh, S.~J.~Huber, K.~Rummukainen and D.~J.~Weir,
``Gravitational waves from the sound of a first order phase transition,''
Phys. Rev. Lett. \textbf{112}, 041301 (2014)
[arXiv:1304.2433 [hep-ph]].

\bibitem{Giblin:2013kea}
J.~T.~Giblin, Jr. and J.~B.~Mertens,
``Vacuum Bubbles in the Presence of a Relativistic Fluid,''
JHEP \textbf{12}, 042 (2013)
[arXiv:1310.2948 [hep-th]].

\bibitem{Giblin:2014qia}
J.~T.~Giblin and J.~B.~Mertens,
``Gravitional radiation from first-order phase transitions in the presence of a fluid,''
Phys. Rev. D \textbf{90}, no.2, 023532 (2014)
[arXiv:1405.4005 [astro-ph.CO]].

\bibitem{Hindmarsh:2015qta}
M.~Hindmarsh, S.~J.~Huber, K.~Rummukainen and D.~J.~Weir,
``Numerical simulations of acoustically generated gravitational waves at a first order phase transition,''
Phys. Rev. D \textbf{92}, no.12, 123009 (2015)
[arXiv:1504.03291 [astro-ph.CO]].

\bibitem{Caprini:2006jb}
C.~Caprini and R.~Durrer,
``Gravitational waves from stochastic relativistic sources: Primordial turbulence and magnetic fields,''
Phys. Rev. D \textbf{74}, 063521 (2006)
[arXiv:astro-ph/0603476 [astro-ph]].

\bibitem{Kahniashvili:2008pf}
T.~Kahniashvili, A.~Kosowsky, G.~Gogoberidze and Y.~Maravin,
``Detectability of Gravitational Waves from Phase Transitions,''
Phys. Rev. D \textbf{78}, 043003 (2008)
[arXiv:0806.0293 [astro-ph]].

\bibitem{Kahniashvili:2008pe}
T.~Kahniashvili, L.~Campanelli, G.~Gogoberidze, Y.~Maravin and B.~Ratra,
``Gravitational Radiation from Primordial Helical Inverse Cascade MHD Turbulence,''
Phys. Rev. D \textbf{78}, 123006 (2008)
[erratum: Phys. Rev. D \textbf{79}, 109901 (2009)]
[arXiv:0809.1899 [astro-ph]].

\bibitem{Kahniashvili:2009mf}
T.~Kahniashvili, L.~Kisslinger and T.~Stevens,
``Gravitational Radiation Generated by Magnetic Fields in Cosmological Phase Transitions,''
Phys. Rev. D \textbf{81}, 023004 (2010)
[arXiv:0905.0643 [astro-ph.CO]].

\bibitem{Caprini:2009yp}
C.~Caprini, R.~Durrer and G.~Servant,
``The stochastic gravitational wave background from turbulence and magnetic fields generated by a first-order phase transition,''
JCAP \textbf{12}, 024 (2009)
[arXiv:0909.0622 [astro-ph.CO]].

\bibitem{Kisslinger:2015hua}
L.~Kisslinger and T.~Kahniashvili,
``Polarized Gravitational Waves from Cosmological Phase Transitions,''
Phys. Rev. D \textbf{92}, no.4, 043006 (2015)
[arXiv:1505.03680 [astro-ph.CO]].

\bibitem{Ellis:2020awk}
J.~Ellis, M.~Lewicki and J.~M.~No,
``Gravitational waves from first-order cosmological phase transitions: lifetime of the sound wave source,''
JCAP \textbf{07}, 050 (2020)
[arXiv:2003.07360 [hep-ph]].

\bibitem{Guo:2020grp}
H.~K.~Guo, K.~Sinha, D.~Vagie and G.~White,
``Phase Transitions in an Expanding Universe: Stochastic Gravitational Waves in Standard and Non-Standard Histories,''
JCAP \textbf{01}, 001 (2021)
[arXiv:2007.08537 [hep-ph]].

\bibitem{Espinosa:2010hh}
J.~R.~Espinosa, T.~Konstandin, J.~M.~No and G.~Servant,
``Energy Budget of Cosmological First-order Phase Transitions,''
JCAP \textbf{06}, 028 (2010)
[arXiv:1004.4187 [hep-ph]].

\bibitem{Leitao:2015fmj}
L.~Leitao and A.~Megevand,
``Gravitational waves from a very strong electroweak phase transition,''
JCAP \textbf{05}, 037 (2016)
[arXiv:1512.08962 [astro-ph.CO]].

\bibitem{Athron:2022mmm}
P.~Athron, C.~Bal{\'a}zs and L.~Morris,
``Supercool subtleties of cosmological phase transitions,''
JCAP \textbf{03}, 006 (2023)
[arXiv:2212.07559 [hep-ph]].

\bibitem{Ellis:2018mja}
J.~Ellis, M.~Lewicki and J.~M.~No,
``On the Maximal Strength of a First-Order Electroweak Phase Transition and its Gravitational Wave Signal,''
JCAP \textbf{04}, 003 (2019)
[arXiv:1809.08242 [hep-ph]].

\bibitem{Du:2024mry}
X.~Du and F.~Wang,
``Positive definiteness constraints of effective scalar potential in Georgi{\textendash}Machacek model,''
Eur. Phys. J. C \textbf{86}, no.1, 40 (2026)
[arXiv:2409.20198 [hep-ph]].

\bibitem{Machacek:1983fi}
M.~E.~Machacek and M.~T.~Vaughn,
``Two Loop Renormalization Group Equations in a General Quantum Field Theory. 2. Yukawa Couplings,''
Nucl. Phys. B \textbf{236}, 221-232 (1984)

\bibitem{Hamada:2015bra}
Y.~Hamada, K.~Kawana and K.~Tsumura,
``Landau pole in the Standard Model with weakly interacting scalar fields,''
Phys. Lett. B \textbf{747}, 238-244 (2015)
[arXiv:1505.01721 [hep-ph]].

\bibitem{Espinosa:1992kf}
J.~R.~Espinosa, M.~Quiros and F.~Zwirner,
``On the nature of the electroweak phase transition,''
Phys. Lett. B \textbf{314}, 206-216 (1993)
[arXiv:hep-ph/9212248 [hep-ph]].

\end{thebibliography}
\end{document}